\documentclass[nofootinbib,preprintnumbers,amsmath,amssymb]{revtex4}
\newcommand{\PRE}[1]{}       % Use if journal style
\usepackage{epsfig}

\newcommand{\postscript}[2]{\setlength{\epsfxsize}{#2\hsize}
   \centerline{\epsfbox{#1}}}

\def\eslt{\not\!\!{E_T}}
\def\to{\rightarrow}

\def\bi{\begin{itemize}}
\def\ei{\end{itemize}}
\def\tchi{\tilde\chi}
\def\te{\tilde e}

\def\tu{\tilde u}
\def\ta{\tilde a}

\def\tb{\tilde b}

\def\tst{\tilde t}

\def\tmu{\tilde \mu}
\def\tg{\tilde g}

\def\tell{\tilde\ell}
\def\tq{\tilde q}
\def\tw{\widetilde\chi}
\def\tz{\widetilde\chi^0}
\def\alt{\stackrel{<}{\sim}}
\def\agt{\stackrel{>}{\sim}}
\def\be{\begin{equation}}  
\def\ee{\end{equation}}  
\def\bea{\begin{eqnarray}}  
\def\eea{\end{eqnarray}}  
\newcommand\plb[3]{{Phys.\ Lett.\ }{\bf B #1} (#2) #3}

\begin{document}

\preprint{OU-HEP-191231}

\title{
\PRE{\vspace*{1.5in}}
Midi-review: Status of weak scale supersymmetry\\ 
after LHC Run 2 and ton-scale noble liquid WIMP searches
\PRE{\vspace*{0.3in}}
}

\author{Howard Baer}
\affiliation{Dept. of Physics and Astronomy,
University of Oklahoma, Norman, OK, 73019, USA
\PRE{\vspace*{.1in}}
}
\author{Vernon Barger}
\affiliation{Dept. of Physics,
University of Wisconsin, Madison, WI 53706, USA
\PRE{\vspace*{.1in}}
}
\author{Dibyashree Sengupta}
\affiliation{Dept. of Physics and Astronomy,
University of Oklahoma, Norman, OK, 73019, USA
\PRE{\vspace*{.1in}}
}
\author{Shadman Salam}
\affiliation{Dept. of Physics and Astronomy,
University of Oklahoma, Norman, OK, 73019, USA
\PRE{\vspace*{.1in}}
}
\author{Kuver Sinha}
\affiliation{Dept. of Physics and Astronomy,
University of Oklahoma, Norman, OK, 73019, USA
\PRE{\vspace*{.1in}}
}

%\date{October 15, 2011}

\begin{abstract}
\noindent
After completion of LHC Run 2, the ATLAS and CMS experiments had collected of 
order 139 fb$^{-1}$ of data at $\sqrt{s}=13$ TeV. 
While discovering a very Standard Model-like Higgs boson of mass 
$m_h\simeq 125$ GeV, no solid signal for physics beyond the Standard Model 
has emerged so far at LHC. 
In addition, no WIMP signals have emerged so far at ton-scale noble liquid
WIMP search experiments.
For the case of weak scale supersymmetry (SUSY), 
which is touted as a simple and elegant solution to the gauge hierarchy problem
and likely low energy limit of compactified string theory,
LHC has found rather generally that gluinos are beyond about 2.2 TeV whilst
top squark must lie beyond 1.1 TeV. 
These limits contradict older simplistic notions of naturalness that 
emerged in the 1980s-1990s, leading to the rather pessimistic view that
SUSY is now excluded except for perhaps some remaining narrow corners of 
parameter space. 
Yet, this picture ignores several important developments in SUSY/string 
theory that emerged in the 21st century: 
1. the emergence of the string theory landscape and its solution to the 
cosmological constant problem, 2. a more nuanced view of naturalness 
including the notion of ``stringy naturalness'', 
3. the emergence of anomaly-free discrete $R$-symmetries and their 
connection to $R$-parity, Peccei-Quinn symmetry, the SUSY $\mu$ problem 
and proton decay and 4. the importance of including a solution to 
the strong CP problem.
Rather general considerations from the string theory landscape favor
large values of soft terms, subject to the vacuum selection criteria 
that electroweak symmetry is properly broken (no charge and/or color breaking (CCB)
 minima) and the
resulting magnitude of the weak scale is not too far from our measured value.
Then stringy naturalness predicts a Higgs mass $m_h\sim 125$ GeV whilst 
sparticle masses are typically lifted beyond present LHC bounds.
In light of these refinements in theory perspective confronted 
by LHC and dark matter search results,
we review the most likely LHC, ILC and dark matter signatures that are
expected to arise from weak scale SUSY as we understand it today.
\end{abstract}
%12.60.-i   Models beyond the standard model
%95.35.+d   Dark matter

\maketitle
\newpage
\vspace{-1.2cm}
\tableofcontents
\newpage
\section{Introduction}
\label{sec:intro}

\subsection{Why SUSY?}

\begin{figure}[bp]
\begin{center}
\includegraphics[height=0.25\textheight]{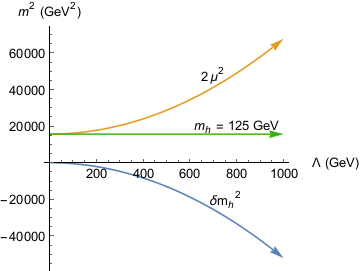}
\caption{Plot of measured Higgs mass squared along with radiative
correction and tree-level term $2\mu^2$. For a given value of 
$\Lambda$, the $\mu^2$ term must be adjusted (fine-tuned) to 
guarantee that $m_h=125$ GeV.
\label{fig:mhSM}}
\end{center}
\end{figure}

The discovery in 2012 of the Higgs boson with mass $m_h\simeq 125$ GeV 
by the ATLAS\cite{atlas_h} and CMS\cite{cms_h} collaborations
at LHC seemingly completes the Standard Model (SM),
and yet brings with it a puzzle.
It was emphasized as early as 1978 by Wilson and Susskind\cite{techni1}
that fundamental scalar particles are unnatural in quantum field theory.
In the case of the SM Higgs boson with a doublet of Higgs scalars $\phi$
and Higgs potential given by
\be
V=-\mu^2\phi^\dagger\phi +\lambda (\phi^\dagger\phi )^2 ,
\ee
one expects a physical Higgs boson mass value
\be
m_h^2\simeq 2\mu^2 +\delta m_h^2 ,
\label{eq:mhSM}
\ee
where the leading radiative correction is given by
\be
\delta m_h^2\simeq \frac{3}{4\pi^2}\left(-\lambda_t^2+\frac{g^2}{4}+
\frac{g^2}{8\cos^2\theta_W} +\lambda \right)\Lambda^2 .
\label{eq:radcor}
\ee
In the above expression, $\lambda_t$ is the top quark Yukawa coupling,
$g$ is the $SU(2)$ gauge coupling and $\lambda$ is the Higgs field
quartic coupling. The quantity $\Lambda$ is the UV energy cutoff to
otherwise divergent loop integrals. Taking $\Lambda$ as high as
the reduced Planck mass $m_P\simeq 2.4\times 10^{18}$ GeV would require
a tuning of $\mu^2$ to $30$ decimal places to maintain the
measured value of $m_h^2$. 
Alternatively, the notion of: 
\begin{quotation}
{\bf practical naturalness:}
that {\it independent} contributions to any observable ${\cal O}$ 
be comparable to or less than ${\cal O}$,
\end{quotation}
then requires that loop integrals be truncated at $\Lambda\sim 1$ TeV.
The situation is plotted in Fig. \ref{fig:mhSM}: as $\Lambda$ increases,
then the free parameter $\mu^2$ must be finely-tuned to large
opposite-sign values so as to maintain $m_h$ at its measured value.
Such fine-tunings are regarded as symptomatic of some missing
ingredient in the theory which, were it present, would render the
theory natural.

In Eq. \ref{eq:radcor}, various divergences appear involving the 
various fermion Yukawa couplings, the electroweak (EW) gauge couplings 
and the Higgs self-coupling $\lambda$. 
The unique solution which tames all these divergences at once
is the inclusion of $N=1$ supersymmetry (SUSY) into the theory\cite{witten}. 
SUSY extends the Poincare spacetime group of symmetries to its more
general structure, the super-Poincare group, which includes anti-commutation
relations as well as commutators. Under SUSY, fields are elevated to 
superfields which then express the Fermi-Bose symmetry inherent in the theory.
Supersymmetrization of the SM to the well-behaved Minimal Supersymmetric
Standard Model\cite{WSS} (MSSM) requires an additional Higgs doublet to 
cancel triangle anomalies and to give mass to all the SM 
quarks and leptons under EW symmetry breaking. 
In the MSSM, then all quadratic divergences neatly cancel,
leaving only log divergences. 
Since the log of a large number can be a small number, the Higgs mass 
instability is tamed and the weak scale can co-exist with higher mass
scales: $m_{PQ}$, $m_{GUT}$, $m_{string}$ etc. Inclusion of soft SUSY
breaking terms can lift the predicted sparticles to the TeV scale 
in accord with constraints from collider searches. 
Under inclusion of $R$-parity conservation, the lightest SUSY
particle (LSP) is stable and if it is electrically and color neutral, 
then it may be a good weakly interacting massive particle (WIMP)
dark matter candidate.
The MSSM with global, broken SUSY is expected to be the low energy effective
theory of more encompassing local SUSY (supergravity) theories which in
turn are the low energy effective theory expected from 
compactified string theory.

\begin{figure}[tbp]
\includegraphics[height=0.23\textheight]{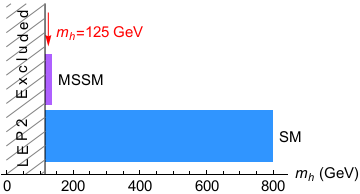}
\caption{
Range of Higgs mass $m_h$ predicted in the Standard Model compared to range
of Higgs mass predicted by the MSSM.
We also show the measured value of the Higgs mass by the arrow.
The left-most region had been excluded by LEP2 searches prior to
the LHC8 run.
\label{fig:mh}}
\end{figure}

While SUSY elegantly solves the gauge hierarchy problem, it is 
actually supported by four sets of data via radiative corrections.
\bi
\item The measured values of the three SM gauge couplings, when extrapolated
to the grand unification scale $m_{GUT}\sim 2\times 10^{16}$ GeV, 
meet at a point under renormalization group (RG) evolution\cite{RGEs}; 
this is not so in the SM or other beyond-the-SM (BSM) extensions.
\item In the MSSM at the weak scale, EW symmetry is not expected to be
broken using generic values for the soft SUSY breaking terms. Under RG
evolution from some high scale (such as $m_{GUT}$), then the large value 
of the top Yukawa coupling drives the soft term $m_{H_u}^2$ to negative
values causing EW symmetry to appropriately break\cite{rewsb}. 
This would not happen if the top mass $m_t\alt 100$ GeV.
\item The value of the newly discovered Higgs boson $m_h\simeq 125$ GeV
falls neatly within the narrow allowed window of MSSM values
$115$ GeV $<m_h\alt 135$ GeV, but only
if radiative corrections from the top-squark sector are 
large enough\cite{mhiggs}. 
Such a high value of $m_h$ is consistent with highly mixed 
TeV-scale top squarks which are beyond current LHC reach.
In the SM, no particular range of $m_h$ is preferred other than that
$m_h\alt 1$ TeV from unitarity constraints: see Fig. \ref{fig:mh}.
\item Precision EW calculation of $m_W$ vs. $m_t$ actually prefer 
the MSSM with heavy ($\agt 1$ TeV) SUSY particles over the SM\cite{sven}.
\ei

It is hard to believe the consistency of all these radiative effects with 
the existence of weak scale SUSY (WSS) is just a coincidence.
Historically, radiative corrections have been a reliable guide
to new physics.
It is important to remember that many new particles 
($W$, $Z$ bosons, top quark, Higgs boson etc.) have been reliably 
presaged by radiative corrections well before actual discovery: 
so may it be with SUSY.

\subsection{Where are the sparticles? LHC Run 2 SUSY search results}

\begin{figure}[tbp]
\postscript{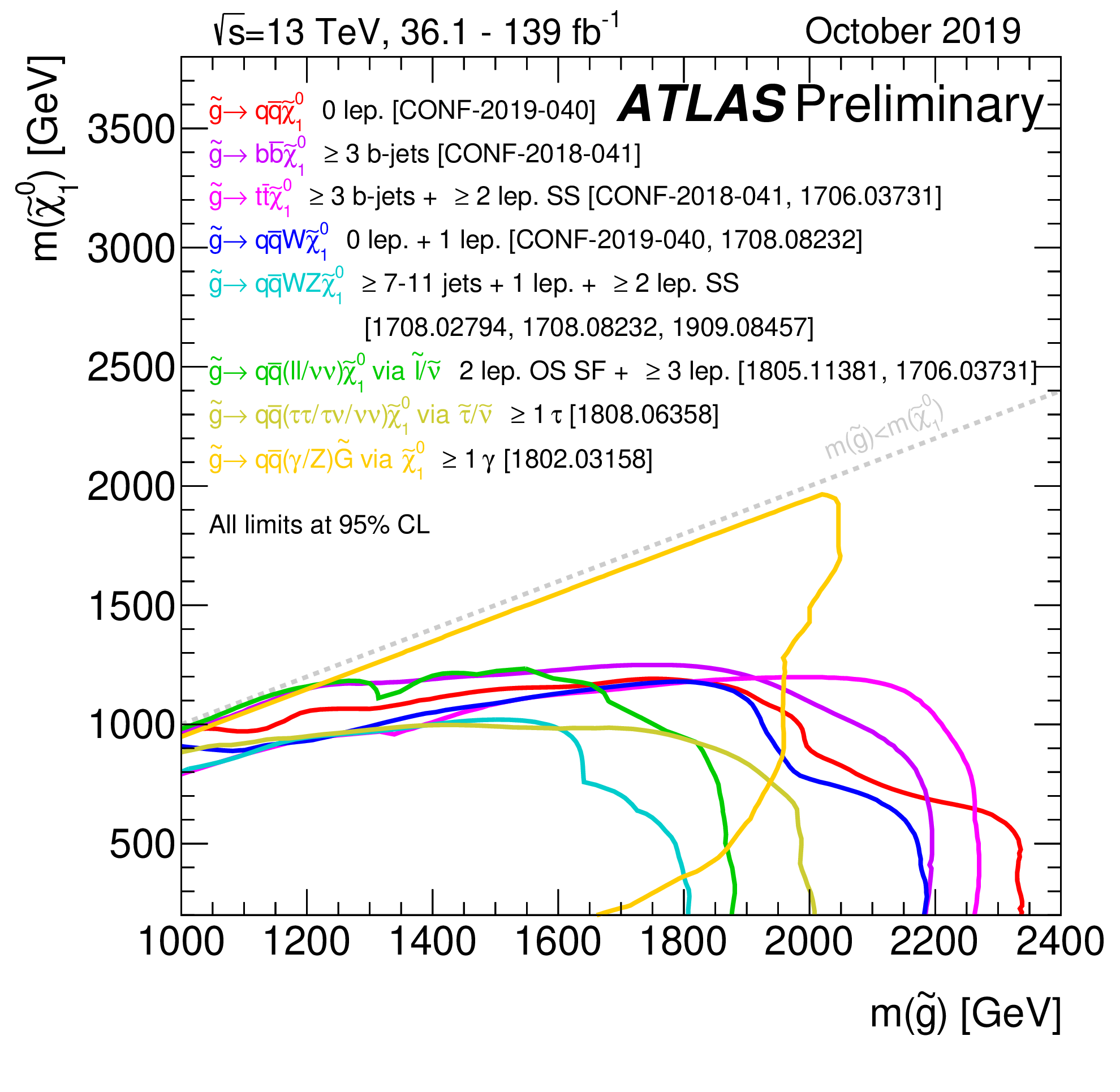}{0.8}
\caption{Results of ATLAS searches for gluino pair production in SUSY
for various simplified models with up to 139 fb$^{-1}$ of data
at $\sqrt{s}=13$ TeV.
\label{fig:Atlas_mgl}}
\end{figure}

The question du jour is then: where are the predicted sparticles and
where are the expected WIMPs?
In Fig. \ref{fig:Atlas_mgl}, we show recent 95\% CL search limits for gluino
pair production within various simplified models as deduced by the
ATLAS experiment\cite{atlas_mgl}. 
The data sets vary from 36-139 fb$^{-1}$ at
$\sqrt{s}=13$ TeV. 
The plot is made in the $m_{\tg}$ vs. $m_{\tz_1}$ mass plane. 
From the plot, we see that for relatively light values of
$m_{\tz_1}\alt 500$ GeV, then the approximate bound from LHC
searches is that $m_{\tg}\agt 2.2$ TeV. 
Limits from CMS are comparable\cite{cms_mgl}.

\begin{figure}[tbp]
\postscript{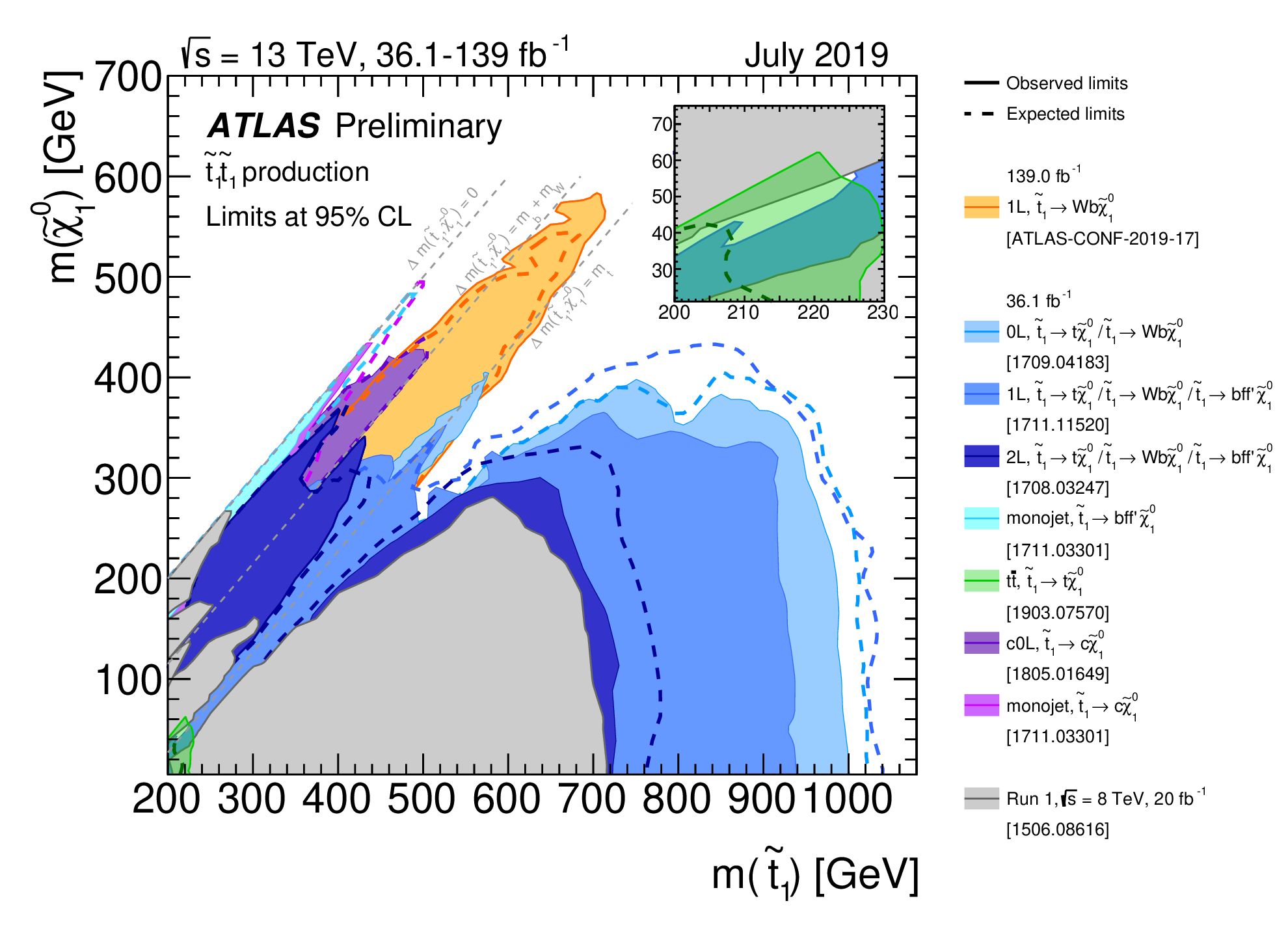}{0.8}
\caption{Results of ATLAS searches for top squark pair production in SUSY
for various simplified models with up to 139 fb$^{-1}$ of data
at $\sqrt{s}=13$ TeV.
\label{fig:Atlas_mt1}}
\end{figure}

In Fig. \ref{fig:Atlas_mt1}, we show similar limits on searches for
top-squark pair production in the $m_{\tst_1}$ vs. $m_{\tz_1}$ plane
for various simplified models with again 36-139 fb$^{-1}$ of
integrated luminosity at $\sqrt{s}=13$ TeV. For $m_{\tz_1}\alt 300$ GeV,
then it is required that $m_{\tst_1}\agt 1$ TeV\cite{atlas_mt1,cms_mt1}.

Many other searches for SUSY particles have been undertaken by ATLAS and CMS.
A recent comprehensive review of LHC SUSY searches 
has been presented by Canepa\cite{Canepa:2019hph}. 
Suffice it to say: so far, no compelling evidence for SUSY has emerged at LHC.

\subsection{Where are the WIMPs?}
\begin{figure}[tbp]
\postscript{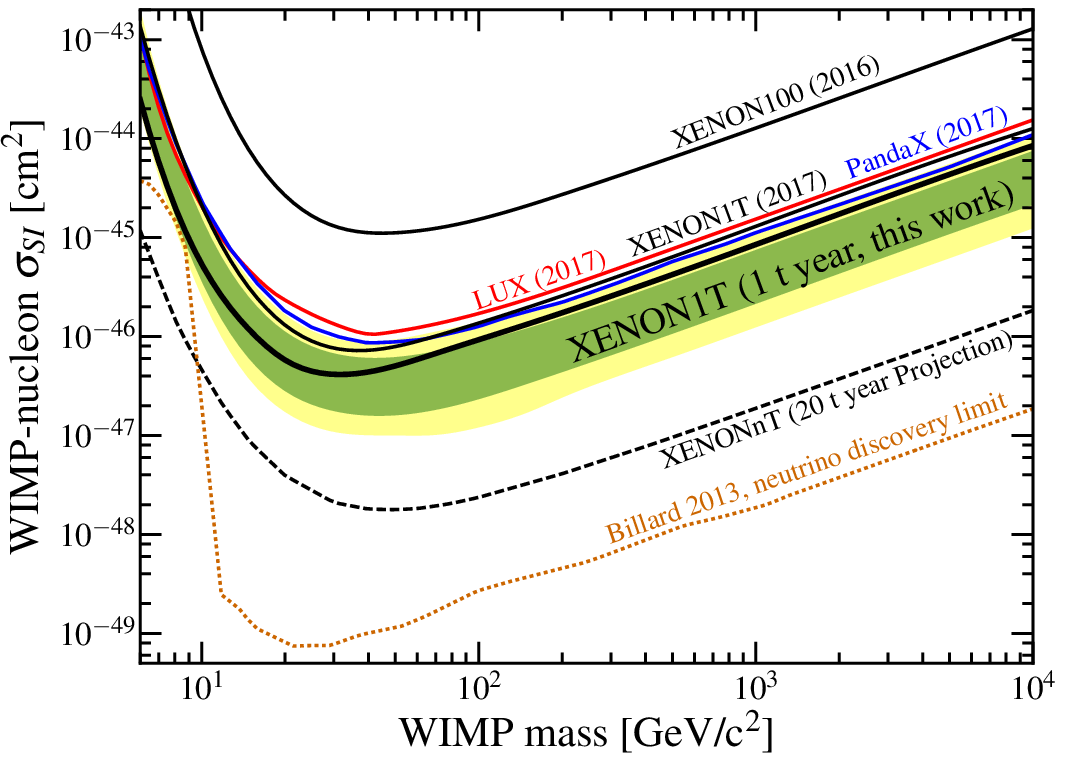}{0.8}
\caption{Results from year-long spin-independent (SI) WIMP-Xe scattering
search by Xe-1ton experiment\cite{Aprile:2018dbl} along with results from 
LUX and PandaX.
\label{fig:Xe1ton}}
\end{figure}

Along with non-appearance of sparticles at LHC, we must also be concerned
with the as-yet non-appearance of WIMPs at direct and/or indirect
WIMP detection experiments. The current limits from the Xe-1ton
experiment are shown in Fig. \ref{fig:Xe1ton}\cite{Aprile:2018dbl}. 
Here, the limits are placed in spin-independent (SI) 
WIMP-nucleon scattering cross section
$\sigma^{SI}(\tz_1 p)$ vs. $m_{\tz_1}$ plane. Limits from Xe100, 
LUX (2017), PandaX (2017), Xe-1ton (2017) and Xe-1ton (1-ton-year exposure)
are shown. At present, the latter limit is strongest and for a 
100 GeV WIMP excludes $\sigma^{SI}(\tz_1p)\agt 10^{-10}$ pb. 
For comparison, the popular hyperbolic branch/focus-point\cite{hb_fp} 
(HB/FP) and many models with well-tempered neutralinos\cite{wtn} predicted
a direct detection cross section $\sigma^{SI}(\tz_1 p)\sim 10^{-8}$ pb,
relatively independent of $m_{\tz_1}\sim 0.1-1$ TeV. 
Thus, these popular models are excluded by 1-2 orders of magnitude
(depending on the value of $m_{\tz_1}$).
\subsection{Comparison to expectations from naturalness}

The concept of naturalness can provide upper bounds on Higgs boson and
sparticle masses. The results depend strongly on the definition of
naturalness which is used. In Table \ref{tab:BG}, we list sparticle mass
bounds derived from Ref. \cite{BG} using the 
$\Delta_{BG}\equiv max_i |\frac{\partial \ln m_Z^2}{\partial\ln p_i}|$ measure
with $\Delta_{BG}<10$, corresponding to $\Delta_{BG}^{-1}=10\%$ 
fine-tuning. The $p_i$ are taken as fundamental parameters of the theory, 
which in this case are the various soft terms and $\mu$ parameter from 
the mSUGRA/CMSSM\cite{cmssm} model. From Table~\ref{tab:BG}, 
we see upper limits
of $m_{\tg}\alt 400$ GeV while most other sparticles are not too far from
the weak scale (defined as $m_{weak}\simeq m_{W,Z,h}\sim 100$ GeV). 
What is immediately of note is that current LHC gluino mass bounds are 
a factor five beyond the naturalness limits. Also, bounds on chargino masses 
from LEP2 ($m_{\tw_1^\pm}>103.5$ GeV) were already barely above the BG 
naturalness bounds. In the Table, we also list 10\% $\Delta_{BG}$ 
bounds on $m_h<115$ GeV from Ref. \cite{CGR}. 
For $m_h\sim 125$ GeV, then $\Delta_{BG}$ rapidly rises to 1000, 
or 0.1\% fine-tuning. 
The final entry in Table \ref{tab:BG} comes from Refs. \cite{PRW} and 
\cite{BKLS}. Using a different measure (labeled in Sec. \ref{sec:nat}
as $\Delta_{HS}$), the authors derive that {\it three} third generation
squarks $\tst_{1,2}$ and $\tb_1$ should all lie below about 500 GeV. 
While one third generation squark might hide in Fig. \ref{fig:Atlas_mt1},
it is hard to envision three hiding on the same plot.
\begin{table}[!htb]
\renewcommand{\arraystretch}{1.2}
\begin{center}
\begin{tabular}{|c|cc|}
\hline
mass & upper limit & source \\
\hline
 $m_{\tg}$ & $<400$ GeV & BG(1987) \\
\hline
$m_{\tu_R}$ & $<400$ GeV & BG(1987) \\
\hline
 $m_{\te_R}$ & $<350$ GeV & BG(1987) \\
\hline
$m_{\tchi_1^\pm}$ & $<100$ GeV & BG(1987) \\
\hline
$m_{\tchi_1^0}$ & $<50$ GeV & BG(1987) \\
\hline
$m_h$ & $<115$ GeV & CGR(2009) \\
\hline
$m_{\tst_{1,2},\tb_1}$ & $<500$ GeV & PRW,BKLS(2011) \\
\hline
\end{tabular}
\caption{Upper bounds on sparticle and Higgs boson masses from 
10\% naturalness using $\Delta_{BG}$ within multi-parameter 
SUSY effective theories, from Ref. \cite{BG} (BG1987) 
and Ref. \cite{CGR} (CGR2009). We also include bounds from
$\Delta_{HS}$ from Refs. \cite{PRW} and \cite{BKLS} (PRW,BKLS2011).
}
\label{tab:BG}
\end{center}
\end{table} 

Taken all together, the first conclusion from comparing LHC Higgs mass 
measurements and sparticle mass limits to Table \ref{tab:BG}, one might
draw a rather pessimistic conclusion regarding SUSY. 
It is that an apparent mass gap has opened up between the weak scale and 
the sparticle mass scale known as the Little Hierarchy problem (LHP):
while SUSY solves the Big Hierarchy problem, a LHP has appeared due to the
strong limits from LHC data. The emergence of the LHP has engendered 
growing skepticism that the common notion of SUSY with weak scale
sparticles is nature's solution to the hierarchy problems.

\subsection{SUSY: from cartoon to paradigm}

In this midi-review (between a mini-review and a review), 
we will argue that the above pessimistic conclusion is too 
strong, and is based on an overly simplistic notion of weak scale SUSY
that is relatively unchanged since the 1980s.
In fact, several developments have emerged since the year $2000$ that have 
changed the paradigm notion of how SUSY might appear. 
These include the following.
\bi
\item Improved scrutiny of the notion of naturalness and naturalness
measures shows that many of the early notions of naturalness are in need
of revision. 
In particular, the model independent electroweak measure $\Delta_{EW}$ 
has emerged\cite{ltr}. Under $\Delta_{EW}$,
then a modified SUSY paradigm arises with higgsinos rather than gauginos
as the lightest electroweakinos. 
Under $\Delta_{EW}$, other sparticle mass limits are lifted by 
factors of 2-50 beyond the early projections from Table \ref{tab:BG}.
This has important consequences for collider searches and for the picture of
SUSY dark matter. An updated discussion of naturalness is the topic of 
Sec. \ref{sec:nat}.
\item The intertwining of the SUSY $\mu$ problem\cite{mu}, the strong CP
problem and the role of the axion in SUSY theories forms the topic of
Sec. \ref{sec:mu}. 
The role of discrete $R$-symmetries\cite{lrrrssv2} is discussed 
which helps to simultaneously solve the SUSY $\mu$ problem and proton-decay problem. 
In addition, both $R$-parity and the global Peccei-Quinn (PQ) 
$U(1)_{PQ}$ needed for an axionic solution 
to the strong CP problem can emerge from the strongest of these, 
a ${\bf Z}_{24}^R$. In this case, then dark matter would be composed of
two particles: a mixture of higgsino-like WIMPs and DFSZ-like axions
with suppressed couplings to photons.
\item Starting in 2001, it was realized that the multitude of
string theory vacua\cite{BP,Susskind:2003kw} provided a setting for 
Weinberg's anthropic solution to the cosmological constant problem\cite{Weinberg:1987dv}. 
Rather general stringy considerations of the so-called ``landscape'' 
of vacua solutions also suggest a statistical preference for 
{\it large soft terms} from the multiverse\cite{Douglas:2004qg}. 
This stringy statistical draw must be compensated for
by requiring that the derived value for the weak scale in each pocket universe 
of the multiverse be not too far from our measured value, so that 
complex nuclei and hence atoms arise in any anthropically allowed pocket 
universe\cite{Agrawal:1997gf}. 
By combining these notions, then it is seen that the Higgs mass
$m_h\sim 125$ GeV is statistically favored while sparticle masses are
drawn beyond LHC search limits\cite{land}. 
Under such a {\it stringy natural} setting,
a 3 TeV gluino is more natural than a 300 GeV gluino\cite{Baer:2019cae}.
\ei
We compare the predictions of landscape SUSY sparticle mass spectra to
those of several other prominent string phenomenology constructs
in Sec. \ref{sec:compare}.

After addressing the above issues, then we briefly summarize the conclusions
as to how SUSY is likely to arise at LHC upgrades and ILC in Sec. \ref{sec:colliders}.
In Sec. \ref{sec:DM}, we summarize expectations for mixed axion/WIMP 
dark matter and explain why so far  no WIMPs have emerged at 
direct/indirect detection experiments. 
In Sec. \ref{sec:baryo}, we briefly summarize several
compelling scenarios for baryogenesis in SUSY models.
Our overall summary and big picture is presented in Sec. \ref{sec:conclude}.

\section{Naturalness re-examined}
\label{sec:nat}

In this Section, we make a critical assessment of several common naturalness
measures found in the literature.\footnote{
Some recent model scans of $\Delta_{BG}$ and $\Delta_{EW}$ and associated
DM and collider phenomenology can be found in Refs.~\cite{vanBeekveld:2016hug}.} 
We then follow up with revised upper bounds on sparticle masses
arising from clarification of electroweak naturalness in SUSY models.

\subsection{$\Delta_{EW}$: electroweak naturalness}

The simplest naturalness measure $\Delta_{EW}$\cite{ltr,rns} 
arises from the form of the 
Higgs potential in the MSSM. 
By minimizing the weak-scale SUSY Higgs potential, including radiative
corrections, one may relate the measured value of the $Z$-boson mass
to the various SUSY contributions:
\bea
m_Z^2/2 &=&\frac{m_{H_d}^2+\Sigma_d^d-(m_{H_u}^2+\Sigma_u^u)\tan^2\beta}
{\tan^2\beta -1}-\mu^2\\ \nonumber
&\simeq &-m_{H_u}^2-\mu^2-\Sigma_u^u(\tst_{1,2}) .
\label{eq:mzs}
\eea
The measure
\begin{equation}
\Delta_{EW}=|(max\ RHS\ contribution)|/(m_Z^2/2)
\label{eq:DEW}
\end{equation}
is then low provided all {\it weak-scale} contributions to $m_Z^2/2$
are comparable to or less than $m_Z^2/2$, in accord with practical naturalness.
The $\Sigma_u^u$ and $\Sigma_d^d$ contain over 40 radiative corrections
which are listed in the Appendix of Ref.~\cite{rns}.
The conditions for natural SUSY (for {\it e.g.} $\Delta_{EW}<30$)\footnote{
The onset of finetuning for $\Delta_{EW}\agt 30$ is visually displayed in
Fig. 1 of Ref. \cite{upper}.}
can then be read off from Eq. \ref{eq:mzs}:
\begin{itemize}
\item The superpotential $\mu$ parameter has magnitude not too far from the
weak scale, $|\mu |\alt 300$ GeV\cite{ccn,Baer:2011ec}.
This implies the existence of light higgsinos $\tz_{1,2}$ and
$\tw_1^\pm$ with $m(\tz_{1,2},\tw_1^\pm)\sim 100-300$ GeV.
\item $m_{H_u}^2$ is radiatively driven from large high scale values to
{\it small} negative values at the weak scale
(this is SUSY with {\it radiatively-driven naturalness} or RNS\cite{ltr}).
\item Large cancellations occur in the $\Sigma_u^u (\tst_{1,2})$ terms for 
large $A_t$ parameters which then allow for $m_{\tst_1}\sim 1-3$ TeV
for $\Delta_{EW}<30$.
The large $A_t$ term gives rise to large mixing in the top-squark sector 
and thus lifts the Higgs mass $m_h$ into the vicinity of 125 GeV.
The gluino contribution to the weak scale is at two-loop order so its mass can
range up to $m_{\tg}\alt 6$ TeV with little cost to 
naturalness\cite{rns,upper,lhc27}.
\item Since first/second generation squarks and sleptons contribute to the
weak scale at one-loop through (mainly cancelling) $D$-terms and at 
two-loops via RGEs, they can range up to 10-30 TeV with 
little cost to naturalness (thus helping to alleviate the 
SUSY flavor and CP problems)\cite{maren,flavor}.
\end{itemize}
Since $\Delta_{EW}$ is determined by the weak scale SUSY parameters,
then different models which give rise to exactly the same sparticle mass
spectrum will have the same fine-tuning value (model independence).
Using the naturalness measure $\Delta_{EW}$, then it has been shown
that plenty of SUSY parameter space remains natural even in the face of
LHC Run 2 Higgs mass measurements and sparticle mass limits\cite{rns}.

\subsection{$\Delta_{HS}$: tuning dependent contributions}

It is also common in the literature to apply practical naturalness to the
Higgs mass:
\begin{equation}
m_h^2\simeq m_{H_u}^2(weak)+\mu^2(weak)+mixing+rad.\ corr.
\label{eq:mhs}
\end{equation}
where the mixing and radiative corrections are both comparable to $m_h^2$.
Also, in terms of some high energy cut-off scale (HS) $\Lambda$, then 
$m_{H_u}^2(weak)=m_{H_u}^2(\Lambda )+\delta m_{H_u}^2$
where it is common to estimate $\delta m_{H_u}^2$ using its renormalization
group equation (RGE) by setting several terms in $dm_{H_u}^2/dt$
(with $t=\log Q^2$) to zero so as to integrate in a single step:
\begin{equation}
\delta m_{H_u}^2\sim -\frac{3f_t^2}{8\pi^2}(m_{Q_3}^2+m_{U_3}^2+A_t^2)
\ln\left( \Lambda^2/m_{soft}^2\right).
\end{equation}
Taking $\Lambda\sim m_{GUT}$ and requiring the high scale measure
\begin{equation}
\Delta_{HS}\equiv\delta m_{H_u}^2/m_h^2
\label{eq:DHS}
\end{equation}
$\Delta_{HS}\alt 1$ then requires three third generation squarks lighter
than 500 GeV\cite{PRW,BKLS} (now highly excluded by LHC top-squark searches)
and small $A_t$ terms (whereas $m_h\simeq 125$ GeV typically requires
large mixing and thus multi-TeV values of $A_0$\cite{mhiggs,h125}).
The simplifications made in this calculation ignore the fact that
$\delta m_{H_u}^2$ is highly dependent on $m_{H_u}^2(\Lambda )$
(which is set to zero in the simplification)\cite{dew,seige,arno}.
In fact, the larger one makes $m_{H_u}^2(\Lambda )$,
then the larger becomes the cancelling correction $\delta m_{H_u}^2$.
Thus, these terms are {\it not independent}:
one cannot tune $m_{H_u}^2(\Lambda )$
against a large contribution $\delta m_{H_u}^2$.
Thus, weak-scale top squarks and small $A_t$ are not required by naturalness.

\subsection{$\Delta_{BG}$: the problem with parameters}

The more traditional measure $\Delta_{BG}$ was proposed by 
Ellis {\it et al.}\cite{eenz}
and later investigated more thoroughly by Barbieri and Giudice\cite{BG}.
The starting point is to express $m_Z^2$ in terms of weak scale SUSY parameters
as in Eq. \ref{eq:mzs}:
\be
m_Z^2 \simeq -2m_{H_u}^2-2\mu^2
\label{eq:mZsapprox}
\ee
where the partial equality obtains for moderate-to-large $\tan\beta$ 
values and where we assume for now that the radiative corrections are small.
An advantage of $\Delta_{BG}$ over the previous large-log measure is that 
it maintains the correlation between $m_{H_u}^2(\Lambda )$ and 
$\delta m_{H_u}^2$ by replacing $m_{H_u}^2 (m_{weak})= 
\left( m_{H_u}^2(\Lambda )+\delta m_{H_u}^2\right)$ by its expression in
terms of high scale parameters.
To evaluate $\Delta_{BG}$, one needs to know the explicit dependence of 
$m_{H_u}^2$ and $\mu^2$ on the fundamental parameters.
Semi-analytic solutions to the one-loop renormalization group equations
for $m_{H_u}^2$ and $\mu^2$ can be found for instance in Refs. \cite{munoz}.
For the case of $\tan\beta =10$, then\cite{abe,martin,feng}
\bea
m_Z^2& \simeq & -2.18\mu^2 + 3.84 M_3^2+0.32M_3M_2+0.047 M_1M_3 \nonumber \\
& & -0.42 M_2^2+0.011 M_2M_1-0.012M_1^2-0.65 M_3A_t \nonumber \\
& & -0.15 M_2A_t-0.025M_1 A_t+0.22A_t^2+0.004 M_3A_b\nonumber \\
& &-1.27 m_{H_u}^2 -0.053 m_{H_d}^2\nonumber \\
& &+0.73 m_{Q_3}^2+0.57 m_{U_3}^2+0.049 m_{D_3}^2 -0.052 m_{L_3}^2+0.053 m_{E_3}^2\nonumber \\
& &+0.051 m_{Q_2}^2-0.11 m_{U_2}^2+0.051 m_{D_2}^2 -0.052 m_{L_2}^2+0.053 m_{E_2}^2\nonumber \\
& &+0.051 m_{Q_1}^2-0.11 m_{U_1}^2+0.051 m_{D_1}^2 -0.052 m_{L_1}^2+0.053 m_{E_1}^2 ,
\label{eq:mZsparam}
\eea
where all terms on the right-hand-side are understood to be 
$GUT$ scale parameters.

Then, the proposal is that the variation in $m_Z^2$ with respect to
parameter variation be small:
\be
\Delta_{BG}\equiv max_i\left[ c_i\right]\ \ {\rm where}\ \
c_i=\left|\frac{\partial\ln m_Z^2}{\partial\ln p_i}\right|
=\left|\frac{p_i}{m_Z^2}\frac{\partial m_Z^2}{\partial p_i}\right|
\label{eq:DBG}
\ee
where the $p_i$ constitute the fundamental parameters of the model.
Thus, $\Delta_{BG}$ measures the fractional change in $m_Z^2$ 
due to fractional variation in the high scale parameters $p_i$.
The $c_i$ are known as {\it sensitivity coefficients}\cite{feng}.

The requirement of low $\Delta_{BG}$ is then equivalent to the 
requirement of no large cancellations on the right-hand-side of 
Eq. \ref{eq:mZsparam} since (for linear terms)
the logarithmic derivative just picks off coefficients of the 
relevant parameter. 
For instance, $c_{m_{Q_3}^2}=0.73\cdot (m_{Q_3}^2/m_Z^2)$. 
If one allows $m_{Q_3}\sim 3$ TeV (in accord with
requirements from the measured value of $m_h$), 
then one obtains $c_{m_{Q_3}^2}\sim 800$
and so $\Delta_{BG}\ge 800$. 
In this case, SUSY would be electroweak fine-tuned to about 0.1\%.
If instead one sets $m_{Q_3}=m_{U_3}=m_{H_u}\equiv m_0$ 
as in models with scalar mass universality, then the various
scalar mass contributions to $m_Z^2$ largely cancel and 
$c_{m_0^2}\sim -0.017 m_0^2/m_Z^2$:
the contribution to $\Delta_{BG}$ from scalars drops by a factor $\sim 50$.

The above argument illustrates the extreme model-dependence of 
$\Delta_{BG}$ for multi-parameter SUSY models.
The value of $\Delta_{BG}$ can change radically from theory to theory 
even if those theories generate exactly the same weak scale 
sparticle mass spectrum: see Table \ref{tab:param}. 
The model dependence of $\Delta_{BG}$ arises due to a violation 
of the definition of practical naturalness: 
one must combine dependent terms into independent quantities
before evaluating EW fine-tuning\cite{dew,mt,seige,arno}.
\begin{table}[!htb]
\renewcommand{\arraystretch}{1.2}
\begin{center}
\begin{tabular}{c|c}
model & $\Delta_{BG}$ \\
\hline
nuhm2 & 984 \\
\hline
mSUGRA/CMSSM & 41 \\
\hline
DDSB($m_{3/2}$) & 29.4 \\
\hline
pMSSM & 28.9 \\
\hline
\end{tabular}
\caption{Values of $\Delta_{BG}$ for various hypothetical
effective SUSY theories leading to the exact same weak scale
spectrum. We take $m_0=3500$ GeV, $m_{1/2}=300$ GeV, $A_0=0$
and $\tan\beta =10$ with $\mu =330.6$ GeV and $m_A=3468$ GeV.
The corresponding value of $\Delta_{EW}$ is 32.7.
The DDSB stands for the one-soft-parameter ($=m_{3/2}$) 
dilaton-dominated SUSY breaking model.
}
\label{tab:param}
\end{center}
\end{table}

\subsection{Some natural SUSY models: NUHM2, NUHM3, nGMM and nAMSB}

A fairly reliable prediction of natural SUSY models is that the four 
higgsinos $\tw_1^\pm$ and $\tz_{1,2}$ lie at the bottom of the 
SUSY particle mass spectra with mass values $\sim \mu\alt 200-300$ GeV. 
However, even this prediction can be upset by models with non-universal 
gaugino masses where for instance the gluino is still beyond LHC bounds 
but where the bino mass $M_1$ and/or the wino mass $M_2$ is 
comparable to or lighter than  $\mu$\cite{Baer:2015tva}. 
In addition, there are several theory motivated models which all give
rise to natural SUSY spectra with $\Delta_{EW}\alt 30$. These
include:
\bi
\item The two- or three- extra parameter non-universal Higgs models,
NUHM2 or NUHM3\cite{nuhm2}. 
These models are slight generalizations of the CMSSM/mSUGRA model\cite{cmssm} 
where gaugino masses are unified to $m_{1/2}$ at the GUT scale but
where the soft Higgs masses $m_{H_u}$ and $m_{H_d}$ are instead independent of
the matter scalar soft masses $m_0$. 
This is well justified since the Higgs superfields necessarily 
live in different GUT multiplets than the matter superfields. 
In the NUHM3 model, it is further assumed that the third generation 
matter scalars are split from the first two generation $m_0(1,2)\ne m_0(3)$.
In these models, typically the parameter freedom in $m_{H_u}$ and $m_{H_d}$
is traded for the more convenient weak scale parameters $\mu$ and $m_A$.
\item The original minimal anomaly-mediated SUSY breaking model\cite{amsb} 
(mAMSB) now seems excluded since wino-only dark matter should have been
detected by indirect dark matter 
searches\cite{Cohen:2013ama,Fan:2013faa,Baer:2016ucr}. Also, in mAMSB 
the anomaly-mediated contribution to the trilinear soft term $A$ is usually
too small to boost the Higgs mass $m_h\to 125$ Gev unless stop masses
lie in the hundred-TeV range. Finally, the mAMSB model typically 
has a large $\mu$ term. The latter two situations lead to mAMSB being highly
unnatural, especially if $m_h\simeq 125$ GeV is required.

In the original Randall-Sundrum paper, the authors suggest additional
bulk contributions to scalar masses to solve the problem of tachyonic 
sleptons. If the bulk contributions to $m_{H_u}^2$ and $m_{H_d}^2$ are 
non-universal with the matter scalars, then one can allow for a small 
natural $\mu$ term. Also, if bulk contributions to the $A$ terms are allowed,
(as suggested in the Randall-Sundrum paper), 
then large stop mixing can occur which both reduces the $\Sigma_u^u(\tst_{1,2})$
terms in Eq. \ref{eq:mzs} while lifting $m_h\to 125$ GeV. 
In that case, natural AMSB models can be generated with small 
$\Delta_{EW}<30$ and with $m_h\simeq 125$ GeV\cite{nAMSB}.
The phenomenology of {\it natural} AMSB (nAMSB) is quite different from
mAMSB: in nAMSB, the higgsinos are the lightest electroweakinos so one
has a higgsino-like LSP even though the winos are still the lightest gauginos. 
Axions are assumed to make up the bulk of dark matter\cite{Bae:2015rra}.
\item The scheme of mirage-mediation (MM) posits soft SUSY breaking terms
which are suppressed compared to the gravitino mass $m_{3/2}$ so that
moduli/gravity mediated contributions to soft terms are comparable to
AMSB contributions\cite{choi}. The original MM calculation of soft terms
within the context of KKLT moduli stabilization with a single K\"ahler 
modulus (stabilized by non-perturbative contributions) in type-IIB string 
models with D-branes depended on discrete choices for modular weights.
These original MM models have been shown to be unnatural under LHC Higgs
mass and sparticle limit constraints\cite{seige}.
However, in more realistic compactifications with many K\"ahler moduli, 
then a more general framework where the discrete modular weights are replaced
by continuous parameters is called for. The resulting 
generalized mirage-mediation model (GMM) maintains the phenomena of
mirage unification of gaugino masses while allowing the flexibility of
generating $m_h\simeq 125$ GeV while maintaining naturalness in the face
of LHC sparticle mass limits. In {\it natural} GMM models (nGMM)\cite{nGMM}, 
the gaugino spectrum is still compressed as in usual MM, but now 
the higgsinos lie at the bottom of the spectra. Consequently, 
the collider and dark matter phenomenology is modified from 
previous expectations. In the nGMM$^{\prime}$ model, the continuous
parameters $c_{H_u}$ and $c_{H_d}$ 
(which used to depend on discrete modular weights) can be traded as in NUHM2,3
for the more convenient weak scale parameters $\mu$ and $m_A$.
\ei

\begin{figure}[tbp]
\includegraphics[height=0.4\textheight]{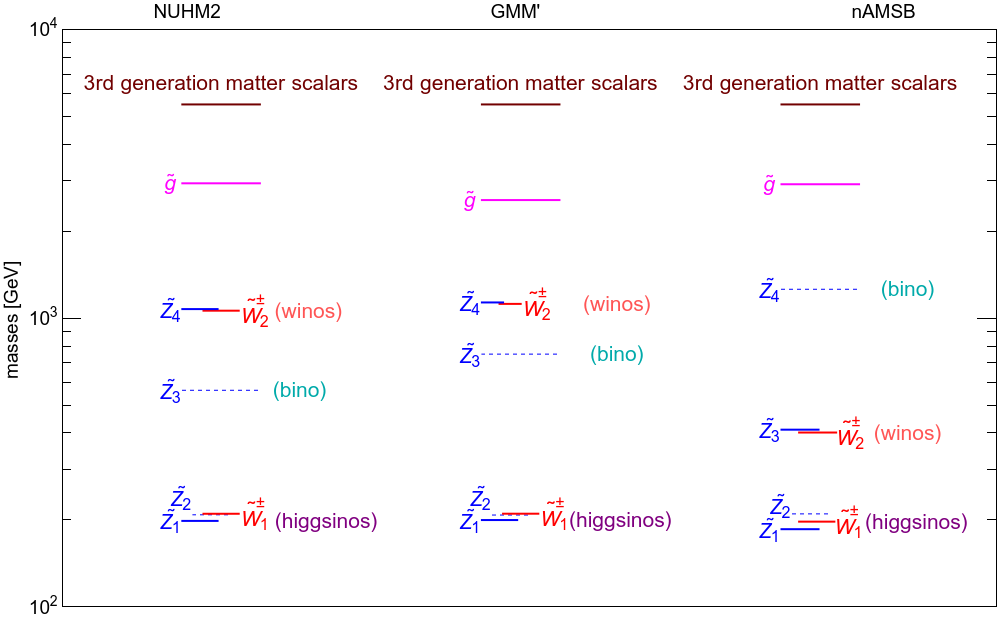}
\caption{Typical mass spectra from natural SUSY in the case of
NUHM2 (with gaugino mass unification), nGMM with mirage unification
and compressed gauginos and natural AMSB where the wino is the lightest
gaugino. In all cases, the higgsinos lie at the bottom of the spectra.
\label{fig:3models}}
\end{figure}

A schematic sketch of the three spectra fron NUHM2, nGMM$^{\prime}$ and
nAMSB is shown in Fig. \ref{fig:3models}. 
The models are hardwired in the Isajet SUSY spectrum generator 
Isasugra\cite{isajet}. 

\subsection{Conclusions on naturalness}

\begin{figure}[tbp]
\begin{center}
\includegraphics[height=0.3\textheight]{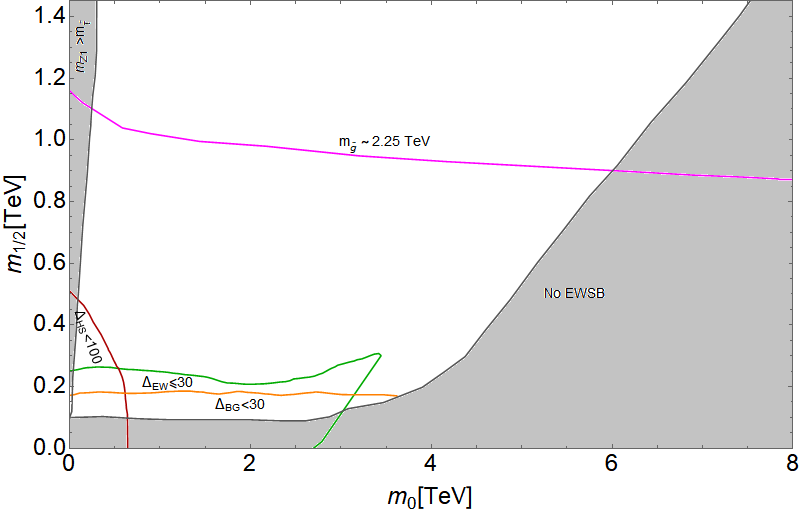}
\caption{The $m_0$ vs. $m_{1/2}$ plane of the mSUGRA/CMSSM model
with $A_0=0$ and $\tan\beta =10$. In this parameter space $m_h < 122 $ GeV.
We show contours of various finetuning measures along with LEP2 and LHC Run 2 search limits (from Ref. \cite{Baer:2019cae}).
\label{fig:cmssm}}
\end{center}
\end{figure}

In Fig. \ref{fig:cmssm}, we compare the three aforementioned fine-tuning 
measures in the $m_0$ vs. $m_{1/2}$ plane of the mSUGRA/CMSSM model for
$A_0=0$ and $\tan\beta =10$. In this plane, the Higgs mass $m_h$ is
always well below $125$ GeV unless one proceeds to far larger values of
$m_0$ and $m_{1/2}$. Also, the $\mu$ parameter is always large except in the
HB/FP region near the edge of the right-side ``no EWSB'' disallowed 
region. The contour $\Delta_{HS}<100$  favors the low
$m_0$ and $m_{1/2}$ corner and disallows $m_0\agt 0.7$ TeV. 
The BG measure $\Delta_{BG}<30$ boundary is roughly flat with $m_0$ 
variation which shows that heavy squarks, including top-squarks, 
can still be natural under this measure. The $\Delta_{EW}<30$ region is
denoted by the green contour and is roughly flat with $m_0$ variation since
the contours of fixed $\mu$ values (not shown) are also flat with $m_0$ 
variation. The curve cuts off around $m_0\sim 3$ TeV when the 
radiative corrections $\Sigma_u^u(\tst_{1,2})$ become large. 
Note that all measures favor small $m_0$ and $m_{1/2}$ (in contrast to
stringy naturalness introduced in Sec. \ref{sec:land}). For comparison, we
show the LHC contour $m_{\tg}=2.25$ TeV (magenta) where the region
below the contour is excluded by LHC gluino pair searches. This picture
presents a rather pessimistic view of SUSY. 
However, one must remember for such parameter choices within the mSUGRA
model even the Higgs mass doesn't match its measured value.

\begin{figure}[tbp]
\begin{center}
\includegraphics[height=0.3\textheight]{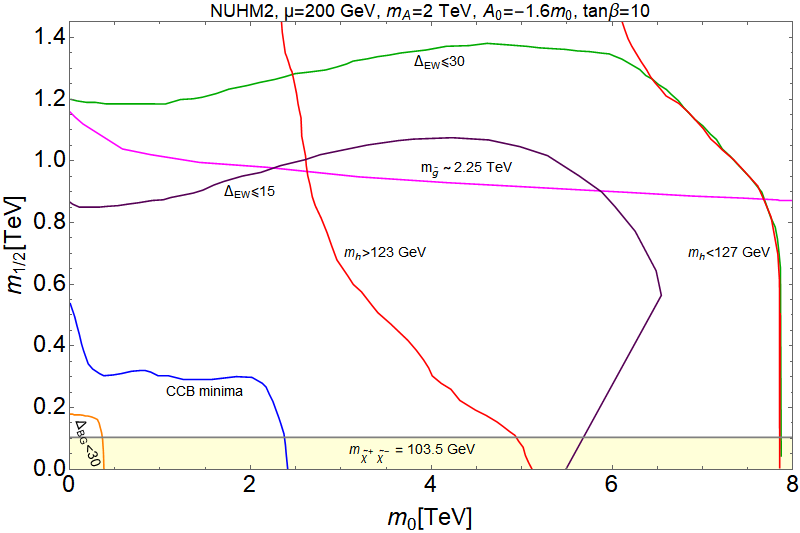}
\caption{The $m_0$ vs. $m_{1/2}$ plane of the NUHM2 model with 
$A_0=-1.6 m_0$, $\tan\beta =10$, $\mu =200$ GeV and $m_A=2$ TeV. 
We show contours of various finetuning measures along with Higgs mass
contours and LEP2 and LHC Run 2 search limits (from Ref. \cite{Baer:2019cae}).
\label{fig:nuhm2}}
\end{center}
\end{figure}

In Fig. \ref{fig:nuhm2}, we instead show the various fine-tuning measures
in the $m_0$ vs. $m_{1/2}$ plane but this time in the 
two-extra-parameter non-universal Higgs model where $m_{H_u}^2$ and
$m_{H_d}^2$ are not set to the matter scalar masses $m_0$. This is sensible
since the Higgs multiplets necessarily live in different GUT multiplets
than matter scalars. The added parameter freedom always allows for 
the possibility of small $\mu$ parameter since $m_{H_u}^2$ and $m_{H_d}^2$
can be traded for weak scale free parameters $\mu$ and $m_A$ via 
the scalar potential minimization conditions. 
For this figure, we choose large $A_0=-1.6 m_0$ and $\tan\beta =10$ but 
with $\mu =200$ GeV and $m_A=2$ TeV. In this case, a wide swath of parameter
space between the red contours admits a Higgs mass 
$123\ {\rm GeV}<m_h<127$ GeV in accord with measured values.

In Fig. \ref{fig:nuhm2},
the $\Delta_{BG}$ measure is squeezed into the lower-left corner which actually
turns out to be a region of charge-and-color breaking (CCB) minima of the
Higgs potential. The $\Delta_{HS}$ measure cannot be plotted since it would
live in the CCB region. However, in this case the $\Delta_{EW}<30$ contour
now appears at very large $m_0$ and $m_{1/2}$ values (green contour) 
and extends well beyond the LHC gluino mass limit. 
Thus, under the model-independent $\Delta_{EW}$ measure, plenty of parameter
space remains beyond current LHC search limits and with the proper value of
light Higgs mass $m_h\sim 125$ GeV.
In fact, scans over many SUSY models with $m_h\sim125$ GeV including 
mSUGRA/CMSSM, GMSB, AMSB and various mirage mediation models with
discrete values of modular weights all turn out to be highly fine-tuned
under $\Delta_{EW}$\cite{seige}. 
Thus, these models would be excluded by LHC as being {\it unnatural}\cite{dew}.
On the other hand, NUHM2  and NUHM3, 
generalized mirage mediation (with continuous rather than discrete 
parameters)\cite{nGMM}, natural AMSB\cite{nAMSB} are all allowed since sizable
natural regions of parameter space remain beyond LHC limits and with
$m_h\sim 125$ GeV.

By scanning the natural SUSY models over all parameter space and requiring
$m_h=125\pm 2$ GeV and $\Delta_{EW}<30$, then new upper bounds can be found
for sparticle masses\cite{rns,upper,lhc27}. These are listed in Table 
\ref{tab:EW} along with the older bounds from Refs. \cite{BG,DG} with
$\Delta_{BG}<30$. 
From Table \ref{tab:EW}, we see that the upper bound on the $\mu$ parameter
is $\mu<350$ GeV for both measures. However, the naturalness upper bound on
$m_{\tg}$ has increased from the old value of $m_{\tg}\alt 0.4-0.6$ TeV to
the new bound $m_{\tg}\alt 6$ TeV: well beyond present LHC bounds and even
well beyond projected search limits for high-luminosity (HL) LHC (which extend
to $m_{\tg}\sim 2.7$ TeV)\cite{Baer:2016wkz}. The old bounds for top squarks
were $m_{\tst_1}\alt 0.45$ TeV, but under $\Delta_{EW}$ these extend to
$m_{\tst_1}<3$ TeV, again well-beyond the reach of HL-LHC.
And whereas before first/second generation squarks and sleptons were 
required to lie $m_{\tq,\tell}<0.55-0.7$ TeV, now using $\Delta_{EW}$
we find $m_{\tq,\tell}\alt 10-30$ TeV (allowing for a mixed 
decoupling/degeneracy solution to the SUSY flavor and CP 
problems\cite{flavor}). 
Thus, we find that under a clarified notion of naturalness, 
plenty of parameter space for weak scale SUSY remains natural and with
$m_h\simeq 125$ GeV.
\begin{table}[!htb]
\renewcommand{\arraystretch}{1.2}
\begin{center}
\begin{tabular}{c|cc}
mass & $BG/DG$ & $\Delta_{EW}$ \\
\hline
 $\mu$ & $<350$ GeV & $<350$ GeV \\
\hline
 $m_{\tg}$ & $<400-600$ GeV & $<6$ TeV \\
\hline
$m_{\tst_1}$ & $<450$ GeV & $<3$ TeV \\
\hline
 $m_{\tq ,\tell}$ & $<550-700$ GeV & $<10-30$ TeV \\
\hline
\end{tabular}
\caption{Upper bounds on sparticle masses from 3\% naturalness
using $\Delta_{BG}$ within multi-parameter SUSY effective theories,
from Refs. \cite{BG,DG} and Refs. \cite{upper,lhc27}.
}
\label{tab:EW}
\end{center}
\end{table} 

\begin{figure}[tbp]
\includegraphics[height=0.5\textheight]{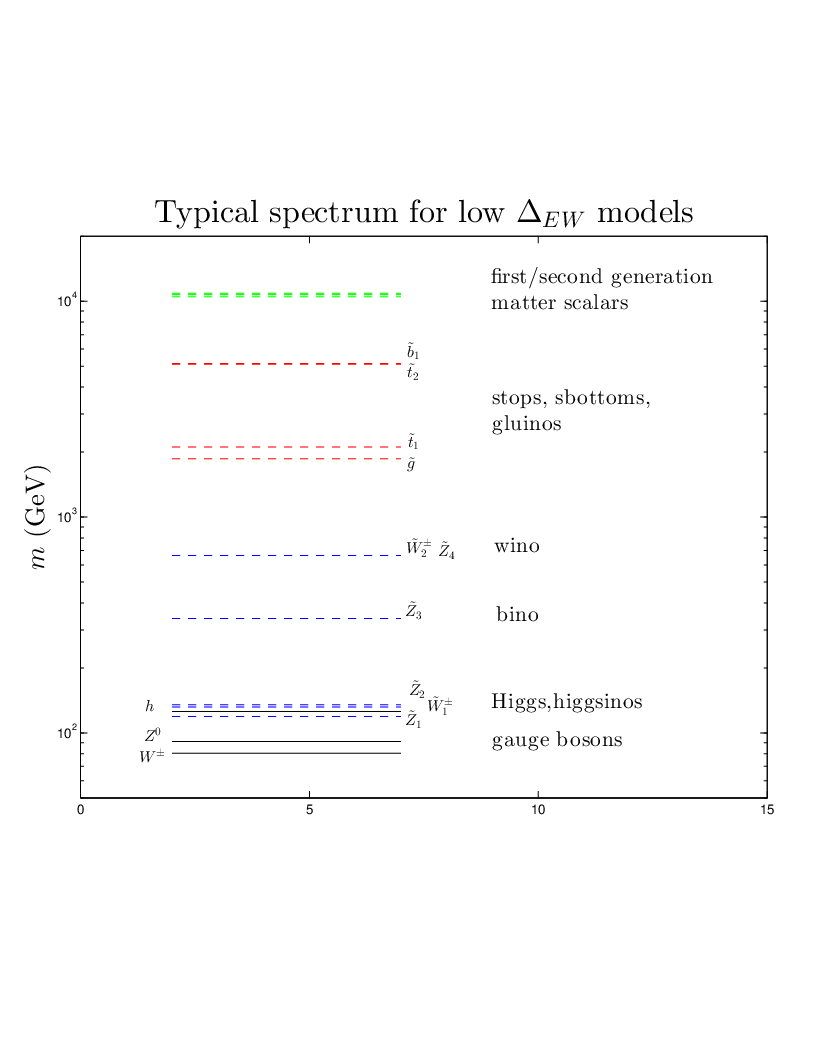}
\caption{Typical mass spectra from natural SUSY where four light higgsinos
lie at the lowest rungs of the anticipated mass spectra.
\label{fig:mass}}
\end{figure}

A pictorial representation of the natural SUSY spectra is shown in 
Fig. \ref{fig:mass}. Here, we see that four light higgsinos 
$\tz_{1,2}$ and $\tw_1^\pm$ are at the bottom of the spectra with mass
$m(higgsinos)\sim \mu$ and with mass splittings of order $5-15$ GeV:
highly compressed. The other gauginos and stops and sbottoms can now live 
in the multi-TeV region safely beyond current LHC bounds while first/second
generation squarks and sleptons inhabit the tens of TeV regime.
The LSP is now the lightest higgsino which is very different from
older expectations. The natural mass ordering brings in new SUSY search
strategies for LHC and new expectations for SUSY dark matter.

\section{QCD naturalness, Peccei-Quinn symmetry, the $\mu$ problem
and discrete symmetries}
\label{sec:mu}

\subsection{QCD naturalness, PQ and axions}

While we require naturalness in the electroweak sector, it is important 
to recall that there is also a naturalness problem in the QCD sector
of the SM.
In the early days of QCD, it was a mystery why the two-light-quark 
chiral symmetry $U(2)_L\times U(2)_R$
gave rise to three and not four light pions\cite{U1}.
The mystery was resolved by 't Hooft's discovery of the 
QCD theta vacuum which allows for the emergence of three pseudo-Goldstone 
bosons-- the pion triplet-- from the spontaneously broken 
global $SU(2)_{axial}$ symmetry, but that didn't respect the 
remaining $U(1)_A$ symmetry\cite{tHooft}.
As a consequence of the theta vacuum, one expects the presence of a term
\be
{\cal L}\ni \frac{\bar{\theta}}{32\pi^2}F_{A\mu\nu}\tilde{F}_A^{\mu\nu}
\label{eq:FFdual}
\ee
in the QCD Lagrangian (where $\bar{\theta}=\theta+arg(det({\cal M}))$ 
and ${\cal M}$ is the quark mass matrix). 
Measurements of the neutron EDM constrain $\bar{\theta}\alt 10^{-10}$
leading to an enormous fine-tuning in $\bar{\theta}$: 
the so-called strong CP problem\cite{peccei_rev}.
The strong CP problem is elegantly solved via the PQWW\cite{pqww} introduction 
of PQ symmetry and the concomitant (invisible\cite{ksvz,dfsz}) axion:
the offending term can dynamically settle to zero.
The axion $a$ is a valid dark matter candidate in its own right\cite{axdm}.

Introducing the axion in a SUSY context solves the strong CP problem 
but also offers an elegant solution to the SUSY $\mu$ problem\cite{KN}. 
The SUSY $\mu$ problem consists of two parts. First, the superpotential
$\mu$ term $W\ni \mu H_uH_d$ is SUSY conserving and so one expects
$\mu$ of order the Planck scale $\mu\sim m_P$. Thus, it must be at first
forbidden, perhaps by some symmetry. Second, the $\mu$ term must be generated,
perhaps via symmetry breaking, such that $\mu$ obtains a natural value of 
order the weak scale $\mu\sim m_{weak}$. A recent review of twenty solutions
to the SUSY $\mu$ problem is presented in Ref. \cite{mu}.

The most parsimonius implementation of the strong CP solution
involves introducing a single MSSM singlet superfield $S$ carrying 
PQ charge $Q_{PQ}=-1$ while the Higgs fields both carry $Q_{PQ}=+1$. 
The usual $\mu$ term is forbidden by the global $U(1)_{PQ}$ symmetry, 
but then we have a superpotential\cite{susydfsz}
\be
W_{DFSZ}\ni \lambda\frac{S^2}{m_P}H_uH_d .
\ee
If PQ symmetry is broken and $S$ receives a VEV $\langle S\rangle\sim f_a$, 
then a weak scale $\mu$ term
\be
\mu\sim \lambda f_a^2/m_P
\ee
is induced which gives $\mu\sim m_Z$ for $f_a\sim 10^{10}$ GeV. 
While Kim-Nilles sought to relate the PQ breaking scale $f_a$ 
to the hidden sector mass scale $m_{hidden}$\cite{KN}, 
we see now that the Little Hierarchy
\be
\mu\sim m_Z\ll m_{3/2}\sim {\rm multi-TeV}
\ee
could emerge due to a mis-match between the PQ breaking scale and 
hidden sector mass scale $f_a\ll m_{hidden}$.

The PQ solution has for long been seen as straddling dangerous ground.
The global $U(1)_{PQ}$ at the heart of the PQ solution is understood
to be inconsistent with the inclusion of gravity in particle 
physics\cite{gravPQ,KM_R}.
If PQ is to work, then the underlying $U(1)_{PQ}$ global symmetry ought to
emerge as an accidental, approximate symmetry arising from some more 
fundamental gravity-safe symmetry, much as baryon and lepton number
conservation arise in the SM accidentally as a consequence of the 
more fundamental gauge symmetry. The fundamental gravity-safe symmetry 
must be especially sharp: if any PQ violating non-renormalizable terms occur in the
PQ sector scalar potential that are suppressed by fewer powers than
$(1/m_P)^8$, then they will cause a shift in the vacuum value such that
$\bar{\theta}>10^{-10}$\cite{KM_R}.

In addition, other problematic terms may arise in the superpotential.
Based upon gauge invariance alone, one expects the MSSM superpotential to
be of the form
\bea
W_{MSSM} &\ni &  \mu H_u H_d+\kappa_i L_i H_u+m_N^{ij}N^c_iN^c_j \label{eq:W} \\
 +f_e^{ij}L_iH_dE_j^c &+& f_d^{ij}Q_iH_dD_j^c+f_u^{ij}Q_iH_uU_j^c
+f_\nu^{ij}L_iH_uN^c_j\nonumber \\
+\lambda_{ijk}L_iL_jE_k^c & +&\lambda^{\prime}_{ijk}L_iQ_jD_k^c+\lambda_{ijk}^{\prime\prime}U_i^cD_j^cD_k^c\nonumber \\
&+& \frac{\kappa_{ijkl}^{(1)}}{m_P}Q_iQ_jQ_kL_l+
\frac{\kappa_{ijkl}^{(2)}}{m_P}U_i^cU_j^cD_k^cE_l^c .\nonumber
\eea
The first term on line 1 of Eq.~\ref{eq:W}, if unsuppressed, should lead to
Planck-scale values of $\mu$ while phenomenology (Eq.~\ref{eq:mzs}) 
requires $\mu$ of order the weak scale $\sim 100-350$ GeV. 
The $\kappa_i$, $\lambda_{ijk}$, $\lambda^{\prime}_{ijk}$ and 
$\lambda^{\prime\prime}_{ijk}$ terms violate either baryon number $B$ 
or lepton number $L$ or both and can, if unsuppressed, 
lead to rapid proton decay and an unstable lightest SUSY particle (LSP). 
The $f_{u,d,e}^{ij}$ are the quark and lepton Yukawa couplings and 
must be allowed to give the SM fermions mass via the Higgs mechanism.
The $\kappa_{ijkl}^{(1,2)}$ terms lead to dimension-five
proton decay operators and are required to be either 
highly suppressed or forbidden. 

It is common to implement discrete symmetries to forbid the 
offending terms and allow the required terms in Eq.~\ref{eq:W}. 
For instance, the ${\bf Z}_2^M$ matter parity (or $R$-parity) forbids the
$\kappa_i$ and $\lambda_{ijk}^{(\prime ,\prime\prime)}$ terms but allows
for $\mu$ and the $\kappa_{ijkl}^{(1,2)}$ terms: 
thus, the ad-hoc $R$-parity conservation all by itself is 
insufficient to cure all of the ills of Eq. \ref{eq:W}

One way to deal with the gravity spoliation issue is to assume instead 
a gravity-safe discrete gauge symmetry ${\bf Z}_M$ of order $M$.
The ${\bf Z}_M$ discrete gauge symmetry can forbid the offending terms of
Eq. \ref{eq:W} while allowing the necessary terms\cite{chunlukas}.
Babu, Gogoladze and Wang\cite{bgw2} have proposed a model 
(written previously by Martin\cite{spm} thus labelled MBGW) with
\be
W_{MBGW}\ni \lambda_{\mu}\frac{X^2H_uH_d}{m_P}+\lambda_2\frac{(XY)^2}{m_P}\label{eq:mbgw}
\ee
which is invariant under a ${\bf Z}_{22}$ discrete gauge symmetry.
These ${\bf Z}_{22}$ charge assignments have been shown to be anomaly-free 
under the presence of a Green-Schwarz (GS) term\cite{gs} 
in the anomaly cancellation calculation. The PQ symmetry then arises as
an {\it accidental approximate global symmetry} as a consequence 
of the more fundamental discrete gauge symmetry. 
The PQ charges of the MBGW model are listed in Table \ref{tab:PQ}.
The discrete gauge symmetry ${\bf Z}_M$ might arise if a charge $Q=Me$ 
field condenses and is integrated out of the low energy theory while
charge $e$ fields survive (see Krauss and Wilczek, Ref.~\cite{KW}). 
While the ensuing low energy theory should be gravity safe, 
for the case at hand one might wonder at the 
plausibility of a condensation of a charge 22 object and whether it might 
occupy the so-called {\it swampland}\cite{swamp} of theories not 
consistent with a UV completion in string theory. 
In addition, the charge assignments\cite{bgw2} are not consistent with  
$SU(5)$ or $SO(10)$ grand unification which may be expected at some
level in a more ultimate theory.
Beside the terms in Eq.~\ref{eq:mbgw}, the lowest order 
PQ-violating  term in the superpotential is $\frac{(Y)^{11}}{m_P^8}$: 
thus the MBGW model is gravity safe. 
\begin{table}[!htb]
\renewcommand{\arraystretch}{1.2}
\begin{center}
\begin{tabular}{c|cc}
multiplet & MBGW & GSPQ \\
\hline
$H_u$ & -1 & -1 \\
$H_d$ & -1 & -1 \\
$Q$   & 1 & 1 \\
$L$   & 1 & 1 \\
$U^c$ & 0 & 0 \\
$D^c$ & 0 & 0 \\
$E^c$ & 0 & 0 \\
$N^c$ & 0 & 0 \\
$X$   & 1 & 1 \\
$Y$   &-1 & -3 \\
\hline
\end{tabular}
\caption{PQ charge assignments for various superfields of the 
MBGW and GSPQ (hybrid CCK) models of PQ breaking from SUSY breaking. 
Another gravity-safe (hybrid SPM) model will have the same PQ charges as 
GSPQ except $Q(X)=-1/3$ and $Q(Y)=1$.
}
\label{tab:PQ}
\end{center}
\end{table}

An alternative very compelling approach is to implement a 
discrete $R$ symmetry ${\bf Z}_N^R$ of order $N$.\footnote{
Discrete $R$ symmetries were used in regard to the $\mu$ problem in 
Ref.~\cite{choi_hall} and for the PQ problem in Ref.~\cite{harigaya}.}
Such discrete $R$ symmetries are expected to arise as discrete remnants
from compactification of 10-d (Lorentz symmetric)
spacetime down to 4-dimensions\cite{Kappl:2010yu,Nilles:2017heg} 
and thus should be in themselves gravity safe\cite{Harlow:2018tng}. 
In fact, in Lee {\it et al.} Ref. \cite{lrrrssv1}, 
it was found that the requirement of an anomaly-free discrete symmetry that 
forbids the $\mu$ term and all dimension four- and five- baryon and 
lepton number violating terms in Eq. \ref{eq:W} while allowing the 
Weinberg operator $LH_uLH_u$ and that commutes with $SO(10)$ 
(as is suggested by the unification of each family
into the 16 of $SO(10)$) has a unique solution: a ${\bf Z}_4^R$ $R$-symmetry.
If the requirement of commutation with $SO(10)$ is weakened to commutation with
$SU(5)$, then further discrete ${\bf Z}_N^R$ symmetries with
$N$ being an integral divisor of 24 are allowed\cite{lrrrssv2}: 
$N=4,6,8,12$ and 24. 
Even earlier\cite{bgw1}, the ${\bf Z}_4^R$ was found to be the simplest
discrete $R$-symmetry to realize $R$-parity conservation whilst 
forbidding the $\mu$ term. 
In that reference, the $\mu$ term was regenerated using 
Giudice-Masiero\cite{GM} which would generate $\mu\sim m_{soft}$ (too large).

$R$-symmetries are characterized by the fact that superspace co-ordinates 
$\theta$ carry non-trivial $R$-charge: 
in the simplest case, $Q_R(\theta )=+1$ so that $Q_R( d^2\theta ) =-2$. 
For the Lagrangian ${\cal L}\ni \int d^2\theta W$ to be invariant under 
$R$-symmetry, then the superpotential $W$ must carry $Q_R(W)= 2$. 
%\textbf{Discrete $R$ symmetries should be gravity-safe since they are expected to 
%emerge as remnants of 10-$d$ Lorentz symmetry under compactification of 
%extra dimensions in superstring theory}.
The ${\bf Z}_N^R$ symmetry gives rise to a universal gauge 
anomaly $\rho$ mod $\eta$ where the remaining contribution $\rho$ 
is cancelled by the Green-Schwarz axio-dilaton shift and
$\eta =N$ ($N/2$) for $N$  odd (even).  
The anomaly free $R$ charges of various MSSM fields are listed in 
Table \ref{tab:R} for $N$ values consistent with grand unification.
\begin{table}[!htb]
\renewcommand{\arraystretch}{1.2}
\begin{center}
\begin{tabular}{c|ccccc}
multiplet & ${\bf Z}_{4}^R$ & ${\bf Z}_{6}^R$ & ${\bf Z}_{8}^R$ & ${\bf Z}_{12}^R$ & ${\bf Z}_{24}^R$ \\
\hline
$H_u$ & 0  & 4  & 0 & 4 & 16 \\
$H_d$ & 0  & 0  & 4 & 0 & 12 \\
$Q$   & 1  & 5 & 1 & 5  & 5 \\
$U^c$ & 1  & 5 & 1 & 5  & 5 \\
$E^c$ & 1  & 5 & 1 & 5  & 5 \\
$L$   & 1  & 3 & 5 & 9  & 9 \\
$D^c$ & 1  & 3 & 5 & 9  & 9 \\
$N^c$ & 1  & 1 & 5 & 1  & 1 \\
\hline
\end{tabular}
\caption{Derived MSSM field $R$ charge assignments for various anomaly-free 
discrete ${\bf Z}_{N}^R$ symmetries which are consistent with $SU(5)$ or 
$SO(10)$ unification (from Lee {\it et al.} Ref.~\cite{lrrrssv2}).
}
\label{tab:R}
\end{center}
\end{table}

In Ref. \cite{Baer:2018avn}, it has been examined whether or not three 
models--CCK\cite{cck}, MSY\cite{msy,radpq} and SPM\cite{spm}-- 
with  radiative PQ breaking which also leads
to generation of the Majorana neutrino see-saw mass scale $M_N$
can be derived from any of the fundamental ${\bf Z}_N^R$ 
symmetries in Table \ref{tab:R}.
In almost all cases, the $hXN^cN^c$ operator is disallowed: then there is no
large Yukawa coupling present to drive the PQ soft term $m_X^2$ negative 
so that PQ symmetry is broken. And since the PQ symmetry does not allow for
a Majorana mass term $M_NN^cN^c$, then no see-saw scale can be developed.
The remaining cases that did allow for a Majorana mass scale were all found 
to be not gravity safe.
Also, the MBGW model was found to not be gravity safe under any of the 
${\bf Z}_N^R$ discrete $R$-symmetries of Table \ref{tab:R}.

Next, a hybrid approach between the radiative breaking models 
and the MBGW model was created by writing a superpotential:
\bea
W&\ni &f_uQH_uU^c+f_dQH_d D^c+f_{\ell}LH_dE^c\nonumber \\
& +&f_{\nu}LH_uN^c + fX^3Y/m_P\nonumber \\
& +&\lambda_\mu X^2 H_uH_d/m_P+M_NN^cN^c/2
\eea
along with PQ charge assignments given under the GSPQ (gravity-safe PQ model)
heading of Table \ref{tab:PQ}.
For this model, we have checked that there is gravity spoliation for
$N=4,\ 6,\ 8$ and 12. But for ${\bf Z}_{24}^R$ and under $R$-charge
assignments $Q_R(X)=-1$ and $Q_R(Y)=5$, then the lowest order
PQ violating superpotential operators allowed are 
$X^8Y^2/m_P^7$, $Y^{10}/m_P^7$ and $X^4Y^6/m_P^7$. 
These operators\footnote{The $X^8Y^2/m_P^7$ term was noted previously 
in Ref. \cite{lrrrssv2}.} 
lead to PQ breaking terms in the scalar potential suppressed by powers
of $(1/m_P)^8$. For instance, the term $\lambda_3X^8Y^2/m_P^7$  leads to
$V_{PQ}\ni 24f\lambda_3^*X^2YX^{*7}Y^{*2}/m_P^8+h.c.$ which is sufficiently 
suppressed by enough powers of $m_P$ so as to be gravity safe\cite{KM_R}.
We have also checked that hybrid model using the MSY $XYH_uH_d/m_P$ 
term is not gravity-safe under any of the discrete $R$-symmetries of
Table \ref{tab:R} but the  
hybrid SPM model with $Y^2H_uH_d/m_P$ and charges $Q_R(X)=5$ and $Q_R(Y)=-13$ 
is gravity-safe under only ${\bf Z}_{24}^R$.

The scalar potential $V_F=|3f\phi_X^2\phi_Y/m_P|^2+|f\phi_X^3/m_P|^2$
of the hybrid CCK model was augmented by the following soft breaking terms
\be
V_{soft}\ni m_X^2|\phi_X|^2+m_Y^2|\phi_Y|^2+(f A_f\phi_X^3\phi_Y/m_P+h.c.)
\ee
and the resultant scalar potential was minimized. 
The minimization conditions are the same as those found 
in Ref. \cite{radpq} Eq's 17-18.
In the case of the GSPQ model, the PQ symmetry isn't broken radiatively, 
but instead can be broken by adopting a sufficiently large negative 
value of $A_f$ (assuming real positive couplings for simplicity).
The scalar potential admits a non-zero minimum in the fields 
$\phi_X$ and $\phi_Y$ for $A_f<0$ (see Fig. 1 of Ref. \cite{Baer:2018avn}
%as shown in Fig. \ref{fig:Vgspq} 
which is plotted for the case of $m_X=m_Y\equiv m_{3/2}=10$ TeV, 
$f=1$ and $A_f=-35.5$ TeV). For these values, it is found that
$v_X=10^{11}$ GeV, $v_Y=5.8\times 10^{10}$ GeV, $v_{PQ}=1.15\times 10^{11}$ GeV
and $f_a=\sqrt{v_X^2+9v_Y^2}=2\times 10^{11}$ GeV. These sorts of numerical 
values lie within the mixed axion/higgsino dark matter sweet spot
of cosmologically allowed values and typically give dominant 
DFSZ axion CDM with $\sim 10\%$ WIMP dark matter\cite{bbc,dfsz2,axpaper}.  
Under these conditions, the model develops a $\mu$ parameter
$\mu =\lambda_\mu v_X^2/m_P$ and for a value $\lambda_\mu =0.036$ then we
obtain a natural value of the $\mu$ parameter at $150$ GeV.
% 
%\begin{figure}[tbp]
%\includegraphics[height=0.2\textheight]{vxy_phix_phiy_hybrid_cck.png}
%\caption{Scalar potential $V_{GSPQ}$ versus $\phi_X$ and $\phi_Y$ for
%$m_X=m_Y\equiv m_{3/2}=10$ TeV, $f=1$ and $A_f=-35.5$ TeV.
%\label{fig:Vgspq}}
%\end{figure}
%

\begin{figure}[tbp]
\includegraphics[height=0.3\textheight]{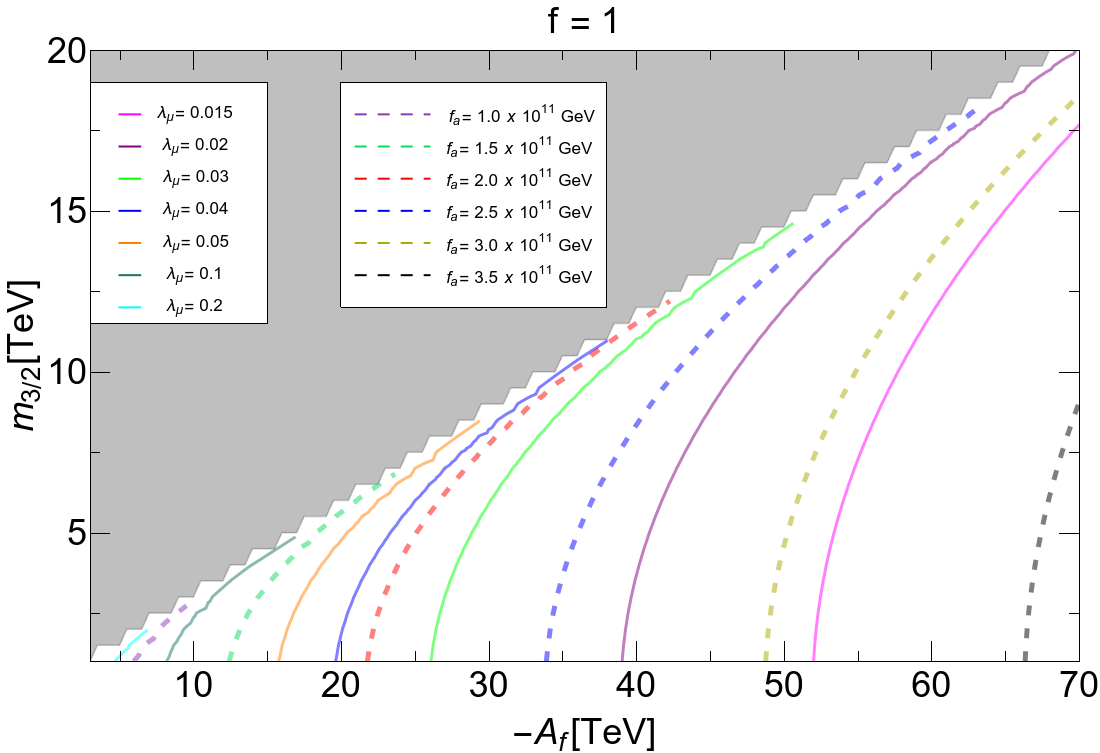}
\caption{Representative values of $\lambda_{\mu}$ required for $\mu =150$ GeV in the $m_{3/2}$ vs. $-A_f$ plane of the GSPQ model for $f=1$. 
We also show several contours of $f_a$ (from Ref. \cite{Baer:2018avn}).
\label{fig:GSPQ}}
\end{figure}

The allowed range of GSPQ model parameter space is shown in Fig. \ref{fig:GSPQ}
where we show contours of $\lambda_{\mu}$ values which lead to $\mu =150$ GeV
in the $m_{3/2}$ vs. $-A_f$ plane for $f=1$. We also show several
representative contours of $f_a$ values.
Values of $\lambda_{\mu}\sim 0.015-0.2$ are generally sufficient for a natural
$\mu$ term and are easily consistent with soft mass 
$m_{soft}\sim m_{3/2}\sim 2-30$ TeV as indicated by LHC searches.
We also note that for $m_{3/2}\sim 5-20$ TeV, then $f_a\sim 10^{11}$ GeV. 
Such high values
of $m_{3/2}$ also allow for a resolution of the early universe gravitino 
problem\cite{linde} 
(at higher masses gravitinos may decay before the onset of 
big bang nucleosynthesis (BBN))
and such high soft masses serve to ameliorate the SUSY flavor and CP 
problems as well\cite{masiero,dine,flavor}. 
They are also expected in several well-known
string phenomenology constructions including compactification of 
$M$-theory on a manifold of $G_2$ holonomy\cite{kane}, the minilandscape of
heterotic strings compactified on orbifolds\cite{mini} and the 
statistical analysis of the landscape of IIB intersecting $D$-brane 
models\cite{land}. 

Thus, the gravity-safe ${\bf Z}_{24}^R$ symmetry\cite{lrrrssv2} 
(which may emerge as a remnant of 10-$d$ Lorentz symmetry 
which is compactified to four spacetime dimensions)
yields an accidental approximate global PQ symmetry as implemented in the
GSPQ model of PQ symmetry breaking as a consequence of SUSY breaking. 
The ${\bf Z}_{24}^R$ (PQ) symmetry breaking leads to 
$\mu\ll m_{soft}$ as required
by electroweak naturalness and to PQ energy scales 
$f_a\sim 10^{11}$ GeV as required by mixed axion-higgsino dark matter. 
The ${\bf Z}_{24}^R$ symmetry also forbids the dangerous dimension-four 
$R$-parity violating terms. 
Dimension-five proton decay 
operators are suppressed to levels well below experimental 
constraints\cite{lrrrssv2}.
Overall, the results of Ref. \cite{Baer:2018avn} show that the axionic solution 
to the strong CP problem is enhanced by the presence of both supersymmetry and 
extra spacetime dimensions which give rise to the gravity-safe 
${\bf Z}_{24}^R$ symmetry from which the required 
global PQ symmetry accidentally emerges. It is rather amusing then that
both the global $U(1)_{PQ}$ and $R$-parity emerge from a single more
fundamental discrete ${\bf Z}_{24}^R$ symmetry.

\section{The string landscape and SUSY}
\label{sec:land}

In Sec.~\ref{sec:nat} we were concerned with naturalness of the EW scale 
while in Sec.~\ref{sec:mu} we were concerned with QCD naturalness involving
the CP-violating $\bar{\theta}$ term. If gravity is included in the SM, then
a third naturalness problem emerges: why is the vacuum energy density 
$\rho_{vac}$ so tiny, or alternatively, why is the cosmological constant (CC) 
$\Lambda$ so tiny when there is no known symmetry to suppress its 
magnitude? Naively, one would expect $\Lambda\sim m_P^4$.

\begin{figure}[tbp]
\begin{center}
\includegraphics[height=0.3\textheight]{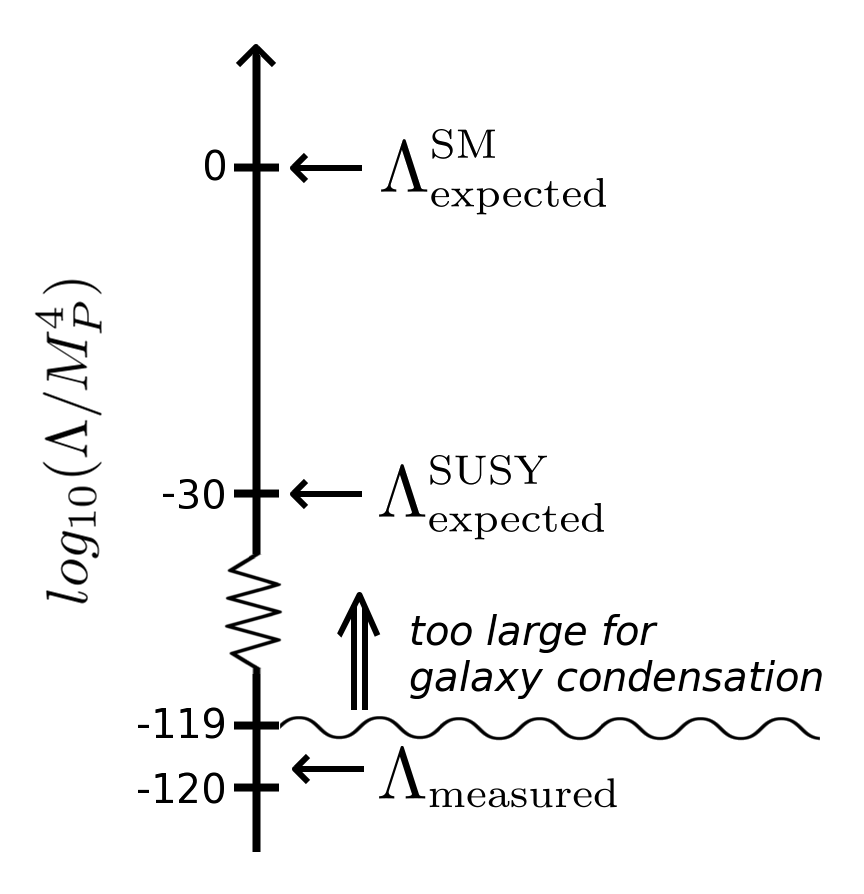}
\caption{
Log portrayal of expected parameter space of the cosmological constant $\Lambda$
from the string theory landscape.
\label{fig:CC}}
\end{center}
\end{figure}

At present, the only plausible solution to the CC problem is the hypothesis
of the landscape: that a vast number of string theory vacua states exist,
each with differing values of physical constants, including $\Lambda$.
Here, our universe is then just one {\it pocket universe} present 
in a vast ensemble of bubble universes contained within the {\it multiverse}.
In this case, a non-zero value of the CC should be present in each 
pocket universe, but if its value is too large, then the universe would expand
too quickly to allow for galaxy condensation and consequently no complex
structure would arise, and no observors would be present to measure $\Lambda$.
This ``anthropic'' explanation for the magnitude of $\Lambda$ met with 
great success by Weinberg who was able to predict its value to within 
a factor of several well before it was actually measured.
The situation is portrayed in Fig. \ref{fig:CC} where it is anticipated that 
within a {\it fertile patch} of the multiverse (all pocket universes
containing the SM as the low energy effective theory but with differing values
of $\Lambda$ spread uniformly across the decades of possible values), the
value of $\Lambda$ is about as large as possible such as to give a 
livable pocket universe.

\begin{figure}[tbp]
\includegraphics[height=0.7\textheight,angle=-90]{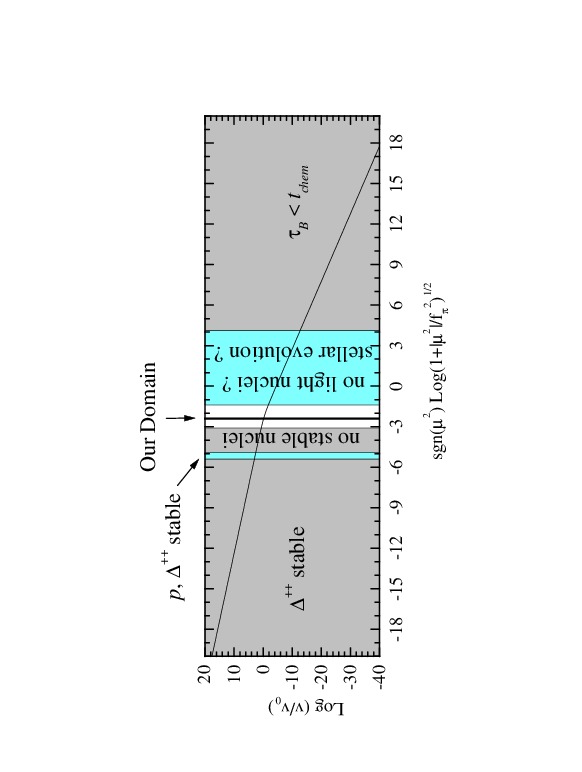}
\caption{Allowed values of $m_{weak}^{PU}$ from
Agrawal {\it et al.} Ref.~\cite{Agrawal:1997gf}.
\label{fig:abds}}
\end{figure}

Can similar reasoning be used to explain the magnitude of other 
mass scales that appear in theories like the SM or the MSSM?
Agrawal {\it et al.}\cite{Agrawal:1997gf} already examined this question 
for the case of the magnitude of the weak scale of the SM in 1997. 
What they found, as depicted in their Fig.~\ref{fig:abds}, 
was that if $m_{weak}\sim m_{W,Z,h}$
of a pocket universe (PU) was larger than our universe's (OU) 
measured value by a factor $m_{weak}^{PU}\agt (2-5)m_{weak}^{OU}$,
then shifts in the light quark masses $m_u$ and $m_d$ 
would imply no stable nuclei and all baryons would exist as just protons. 
Nuclear physics would not be as we know it: 
complex nuclei and consequently atoms as we know them wouldn't form. 
This violates the so-called {\it atomic principle}: 
that atoms as in our pocket universe 
must be present for observors such as ourselves to arise.

The emergence of the string theory landscape\cite{BP,Susskind:2003kw} 
led Douglas\cite{Douglas:2004qg} to consider whether the scale of 
SUSY breaking might arise in a similar fashion. In the landscape, 
then of order $10^{500}$ different vacua states might exist\cite{Ashok:2003gk},
each with different matter content, gauge groups and physical constants.
For a fertile patch of the landscape conatining the MSSM as the 
low energy effective theory, then the differential distribution of
vacua with respect to the hidden sector SUSY breaking scale 
$m_{hidden}^4=\sum_i|F_i|^2+{1\over 2}\sum_{\alpha}D_{\alpha}^2$ 
is expected to be of the form
\be
dN_{vac}[m_{hidden}^2,m_{weak},\Lambda ]=f_{SUSY}(m_{hidden}^2)\cdot
f_{EWSB}\cdot f_{cc}\cdot dm_{hidden}^2
\label{eq:dNvac}
\ee
where the soft SUSY breaking scale $m_{soft}\sim m_{3/2}\sim m_{hidden}^2/m_P$.
In string theory, it is expected that a number of hidden sectors occur with
the overall SUSY breaking scale determined by contributions from various
$F_i$ and $D_{\alpha}$ SUSY breaking fields with non-zero SUSY breaking vevs.
The CC is given here by 
\be
\Lambda =m_{hidden}^4-3e^{K/m_P^2} |W|^2/m_P^2
\ee
where we assume gravity-mediated SUSY breaking and where $K$ is the K\"ahler
potential and $W$ is the superpotential. A small cosmological constant
$\Lambda\sim 0$ can be selected for by scanning over $W$ values distributed 
uniformly as a complex variable independent of the values of $F_i$ and 
$D_{\alpha}$ and hence a small CC has no effect on the distribution of SUSY
breaking scales\cite{Susskind:2004uv,Denef:2004ze,Douglas:2004qg}.

\begin{figure}[tbp]
\begin{center}
\includegraphics[height=0.3\textheight]{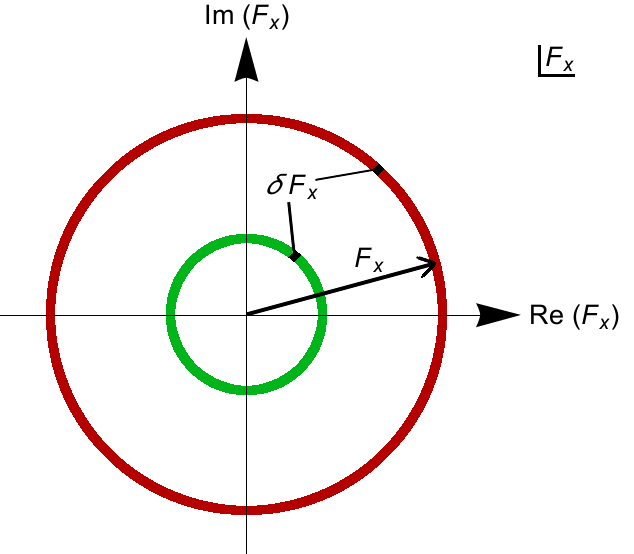}
\caption{
Annuli of the complex $F_X$ plane giving rise to linearly increasing
selection of soft SUSY breaking terms.
\label{fig:f_x}}
\end{center}
\end{figure}

Another key observation from examining flux vacua in IIB string theory 
is that the SUSY breaking $F_i$ and $D_\alpha$ terms
are likely to be uniformly distributed-- in the former case as complex numbers while in the latter case
as real numbers.  
In this case, one then obtains the following distribution of supersymmetry breaking scales
\be 
\label{brokensusydistri}
f_{SUSY}(m_{hidden}^2) \, \sim \, (m^2_{hidden})^{2n_F+n_D - 1}
\ee
where $n_F$ is the number of $F$-breaking fields and $n_D$ is the number of $D$-breaking fields in
the hidden sector\cite{Douglas:2004qg}.
The case of $n_F = 1$ is displayed in Figure~\ref{fig:f_x}. 
We label the visible sector soft term mass scale as $m_{soft}$ where in SUGRA breaking models 
we typically have $m_{soft}\sim m_{hidden}^2/m_P\sim m_{3/2}$. 
Thus, the case of $n_F=1$ $n_D=0$ would give a {\it linearly increasing} probability distribution 
for generic soft breaking terms simply because the area of annuli within the complex plane increases
linearly. We will denote the collective exponent in Eq. (\ref{brokensusydistri}) as 
$n\equiv 2n_F+n_D-1$ so that the case $n_F=1$, $n_D=0$ leads to $n=1$ with 
$f_{SUSY}(m_{soft})\sim m_{soft}^1$. 
The case $n_F=0$ with $n_D=1$ would lead to a uniform distribution in soft terms 
$f_{SUSY}(m_{soft})\sim m_{soft}^0$. For the more general case with an assortment of $F$ and $D$ terms 
contributing comparably to SUSY breaking, then high scale SUSY breaking models 
would be increasingly favored.

\begin{figure}[tbp]
\begin{center}
\includegraphics[height=0.35\textheight]{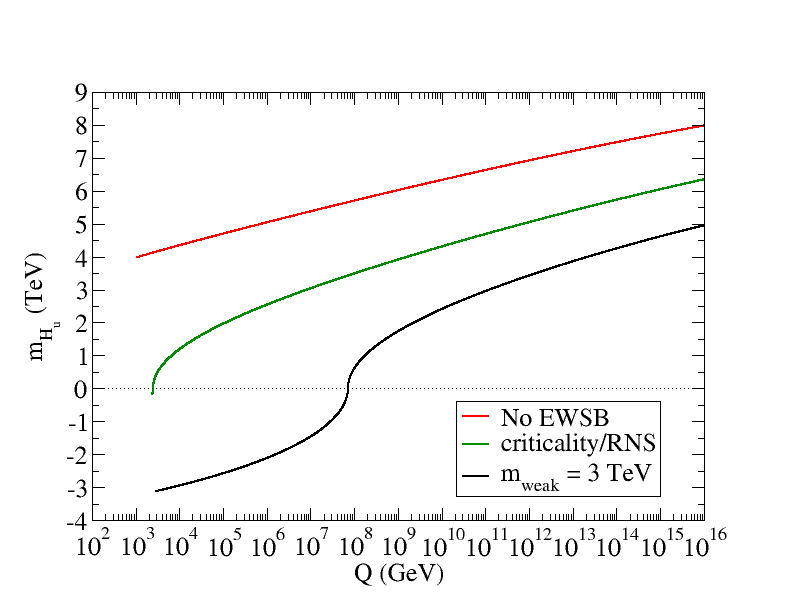}
\caption{Evolution of the soft SUSY breaking mass squared term
$sign(m_{H_u}^2)\sqrt{|m_{H_u}^2|}$ vs. $Q$ for the case of no EWSB (upper),
criticality (middle) as in radiatively-driven natural SUSY (RNS)
and $m_{weak}\sim 3$ TeV (lower).
Most parameters are the same as in Fig. \ref{fig:mHu_mhf}.
\label{fig:mHuQ}}
\end{center}
\end{figure}

\begin{figure}[!htbp]
\begin{center}
\includegraphics[height=0.4\textheight]{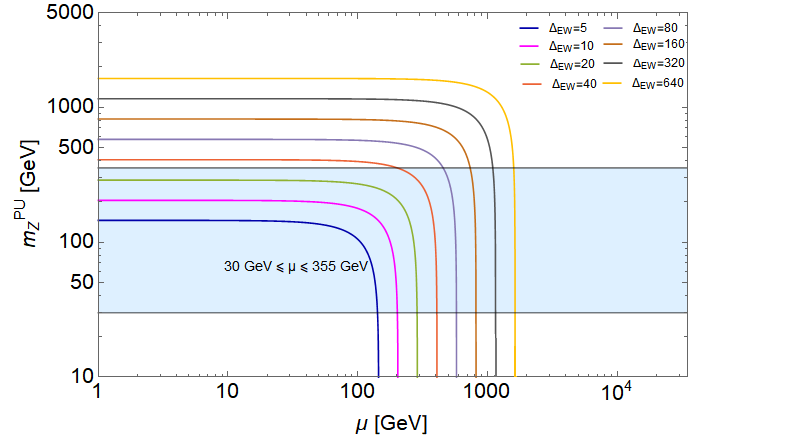}
\caption{The pocket universe value of $m_Z^{PU}$  versus the SUSY  $\mu$ 
parameter for various values of EW finetuning parameter $\Delta_{EW}$.
The anthropic band is shown in blue.
\label{fig:mzPU}}
\end{center}
\end{figure}

An initial guess for $f_{EWFT}$-- the (anthropic) finetuning factor--
was $m_{weak}^2/m_{soft}^2$ which would penalize soft terms which were much
bigger than the weak scale.
This ansatz fails on several points.
\begin{itemize}
\item Many soft SUSY breaking choices will land one into charge-or-color breaking (CCB) minima of the EW scalar potential. Such vacua would likely not lead to a livable
universe and should be vetoed.
\item Other choices for soft terms may not even lead to EW symmetry breaking (EWSB).
For instance, if $m_{H_u}^2(\Lambda )$ is too large, 
then it will not be driven negative to trigger spontaneous EWSB 
(see Fig. \ref{fig:mHuQ}). 
These possibilities also should be vetoed.
\item In the event of appropriate EWSB minima, then sometimes {\it larger} 
high scale soft terms lead to {\it more natural} weak scale soft terms. 
For instance, if $m_{H_u}^2(\Lambda )$ is
large enough that EWSB is {\it barely broken}, then $|m_{H_u}^2(weak)|\sim m_{weak}^2$.
Likewise, if the trilinear soft breaking term $A_t$ is big enough, then there 
is large top squark mixing and the $\Sigma_u^u(\tst_{1,2})$ terms enjoy large
cancellations, rendering them $\sim m_{weak}^2$. 
The same large $A_t$ values lift the Higgs mass $m_h$ up to the 125 GeV regime.
\end{itemize}
Here, we will assume a {\it natural} solution to the SUSY $\mu$ problem\cite{mu}.
As seen in Fig. \ref{fig:mzPU}, a natural value of $\mu$ allows for far more landscape vacua
to generate an anthropically-required value for $m_{weak}$. But once $\mu$ is fixed, 
then we are no longer allowed to use it to tune to  our measured value of $m_Z^{OU}$: 
instead, we must live with the value of $m_Z^{PU}$ generated in each pocket-universe.

Some attractive possibilities for generating $\mu$ are the hybrid CCK or SPM 
models\cite{Baer:2018avn} which are based on the previously-mentioned 
${\bf Z}_{24}^R$ discrete $R$ symmetry which can emerge 
from compactification of extra dimensions in string theory. 
The ${\bf Z}_{24}^R$ symmetry is strong enough to allow a gravity-safe $U(1)_{PQ}$ 
symmetry to emerge (which solves the strong CP problem) while also forbidding 
$R$-parity violating (RPV) terms
(so that WIMP dark matter is generated). 
Thus, both Peccei-Quinn (PQ) and $R$-parity conservation (RPC) 
arise as approximate accidental symmetries similar to the way 
baryon and lepton number conservation emerge accidentally
(and likely approximately) due to the SM gauge symmetries. 
These hybrid models also solve the SUSY $\mu$ problem via a 
Kim-Nilles\cite{KN} operator so that 
$\mu\sim \lambda_\mu f_a^2/m_P$ and $\mu\sim 100-200$ GeV (natural) for $f_a\sim 10^{11}$ GeV
(the sweet zone for axion dark matter). 
The ${\bf Z}_{24}^R$ symmetry also suppresses dimension-5 proton decay operators\cite{lrrrssv2}. 

Once a natural value of $\mu\sim 100-300$ GeV is obtained, then we may invert the usual usage
of Eq. \ref{eq:mzs} to determine the value of the weak scale in various pocket 
universes (with MSSM as low energy effective theory) for a given choice of soft terms.
Based on nuclear physics calculations by Agrawal {\it et al.}\cite{Agrawal:1997gf}, a pocket universe value of $m_{weak}^{PU}$ which deviates from our measured value by a factor 2-5 is likely to lead
to an unlivable universe as we understand it. Weak interactions and fusion processes 
would be highly suppressed and even complex nuclei could not form. 
%The situation is shown in Fig. \ref{fig:mweak}. 
We will adopt a conservative value where the $m^{PU}_{weak}$ should not
deviate by more than a factor four from $m^{OU}_{weak}$. 
This corresponds to a value of $\Delta_{EW}\alt 30$.
Thus, for our final form of $f_{EWSB}$ we will adopt 
\be
f_{EWSB}=\Theta(30-\Delta_{EW})
\ee
while also vetoing CCB or no EWSB vacua.

\begin{figure}[tbp]
\begin{center}
\includegraphics[height=0.25\textheight]{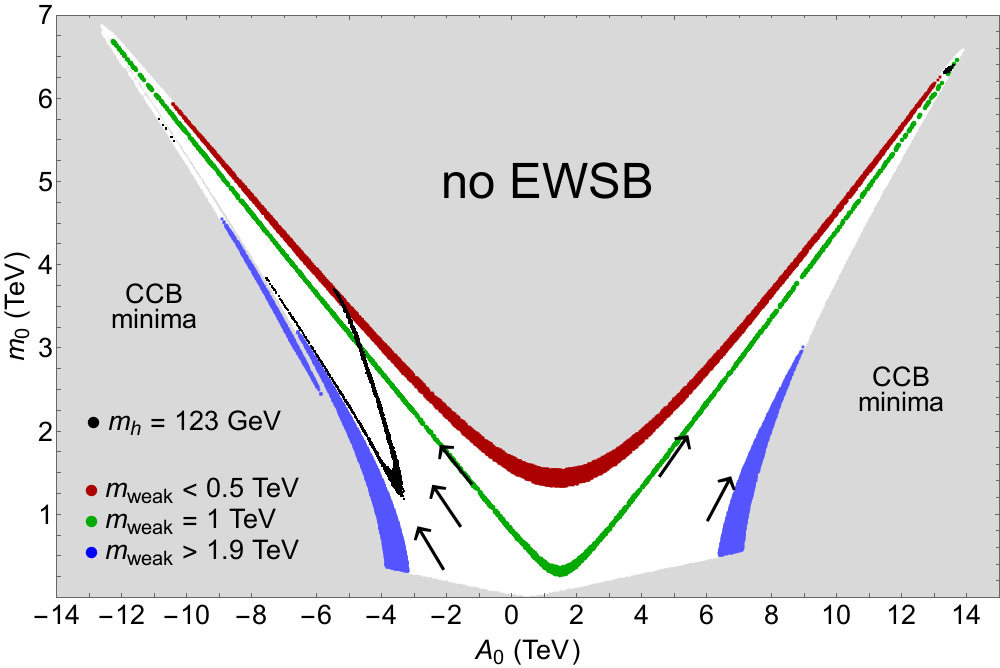}
\caption{Contours of $m_{weak}$ in the $A_0$ vs. $m_0$ plane
for $m_{1/2}=1$ TeV, $m_{H_u}=1.3 m_0$, $\tan\beta =10$ and $m_{H_d}=1$ TeV.
The arrows show the direction of statistical/anthropic pull on soft SUSY breaking terms.
Within the black contour is where $m_h>123$ GeV. 
There is also a slight black contour in the upper-right horn as well.
\label{fig:A0_m0}}
\end{center}
\end{figure}

In Fig. \ref{fig:A0_m0} we show the 
$A_0$ vs. $m_0$ plane for the NUHM2 model with $m_{1/2}$ fixed at 1 TeV, $\tan\beta =10$
and $m_{H_d}=1$ TeV. We take $m_{H_u}=1.3 m_0$. 
The plane is qualitatively similar for different reasonable parameter choices. 
We expect $A_0$ and $m_0$ statistically to be drawn as large as possible
while also being anthropically drawn towards $m_{weak}\sim 100-200$ GeV, labelled as
the red region where $m_{weak}<500$ GeV. The blue region has $m_{weak}>1.9$ TeV and the green
contour labels $m_{weak}=1$ TeV. The arrows denote the combined 
statistical/anthropic pull on the soft terms: towards large soft terms but low $m_{weak}$.
The black contour denotes $m_h=123$ GeV with the regions to the upper left 
(or upper right, barely visible) containing larger values of $m_h$. 
We see that the combined pull on soft terms brings us to
the region where $m_h\sim 125$ GeV is generated. 
This region is characterized by highly mixed TeV-scale top squarks\cite{mhiggs,h125}. 
If instead $A_0$ is pulled too large,
then the stop soft term $m_{U_3}^2$ is driven tachyonic resulting in charge and color
breaking minima in the scalar potential (labelled CCB). 
If $m_0$ is pulled too high for fixed $A_0$, then electroweak symmetry isn't even broken.

\begin{figure}[tbp]
\begin{center}
\includegraphics[height=0.25\textheight]{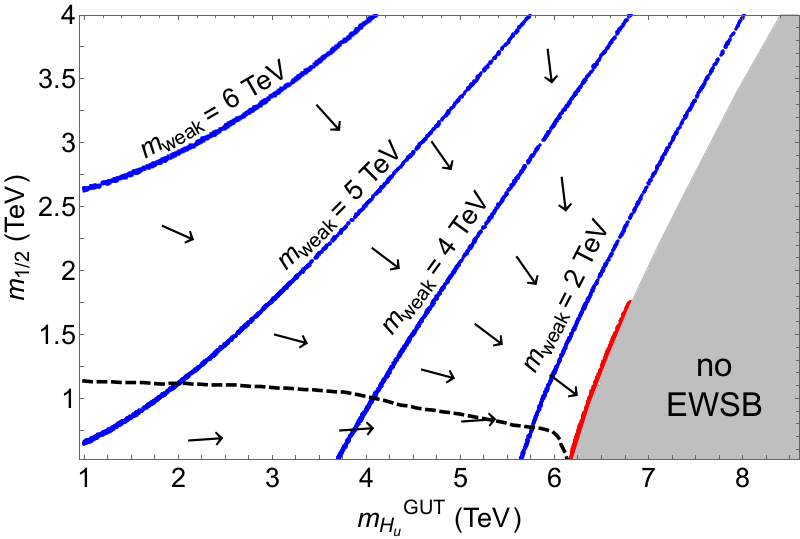}
\caption{Contours of $m_{weak}$ (blue) in the $m_{H_u}$ vs. $m_{1/2}$ plane
for $m_0=5$ TeV, $A_0=-8$ TeV, $\tan\beta =10$ and $m_{H_d}=1$ TeV.
Above the black dashed contour is where $m_h>124$ GeV.
The red region has $m_{weak}<0.5$ TeV.
The arrows show the direction of the statistical/anthropic pull on soft SUSY breaking terms.
\label{fig:mHu_mhf}}
\end{center}
\end{figure}

In Fig. \ref{fig:mHu_mhf}, we show contours of $m_{weak}$ in the $m_{H_u}$ vs. $m_{1/2}$
plane for $m_0=5$ TeV, $A_0=-8$ TeV, $\tan\beta =10$ and $m_{H_d}=1$ TeV. The statistical 
flow is to large values of soft terms but the anthropic flow is towards the red region where
$m_{weak}<0.5$ TeV.
While $m_{1/2}$ is statistically drawn to large values, if it is too large then,
as before, the $\tst_{1,2}$ become too heavy and the $\Sigma_u^u(\tst_{1,2})$ become
too large so that $m_{weak}$ becomes huge.
The arrows denote the direction of the combined statistical/anthropic flow. 
The region above the black dashed contour has $m_h>124$ GeV. 
The value of $m_{H_u}(GUT)$ would like to be statistically as large as possible but if it is too large
then EW symmetry will not break. Likewise, if $m_{H_u}(GUT)$ is not large enough, then it is driven to
large negative values so that $m_{weak}\sim$ the TeV regime and weak interactions are too weak.
The situation is shown in Fig. \ref{fig:mHuQ} where we show the running of 
$sign(m_{H_u}^2)\sqrt{|m_{H_u}^2|}$ versus energy scale $Q$ 
for several values of $m_{H_u}^2(GUT)$ for $m_{1/2}=1$ TeV and with other parameters
the same as Fig. \ref{fig:mHu_mhf}. Too small a value of $m_{H_u}^2(GUT)$ leads to too large
a weak scale while too large a value results in no EWSB. The combined statistical/anthropic
pull is for {\it barely-broken} EW symmetry where soft terms teeter on the 
edge of criticality: between breaking and not breaking EW symmetry. 
This yields the other naturalness condition that $m_{H_u}$ is driven small negative: 
then the weak interactions are of the necessary strength.
These are just the same conditions for supersymmetric models with radiatively-driven 
natural SUSY (RNS)\cite{ltr,rns}.
Such behavior is termed by Ref. \cite{ArkaniHamed:2005yv} as {\it living dangerously} 
in that the landscape statistically pulls parameters towards the edge-- 
(but not all the way) of disaster.\footnote{See also Giudice and Rattazzi, 
Ref. \cite{Giudice:2006sn}.}

\subsection{Probability distributions for Higgs and sparticle 
masses from the landscape}

To gain numerical predictions for Higgs boson and sparticle masses from
the string landscape, we scan over the parameter space of the NUHM3 model
with parameters
\be
m_0(1,2),\ m_0(3),\ m_{1/2},\ A_0,\ \tan\beta,\ \mu,\ {\rm and}\ m_A\ \ (NUHM3)
\ee
with $\mu$ fixed at a natural value $150$ GeV (which arises from an 
assumed natural solution to the SUSY $\mu$ problem) and 
a power law selection on soft terms for $n=0$, 1 and 2.
$\tan\beta$ is scanned as flat from 3-60.

\begin{figure}[tbp]
\begin{center}
\includegraphics[height=0.2\textheight]{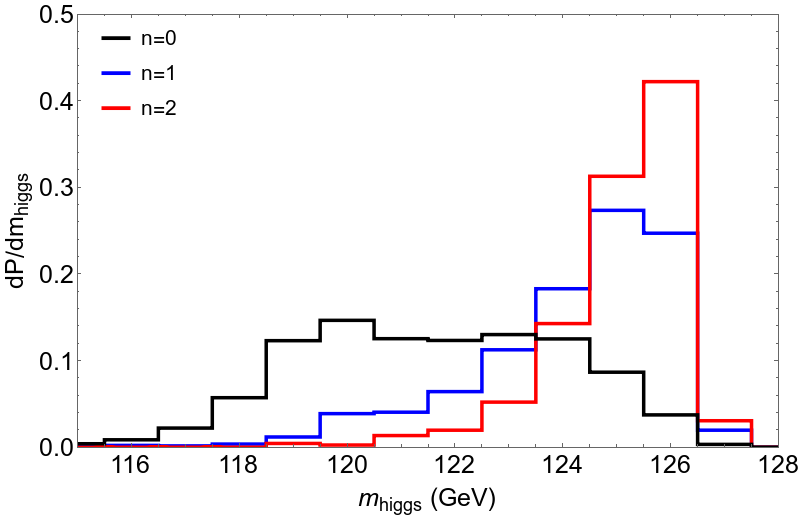}
\caption{Distribution in $m_h$ after requiring 
the anthropic selection of $m_{weak}<350$ GeV.
\label{fig:higgs}}
\end{center}
\end{figure}

In Fig. \ref{fig:higgs}, we show the landscape probability distribution 
$dP/dm_h$ vs. $m_h$ for various $n$ values.
For $n=0$, we find a broad spread of values ranging from $m_h\sim 119-125$ GeV. 
This may be expected for the $n=0$ case since we have a uniform scan 
in soft terms and low $\Delta_{\rm EW}$ can be found for
$A_0\sim 0$ which leads to little mixing in the stop sector and 
hence too light values of $m_h$. 
Taking $n=1$, instead we now see that the distribution in $m_h$ peaks 
at $\sim 125$ GeV with the bulk of probability between 
$123$ GeV $<m_h<$127 GeV-- in solid agreement with the measured value of 
$m_h=125.09\pm 0.24$ GeV\cite{pdg}.\footnote{Here, we rely on the Isajet 7.87 theory evaluation of $m_h$
which includes renormalization group improved 1-loop corrections to $m_h$ along with leading 
two-loop effects. Calculated values of $m_h$ are typically within 1-2 GeV of similar calculations from
the latest FeynHiggs\cite{feynhiggs} and SUSYHD\cite{susyhd} codes.} 
This may not be surprising since the landscape is pulling the various 
soft terms towards large values including large mixing in the Higgs sector 
which lifts up $m_h$ into the 125 GeV range. 
By requiring the $\Sigma_u^u(\tst_{1,2})/(m_Z^2/2)\alt 30$ 
(which would otherwise yield a weak scale in excess of 350 GeV) 
then too large of Higgs masses are vetoed. 
For the $n=2$ case with a stronger draw towards
large soft terms, the $m_h$ distribution hardens with a peak at 
$m_h\sim 126$ GeV.

\begin{figure}[tbp]
\begin{center}
\includegraphics[height=0.2\textheight]{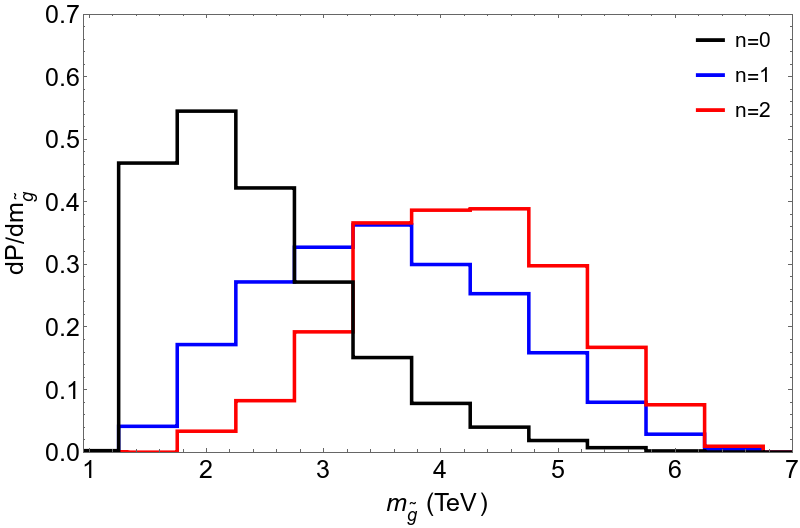}
\caption{Distribution in $m_{\tg}$ after requiring 
the anthropic selection of $m_{weak}<350$ GeV.
\label{fig:gl}}
\end{center}
\end{figure}

In Fig. \ref{fig:gl}, we show the distribution in gluino mass $m_{\tg}$. 
From the figure, we see that the $n=1$ distribution rises to a peak probability
around $m_{\tg}=3.5$ TeV. 
This may be compared to current LHC13 limits which require
$m_{\tg}\agt 2.2$ TeV\cite{lhc_mgl}. 
Thus, it appears LHC13 has {\it only begun} to explore the relevant
string theory predicted mass values. 
The distribution fall steadily such that essentially 
no probability exists for $m_{\tg}\agt 6$ TeV. 
This is because such heavy gluino masses lift the top-squark sector soft terms 
under RG running so that $\Sigma_u^u(\tst_{1,2})/(m_Z^2/2)$ then exceeds 30.
For $n=2$, the distribution is somewhat harder, peaking at around $m_{\tg}\sim 4.5$ TeV.
The uniform $n=0$ distribution peaks around 2 TeV.

\begin{figure}[tbp]
\begin{center}
\includegraphics[height=0.2\textheight]{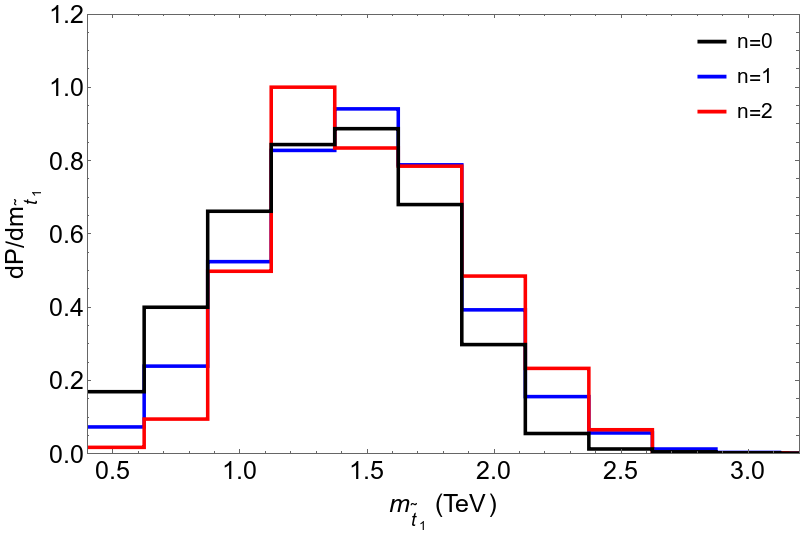}
\caption{Distribution in $m_{\tst_1}$ after requiring 
the anthropic selection of $m_{weak}<350$ GeV.
\label{fig:mt1}}
\end{center}
\end{figure}

In Fig. \ref{fig:mt1}, we show the probability distribution in $m_{\tst_1}$. In this case, all three $n$ values lead to a peak around 
$m_{\tst_1}\sim 1.5$ TeV. 
While this may seem surprising at first, in the case of $n=1,\ 2$ 
we gain large $A_t$ trilinear terms which lead to large mixing 
and a diminution of the eigenvalue $m_{\tst_1}$\cite{ltr} 
even though the soft terms entering the stop mass matrix may be increasing. 
There is not so much probability below 
$m_{\tst_1}=1$ TeV which corresponds to recent LHC13 mass limits\cite{lhc_mt1}. 
Thus, again, LHC13 has only begun to explore the predicted string theory 
parameter space.
The distributions taper off such that hardly any probability is left beyond 
$m_{\tst_1}\sim 2.5$ TeV. 
This upper limit is apparently within reach of high-energy LHC
operating with $\sqrt{s}\sim 27$ TeV where the reach in $m_{\tst_1}$ 
extends to about $2.5-3$ TeV\cite{lhc27}.

\begin{figure}[tbp]
\begin{center}
\includegraphics[height=0.2\textheight]{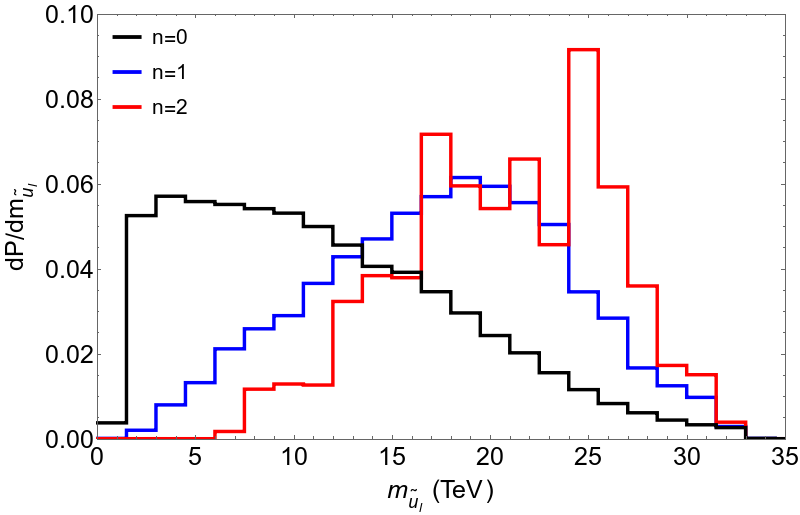}
\caption{Distribution in $m_{\tu_L}$ after requiring 
the anthropic selection of $m_{weak}<350$ GeV.
\label{fig:mul}}
\end{center}
\end{figure}

In Fig. \ref{fig:mul}, we show the distribution $dP/dm_{\tu_L}$ 
versus one of the first generation squark masses $m_{\tu_L}$. 
Here, it is found for $n=1,\ 2$ that the distribution peaks
around $m_{\tq}\sim 20-25$ TeV-- well beyond LHC sensitivity, 
but in the range to provide at least a partial decoupling solution 
to the SUSY flavor and CP problems\cite{flavor}.
It would also seem to reflect a rather heavy gravitino mass 
$m_{3/2}\sim 10-30$ TeV in accord 
with a decoupling solution to the cosmological gravitino problem\cite{linde}. 
The $n=0$  distribution peaks around $m_{\tq}\sim 8$ TeV and drops 
steadily to the vicinity of 40 TeV.
For much heavier squark masses, then two-loop RGE terms tend to drive the 
stop sector tachyonic resulting in CCB minima.

\subsection{Summary of landscape predictions for Higgs and sparticle masses}

From our $n=1,\ 2$ results which favor a value $m_h\sim 125$ GeV, then we also expect
\begin{itemize}
\item $m_{\tg}\sim 4\pm 2$ TeV,
\item $m_{\tst_1}\sim 1.5\pm 0.5$ TeV,
\item $m_A\sim 3\pm 2$ TeV,
\item $\tan\beta \sim 13\pm 7$,
\item $m_{\tw_1,\tz_{1,2}}\sim 200\pm 100$ GeV and
\item $m_{\tz_2}-m_{\tz_1}\sim 7\pm 3$ GeV with
\item $m(\tq ,\tell )\sim 20\pm 10$ TeV (for first/second generation matter scalars).
\end{itemize}
These results can provide some guidance as to SUSY searches at future colliders and also a convincing rationale
for why SUSY has so far eluded discovery at LHC.  
They provide a rationale for why SUSY might contain its own decoupling solution to the
SUSY flavor and CP problems and the cosmological gravitino and moduli problems.
They predict that precision electroweak and Higgs coupling measurements should look very
SM-like until the emergence of superpartners at LHC and/or ILC.
They also help explain why no WIMP signal has been seen: dark matter may be 
a higgsino-like-WIMP plus axion admixture
with far fewer WIMP targets than one might expect under a WIMP-only dark matter 
hypothesis\cite{Baer:2018rhs}.

\subsection{Related works on SUSY from the landscape}

A variety of other issues have been explored in SUSY from the landscape.
Below is a brief summary.
\bi
\item In Ref. \cite{Baer:2019xww}, LHC SUSY and WIMP dark matter search 
constraints confront the string theory landscape. 
In this case, it is seen that landscape SUSY typically lies well beyond 
current LHC search limits as presented for various simplified models.
In addition, the depleted WIMP abundance from landscape SUSY with a 
higgsino-like LSP lies below WIMP direct and indirect detection limits- 
in part because the WIMPs typically make up only 10-20\% of the dark matter
with the remainder consisting of axions.
\item In Ref. \cite{Baer:2019uom}, it is examined whether landscape SUSY
with the gravity-safe hybrid CCK mixed axion-higgsino-like WIMP 
dark sector can provide information on the magnitude of the PQ scale $f_a$.
In this case, since SUSY breaking determines $f_a$, an 
independent draw on PQ sector soft terms pulls $f_a$ beyond its sweet spot 
to yield overproduction of axion dark matter. The overproduction of axions 
cannot be compensated for by small misalignment angle (as suggested in 
Ref. \cite{wilczek}) since also large $f_a$ causes increased WIMP dark matter due to
late-time saxion and axino decays in the early universe. 
It is concluded that PQ sector soft terms must be correlated with visible sector
soft terms and thus lie within the cosmological sweet spot $f_a\sim 10^{11}$ GeV.
\item In Ref. \cite{flavor}, the possibility of a landscape solution to the SUSY
flavor and CP problems is investigated. Since the first and second generation soft terms are pulled to common upper bounds, then it is found that a mixed
decoupling/degeneracy solution emerges from the landscape with $n\ge 1$
so that the SUSY flavor and CP problems are solved.
\item In Ref. \cite{Baer:2019tee}, 
the case of mirage mediation from the landscape is 
examined wherein there is a landscape draw to large moduli-mediated
soft terms as compared to anomaly-mediated soft terms. In this case, for
a given value of $m_{3/2}$ (which can be measured in the MM scenario), 
then probability distributions for the mirage unification scale can be gained:
{\it e.g.} for $m_{3/2}=20$ TeV, then one expects gaugino masses to unify
around $\mu_{mir}\sim 10^{13-14}$ GeV. The overall Higgs and sparticle mass
predictions are similar to NUHM3 except that the gaugino spectrum is 
compressed. 
\ei

\subsection{Stringy naturalness}

For the case of the string theory landscape, in Ref. \cite{Douglas:2004zg} 
Douglas has introduced the concept of {\it stringy naturalness}:
\begin{quotation}
{\bf Stringy naturalness:} the value of an observable ${\cal O}_2$ 
is more natural than a value ${\cal O}_1$ if more 
{\it phenomenologically viable} vacua lead to  ${\cal O}_2$ than to ${\cal O}_1$.
\end{quotation}

We can compare the usual naturalness measures as shown in Fig's \ref{fig:cmssm}
and \ref{fig:nuhm2} against similar $m_0$ vs. $m_{1/2}$ planes under 
stringy naturalness.
We generate SUSY soft parameters
in accord with Eq.~\ref{eq:dNvac} for various values of 
$n=2n_F+n_D-1=1$ and 4.
The more stringy natural regions of parameter space are denoted by the higher
density of sampled points.
\begin{figure}[!htbp]
\begin{center}
\includegraphics[height=0.27\textheight]{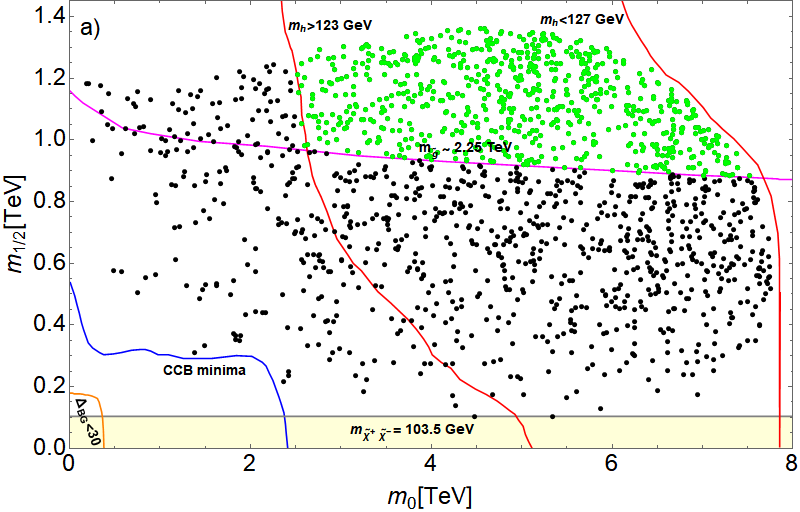}
\caption{The $m_0$ vs. $m_{1/2}$ plane of the NUHM2 model 
with $A_0=-1.6 m_0$, $\mu =200$ GeV and $m_A=2$ TeV and
an $n=1$ draw on soft terms,
The higher density of points denotes greater stringy naturalness.
The LHC Run 2 limit on $m_{\tg}>2.25$ TeV is shown by the magenta curve.
The lower yellow band is excluded by LEP2 chargino pair search limits.
The green points are LHC-allowed while black are LHC-excluded.
\label{fig:m0mhfn1}}
\end{center}
\end{figure}

In Fig. \ref{fig:m0mhfn1}, we show the stringy natural regions for the case
of $n=1$. 
Of course, no dots lie below the CCB boundary since such minima must be vetoed
as they likely lead to an unlivable pocket universe. 
Beyond the CCB contour, the solutions are in accord with livable vacua. 
But now the density of points {\it increases} with increasing 
$m_0$ and $m_{1/2}$ (linearly, for $n=1$), showing that the more stringy 
natural regions lie at the 
{\it highest} $m_0$ and $m_{1/2}$ values which are consistent with 
generating a weak scale within the Agrawal bounds. 
Beyond these bounds, the density of points of course drops to zero 
since contributions to the weak scale exceed its measured value by a factor 4. 
There is some fluidity of this latter bound
so that values of $\Delta_{EW}\sim 20-40$ might also be entertained. 
The result that stringy naturalness for
$n\ge 1$ favors the largest soft terms (subject to $m_Z^{PU}$ not ranging too far from
our measured value) stands in stark contrast to conventional naturalness
which favors instead the lower values of soft terms. 
Needless to say, the stringy natural
favored region of parameter space is in close accord with LHC results in that
LHC find $m_h=125$ GeV with no sign yet of sparticles.

\begin{figure}[!htbp]
\begin{center}
\includegraphics[height=0.27\textheight]{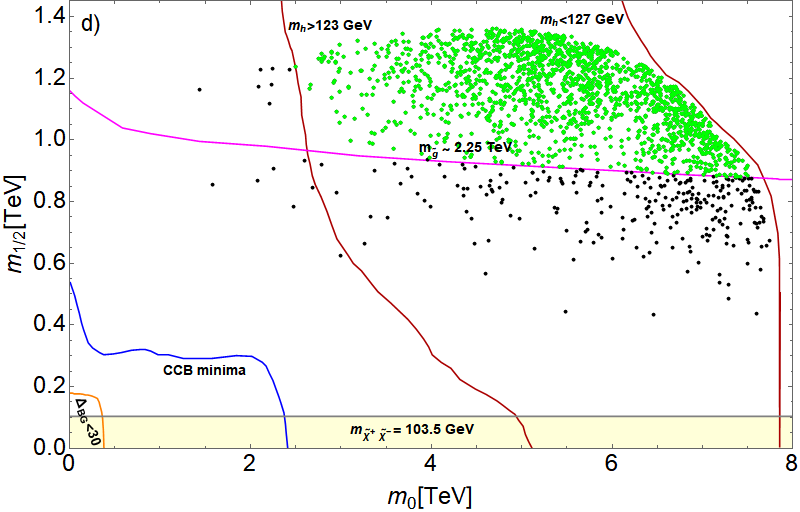}
\caption{The $m_0$ vs. $m_{1/2}$ plane of the NUHM2 model 
with $A_0=-1.6 m_0$, $\mu =200$ GeV and $m_A=2$ TeV and
an $n=4$ draw.
The higher density of points denotes greater stringy naturalness.
The LHC Run 2 limit on $m_{\tg}>2.25$ TeV is shown by the magenta curve.
The lower yellow band is excluded by LEP2 chargino pair search limits.
The green points are LHC-allowed while black are LHC-excluded.
\label{fig:m0mhfn4}}
\end{center}
\end{figure}

In Fig. \ref{fig:m0mhfn4}, we show the same plane under an $n=4$ draw on 
soft terms. In this case, the density of dots is clearly highest 
(corresponding to most stringy natural) at the largest values of 
$m_0$ and $m_{1/2}$ as opposed to Fig. \ref{fig:nuhm2} where the most natural
regions are at low $m_0$ and $m_{1/2}$. 
In this sense, under stringy naturalness, a 3 TeV gluino is more natural 
than a 300 GeV gluino!

\section{Comparison of landscape SUSY with other stringy scenarios}
\label{sec:compare}

\subsection{Mini-landscape}

A very practical avenue for linking string theory to weak scale
physics, known as the mini-landscape, has been investigated at some
length\cite{mini}. 
The methodology of the mini-landscape is to adopt a toy, 
but calculable, compactification onto a particular orbifold which is
engineered to yield a 4-d low energy theory with many of the properties 
of the MSSM. While compactification onto an orbifold may not be
ultimately realistic, it is manageable and can yield important lessons\cite{nv} 
as to how the MSSM might arise in more plausible Calabi-Yau 
compactifications. A key motivation is to aim for a compactification
which includes {\it local} SUSY grand unification\cite{localGUTs}, 
wherein different regions of the compact space exhibit different 
gauge symmetries--
perhaps including $SU(5)$, or better, $SO(10)$-- but where the
intersection of these symmetries leads to just the SM gauge group. 

Motivated by grand unification, the mini-landscape adopts the
$E_8\times E_8$ gauge structure of the heterotic string since one of
the $E_8$ groups automatically contains as sub-groups the grand unified
structures that the SM multiplets and quantum numbers seems to
reflect: $E_8\supset E_6\supset SO(10)\supset SU(5)\supset G_{\rm SM}$
where $G_{\rm SM}\equiv SU(3)_C\times SU(2)_L\times U(1)_Y$.  The other
$E_8$ may contain a hidden sector with $SU(n)$ subgroups which become
strongly interacting at some intermediate scale $\Lambda\sim 10^{13}$
GeV leading to gaugino condensation and consequent supergravity
breaking\cite{inocond}.  Compactification of the heterotic string on a
$Z_6-II$ orbifold\cite{Kobayashi:2004ya,Buchmuller:2005jr} 
can lead to low energy theories which include the MSSM, 
possibly with additional exotic, vector-like matter states (which may
decouple).

A detailed exploration of the mini-landscape has been performed a
number of years ago. In this picture, the properties of the 4-D low
energy theory are essentially determined by the geometry of the
6-D compactified space (orbifold), 
and by the location (geography) of the various superfields on this space. 
The gauge group of the 4-D theory is $G_{\rm SM}$ although the symmetry may be
enhanced for fields confined to fixed points, or to fixed tori, in the
extra dimensions. Examination of the models which lead to MSSM-like
structures revealed the following picture\cite{Nilles:2015qka}.
\begin{enumerate}
\item The first two generations of matter live at orbifold fixed
  points which exhibit the larger $SO(10)$ gauge symmetry 
(the twisted sector); thus, first
  and second generation fermions fill out the 16-dimensional spinor
  representation of $SO(10)$.
\item The Higgs multiplets $H_u$ and $H_d$ live in the untwisted
  sector and are bulk fields that feel just $G_{\rm SM}$.
  As such, the Higgs fields come in incomplete
  GUT multiplets which automatically solves the classic doublet-triplet
  splitting problem. The gauge superfields also live mainly in the bulk
  and thus occur in SM representations as well.
\item The third generation quark doublet and the top singlet also
  reside in the bulk, and thus have large overlap with the Higgs fields
  and correspondingly large Yukawa couplings. The location of other
  third generation matter fields is model dependent.  The small
  overlap of Higgs and first/second generation fields (which do not
  extend into the bulk) accounts for their much smaller Yukawa
  couplings.
\item Supergravity breaking may arise from hidden sector gaugino
  condensation with $m_{3/2}\sim \Lambda^3/m_{\rm P}^2$ with the gaugino
  condensation scale $\Lambda\sim 10^{13}$ GeV. SUSY breaking effects
  are felt differently by the various MSSM fields as these are located
  at different places on the orbifold. Specifically, the Higgs
  and top squark fields in the untwisted sector feel extended
  supersymmetry (at tree level) in 4-dimensions, and are thus more
  protected than the fields on orbifold fixed points which receive protection
  from just $N=1$ supersymmetry \cite{Krippendorf:2012ir}. 
First/second
generation matter scalars are thus expected with masses $\sim m_{3/2}$. 
Third generation and Higgs soft mass parameters (which enjoy the
added protection from extended SUSY) are suppressed by an additional
loop factor $\sim 4\pi^2 \sim \log (m_{\rm Pl}/m_{3/2})$. 
Gaugino masses and trilinear soft terms are
expected to be suppressed by the same factor. 
The suppression of various soft SUSY breaking terms means that 
(anomaly-mediated) loop contributions\cite{amsb} may be comparable 
to modulus- (gravity-) mediated contributions leading to models with mixed
moduli-anomaly mediation\cite{choi} (usually dubbed as 
{\it mirage mediation} or MM for short);
in the MM scenarios, gaugino masses apparently unify at some
intermediate scale
\be
\mu_{\rm mir}\sim m_{\rm GUT}e^{-8\pi^2/\alpha}, 
\label{eq:mumir}
\ee
where $\alpha$ parametrizes the relative amounts of moduli- versus anomaly-mediation. 
\end{enumerate} 

The spectrum of Higgs bosons and superpartners from the 
mini-landscape\cite{Baer:2017cck}
is thus expected to be rather similar to that expected from
the full landscape of MSSM theories provided both invoke a {\it natural}
solution to the SUSY $\mu$ problem with 
$\mu\sim 100-300$ GeV\cite{Baer:2019tee}.

\subsection{SUSY from IIB string models with moduli stabilization}

Upon compactification of string theory to our usual $4-d$ spacetime
along with a compact $6-d$ manifold, then one expects
a $4d$ effective supergravity theory containing at least the 
Standard Model fields along with a plethora of moduli fields-- 
massless gravitationally coupled scalar fields which gain mass from fluxes, 
perturbative corrections to the K\"ahler potential, or non-perturbative effects. 
The moduli-- grouped as to Hodge number $h^{1,1}$
K\"ahler moduli ($T_i$), $h^{1,2}$ complex structure moduli $(U_j)$
and the dilaton $S$-- once stabilized, obtain vevs which determine 
various parameters of the theory such as gauge and Yukawa couplings etc.
Thus, moduli stabilization is one key to making string theory 
predictive from a top-down approach. 
Two prominent scenarios for moduli-stabilization in type II-B 
string theory have emerged.

\subsubsection{KKLT} 

The KKLT\cite{kklt} scenario makes use of flux
compactifications as a route to stabilize the dilaton $S$ and all complex
structure moduli $U_j$ at mass scales of order $m_{string}$.
The SM fields are assumed localized on either a $D3$ or $D7$ brane 
within the compact space. In the original work, a single 
K\"ahler modulus $T$ is assumed, and it is assumed to be stabilized by
non-perturbative effects such as hidden sector gaugino condensation or
the presence of brane instantons leading to a hierarchically smaller 
mass $m_T\ll m_{string}$. Once all moduli are stabilized, then
one is led to a supersymmetric  effective theory with an AdS vacuum. 
The AdS minimum can be uplifted by effects such as adding an anti-$D3$ brane
at the tip of a Klebanov-Strassler throat
which breaks SUSY and generates a (metastable) de Sitter minimum as
required by observation. We note that there has been considerable 
recent debate on these steps in the context of the string swampland 
program\cite{Obied:2018sgi}.

The KKLT model is characterized by a mass hierarchy\cite{choi}
\be
m_T\gg m_{3/2}\gg m_{soft}
\ee
where the relative strengths are related by a factor 
$\log(m_P/m_{3/2})\sim 4\pi^2\sim 40$. Since one expects $m_{soft}\sim 1$ TeV, 
then $m_{3/2}\sim 40$ TeV and $m_T\sim 1600$ TeV. With such a hierarchy, 
then anomaly-mediated contributions to soft terms should be comparable
to moduli/gravity-mediated contributions and hence one is led to
mirage-mediation soft terms\cite{choi}. 
Typically a little hierarchy may arise as 
well between soft scalar masses and gaugino masses/$A$-terms. 
In such a scenario, then one might expect a mini-split mass 
hierarchy\cite{minisplit}
as shown in Table \ref{tab:stringSUSY} with $m_{gauginos}\ll m_{scalars}$.
In such a case, then one expects large $\Sigma_u^u(\tst_{1,2})$ 
contributions to $m_{weak}$ which must be tuned away.

\subsubsection{Large volume scenario (LVS)}

In the LVS\cite{Balasubramanian:2005zx}, again II-B flux compactification 
leads to stabilization of the dilaton and complex structure moduli.
In order to stabilize K\"ahler moduli, a compact manifold of the 
``swiss cheese'' variety is selected containing at least two cycles: one
large which sets the overall volume of the compact manifold 
(the overall size of the cheese), and the other(s)
quite small corresponding to holes in the cheese. Such a set-up leads to
comparable perturbative and non-perturbative contributions to the scalar
potential which allow for K\"ahler moduli stabilization but with an
exponentially large manifold volume leading to an effective theory 
valid up to some intermediate mass scale well below the GUT scale 
(thus perhaps not consistent with gauge coupling unification or GUTs). 
The large volume also leads to a disparity in the scales $m_{3/2}$ and $m_P$.
Unlike in KKLT, for LVS, the AdS vacuum already maintains broken SUSY.
Like KKLT, uplifting is required to gain a scalar potential of de Sitter type. 

The computation of soft terms in the LVS scenario\cite{Choi:2010gm} 
depends on a variety of
factors such as whether or not the SM lives on a D3 or a D7 brane and how 
visible sector moduli are stabilized: non-perturbatively or via $D$-terms.
The various choices lead to LVS models with typically very massive scalars
(leading to electroweak unnatural SUSY models). Computation of
soft terms using nilpotent goldstino fields and anti-$D3$-branes for 
uplifting were performed in Ref. \cite{Aparicio:2015psl}. 
For LVS with the SM located 
on a $D3$-brane, then a version of split SUSY is expected to ensue
with scalar masses in the $10^3-10^{11}$ GeV range but with weak scale 
gauginos. 
For LVS with the SM localized on a $D7$ brane, then high scale SUSY 
may be expected with all soft terms/sparticle masses in the 
$10^3-10^{11}$ GeV range, as detailed in Table~\ref{tab:stringSUSY}.

\subsection{$M$-theory compactified on manifold of $G_2$ holonomy}

In Refs. \cite{Acharya:2006ia,Acharya:2007rc,Acharya:2008zi},\footnote{
For recent reviews, see \cite{Acharya:2012tw} and \cite{Kumar:2015cva}.} 
the authors seek to derive general consequences from
11-dimensional $M$-theory compactified on a manifold of $G_2$ holonomy. 
Such a compactification preserves $N=1$ supersymmetry in the 4-d 
low energy effective theory,  a seemingly necessary phenomenological 
condition to stabilize the mass of the newly discovered Higgs boson.
Then, in the limit of small string coupling and small extra dimensions, 
the low energy limit of the theory is $N=1$, $d=4$ supergravity theory
which of necessity includes the MSSM (plus perhaps other exotic matter)
along with numerous moduli fields $s_i$ 
(gravitationally coupled scalar fields which
parametrize aspects of the compactification such as the size and shape of 
extra dimensions) and associated axion fields $a_i$. 
The low energy theory is assumed valid just below the Kaluza-Klein scale 
$m_{KK}$ which is of order $m_{GUT}\sim 2\times 10^{16}$ GeV.
The low energy effective SUGRA theory is then determined by the holomorphic
superpotential $W$, the holomorphic gauge kinetic function(s) $f_a$ 
(where $a$ labels the gauge group) and the real, non-holomorphic K\"ahler 
potential $K$. The field content of compactified $M$-theory thus contains
the usual matter and gauge superfields, moduli and axions, and possible hidden sector fields. The gravitino gains a mass via hidden sector SUSY breaking so 
that $m_{3/2}=\sqrt{\sum_i\langle F^i F_i\rangle}/\sqrt{3}m_P$ with 
$m_P$ the usual 4-d reduced Planck mass. The extra-dimensional gauge 
symmetry, upon compactification, leads to shift symmetries for the axionic 
fields which restrict the superpotential to exponentially suppressed 
non-perturbative contributions which give rise to suppressed 
(relative to $m_P$) scales $W\sim\Lambda^3\sim e^{-b/\alpha_Q}m_P^3$ 
plus other suppressed contributions from broken shift symmetries.
This results in an exponential hierarchy between $m_{3/2}$ and $m_P$.
With at least two hidden sector gauge groups, then all moduli become 
stabilized. By including hidden sector matter fields, then the AdS vacuum 
state is uplifted to de Sitter.

In the $G_2MSSM$ theory, the lightest modulus mass is determined to be of 
order $m_{3/2}$. To avoid the cosmological modulus problem (moduli 
decaying too late in the universe and thus upsetting BBN predictions) \cite{Kane:2015jia}, 
then $m_{LM}\sim m_{3/2}\sim 30-100$ TeV, where $m_{LM}$ is the mass of
the lightest of the moduli. SUSY breaking scalar mass soft terms are then 
expected to be of order $m_{3/2}$ along with small non-universal
contributions. Trilinear soft terms are also of order $m_{3/2}$. 
Gaugino masses are suppressed from scalar masses by a factor
$\log(m_P/m_{3/2})\sim 30 $ and are thus expected of order $\sim 1$ TeV
for $m_{3/2}\sim 30$ TeV. The suppressed gaugino masses are thus expected 
to have comparable moduli- and anomaly-mediated contributions so that the
gaugino masses are compressed but with a bino-like LSP. An overabundance of
bino-like dark matter is avoided because the light modulus fields alters
the relic density computation; its decay injects late-time entropy into the
early universe thus diluting all relics, but possibly adding to the LSP 
abundance: thus, a hallmark feature of this scenario is a non-thermal
mixture of axions and WIMPs\cite{Acharya:2009zt}. 
The $\mu$ parameter is expected to be 
suppressed by some emergent discrete symmetry but then re-generated
at a suppressed level compared to $m_{3/2}$ with $\mu\sim 1$ 
TeV\cite{Acharya:2011te}. 

Phenomenologically, the above discussion leads to a SUSY spectrum with
scalar masses $m_{\phi}\sim 30-100$ TeV but with a compressed spectrum 
of gauginos around the TeV scale and higgsinos also $\sim 1$ TeV.
Then, the resulting SUSY spectra may be accessible to LHC via gluino
pair production followed by $\tg$ cascade decays\cite{cascade} to
mainly 3rd generation quarks plus either a bino or a higgsino 
LSP\cite{Ellis:2014kla}.
The Higgs mass is expected at $m_h\sim 105-130$ GeV with the
region around 125 GeV preferred\cite{Kane:2011kj}.
In such a set-up, it is hard to understand why the weak scale 
exists at $m_{weak}\sim 100$ GeV whilst the $\mu$ parameter and
the $\Sigma_u^u(\tst_{1,2})$ contributions to $m_{weak}$ are very large
and hence require fine-tuning.

\begin{table}[!htb]
\renewcommand{\arraystretch}{1.2}
\begin{center}
\begin{tabular}{|c|c|c|c|c|c|c|}
\hline
soft term & landscape & mini-landscape & KKLT-D3/D7 & LVS-D3 & LVS-D7 & $G_2$MSSM \\
\hline
 gauginos & 1-1.5 TeV & $\sim 1$ TeV/mirage & $\sim 1$ TeV/mirage & $\sim 1$ TeV
& $\sim 10^{3-11}$ TeV & $\sim 1$ TeV \\
scalars(1,2) & 10-30 TeV & 10-30 TeV & $\sim 30$ TeV & $\sim 10^{3-11}$ TeV 
& $\sim 10^3$ TeV & $\sim 50$ TeV \\
scalars(3) & 1-5 TeV & 1-5 TeV & $\sim 30$ TeV & $\sim 10^3$ TeV & $\sim 10^{3-11}$ TeV & $\sim 50$ TeV \\
$A$-terms & $\sim -1.6m_0(3)$ & 1-5 TeV & $\sim 1$ TeV & $\sim 1$ TeV & $10^{3-11}$ TeV & $\sim 50$ TeV \\
$\mu$ & $0.1-0.3$ TeV & $\sim 0.1-1$ TeV & $\sim 1$ TeV & $\sim 1$ TeV & $\sim 10$ TeV & $\sim 1$ TeV \\
nickname  & natural/mirage & natural/mirage & mini-split/mirage & split & high-scale & mini-split/mirage \\ 
\hline
\hline
\end{tabular}
\caption{Expected mass range for soft terms/sparticles in a variety
of stringy SUSY models along with spectrum nick name.
}
\label{tab:stringSUSY}
\end{center}
\end{table} 

\section{Implications for SUSY collider searches}
\label{sec:colliders}

\subsection{Search for SUSY at LHC}

\subsubsection{LHC gluino pair searches}

In Ref.~\cite{Baer:2016wkz},  the  reach of HL-LHC for
gluino pair production was evaluated, assuming that 
$\tg\rightarrow t\tst_1$ and $\tst_1 \rightarrow b\tilde{\chi}_{1}^+$ 
or $t\tz_{1,2}$ and that the decay products of the higgsinos 
$\tilde{\chi}_{1}^\pm$ and $\tz_2$ are essentially invisible.
For events with $\eslt >900$ GeV, $n(jets)\ge 4$ and at least two tagged
$b$-jets (plus other cuts detailed in Ref. \cite{Baer:2016wkz}), 
it was found that HL-LHC had a $5\sigma$ reach for $m_{\tg}$ of
2.4 (2.6) ((2.8)) TeV for 300 (1000) ((3000)) fb$^{-1}$, respectively.

In Ref.~\cite{Baer:2017pba}, the reach of high energy LHC (HE-LHC, LHC
with $\sqrt{s}=27$ TeV)  
for both gluinos and top-squarks in the light higgsino scenario 
was evaluated but with $\sqrt{s}=33$ TeV. 
These results were updated for HE-LHC with $\sqrt{s}=27$ TeV and 
15 fb$^{-1}$ of integrated luminosity 
in Ref.~\cite{lhc27} where more details can be found.
A combination of Madgraph, Pythia and Delphes 
was used to simulate SUSY signal events and SM backgrounds.
SM backgrounds included $t\bar{t}$, $t\bar{t}b\bar{b}$, $t\bar{t}t\bar{t}$, 
$t\bar{t}Z$, $t\bar{t}h$, $b\bar{b}Z$ and single top production. 
We require at least four hard jets, with two or more tagged  as  $b$-jets,
no isolated leptons and hard MET and $p_T(jet)$ cuts.

\begin{figure}[tbp]
\begin{center}
\includegraphics[width=0.5\textwidth]{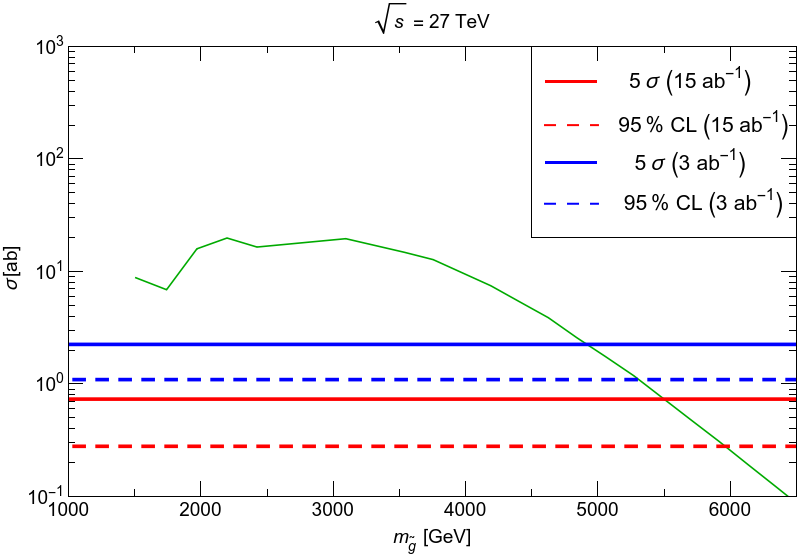}
\caption{Plot of gluino pair production cross section vs. $m_{\tg}$ 
after cuts at HE-LHC with $\sqrt{s}=27$ TeV (green curve). 
We also show the $5\sigma$ reach and 95\% CL exclusion lines
assuming 3 and 15 ab$^{-1}$ of integrated luminosity.
\label{fig:mgl}}
\end{center}
\end{figure}

Our results are shown in Fig. \ref{fig:mgl} where we plot the gluino pair
production signal versus $m_{\tg}$ for a natural NUHM2 model line with 
parameter choice $m_0=5m_{1/2}$, $A_0=-1.6 m_0$, $m_A=m_{1/2}$,
$\tan\beta =10$ and $\mu =150$ GeV with varying $m_{1/2}$. 
We do not expect the results to be sensitive to this precise choice 
as long as first generation squarks are much heavier than gluinos.
From the figure, we see that the $5\sigma$ discovery reach of HE-LHC
extends to $m_{\tg}=4900$ GeV for 3 ab$^{-1}$ and to $m_{\tg}=5500$ GeV
for 15 ab$^{-1}$ of integrated luminosity. The corresponding $95\%$ CL exclusion
reaches extend to $m_{\tg}=5300$ GeV for 3 ab$^{-1}$ and to $m_{\tg}=5900$ GeV
for 15 ab$^{-1}$ of integrated luminosity. For comparison, the $5\sigma$
discovery reach of LHC14 is (2.4) 2.8~TeV for an integrated luminosity
of (300) 3000~fb$^{-1}$~\cite{Baer:2016wkz}.%: see \fig{fig:figreach}  

\subsubsection{LHC top squark pair searches}

In Ref. \cite{atlas_mt1}, the HL-LHC reach for top-squark pair production 
was evaluated assuming LHC14 with 3000 fb$^{-1}$.
The 
%For comparison, the Atlas projected (ATLAS-PHYS-PUB-2018-021) 
$95\%$ CL LHC14 reach with 3000 fb$^{-1}$ 
extends to $m_{\tst_1} \simeq  1700$~GeV.

In Ref.~\cite{Baer:2017pba}, the reach of a 33 TeV LHC upgrade for 
top-squark pair production was investigated. This analysis was repeated 
using the updated LHC energy upgrade $\sqrt{s}=27$ TeV. 
A combination of Madgraph, Pythia and Delphes was again used for
SUSY signal and SM background calculations.
Top-squark pair production events were generated within a
simplified model where $\tst_1\to b\tw_1^+$ at 50\%, and
$\tst_1\to t\tz_{1,2}$ each at 25\% branching fraction, which are typical
of most SUSY models~\cite{Baer:2016bwh} with light higgsinos. 
The higgsino-like electroweakino masses are $m_{\tz_{1,2},\tw_1^\pm}\simeq 150$ GeV.
We required at least two hard $b$-jets, no isolated leptons and 
hard $\eslt$ and $p_T(jet)$ cuts: see \cite{lhc27} for details.

\begin{figure}[tbp]
\begin{center}
\includegraphics[width=0.5\textwidth]{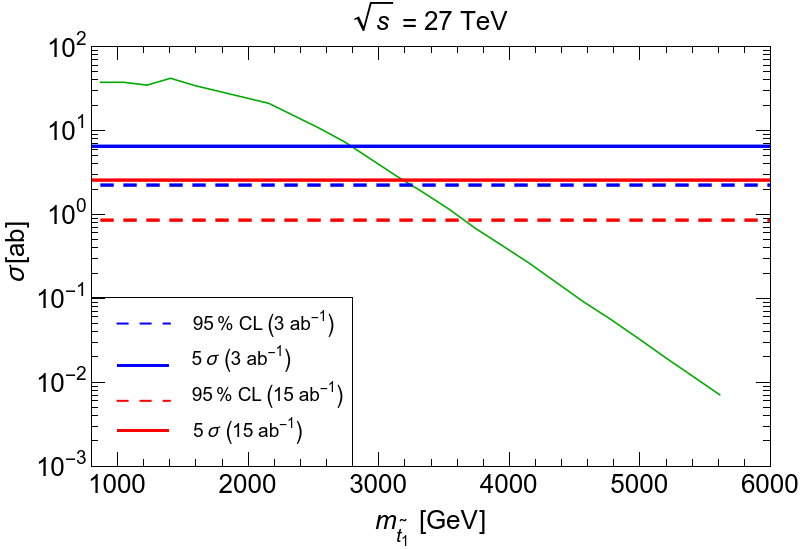}
\caption{Plot of top-squark pair production cross section vs. $m_{\tst_1}$ 
after cuts at HE-LHC with $\sqrt{s}=27$ TeV (green curve). 
We also show the $5\sigma$ reach  and 95\% CL exclusion lines,
assuming  3 and 15 ab$^{-1}$ of integrated luminosity.
\label{fig:mt1prime}}
\end{center}
\end{figure}

Using these background rates for LHC at $\sqrt{s}=27$ TeV, we compute the
$5\sigma$ reach and $95\%$ CL exclusion of HE-LHC for 3 and 15 ab$^{-1}$ of
integrated luminosity using Poisson statistics. 
Our results are shown in Fig. \ref{fig:mt1prime} along with the
top-squark pair production cross section after cuts versus $m_{\tst_1}$.
From the figure, we see the $5\sigma$ discovery reach of HE-LHC
extends to $m_{\tst_1}=2800$ GeV for 3 ab$^{-1}$ and to 3160 GeV for
15 ab$^{-1}$. The 95\% CL exclusion limits extend to $m_{\tst_1}=3250$ GeV
for 3 ab$^{-1}$ and to $m_{\tst_1}=3650$ GeV for 15 ab$^{-1}$. 
We checked that $S/B$ exceeds 0.8 whenever we deem the signal to be 
observable\cite{lhc27}.

\subsubsection{Combined LHC reach for stops and gluinos}

\begin{figure}[tbp]
\begin{center}
\includegraphics[width=0.5\textwidth]{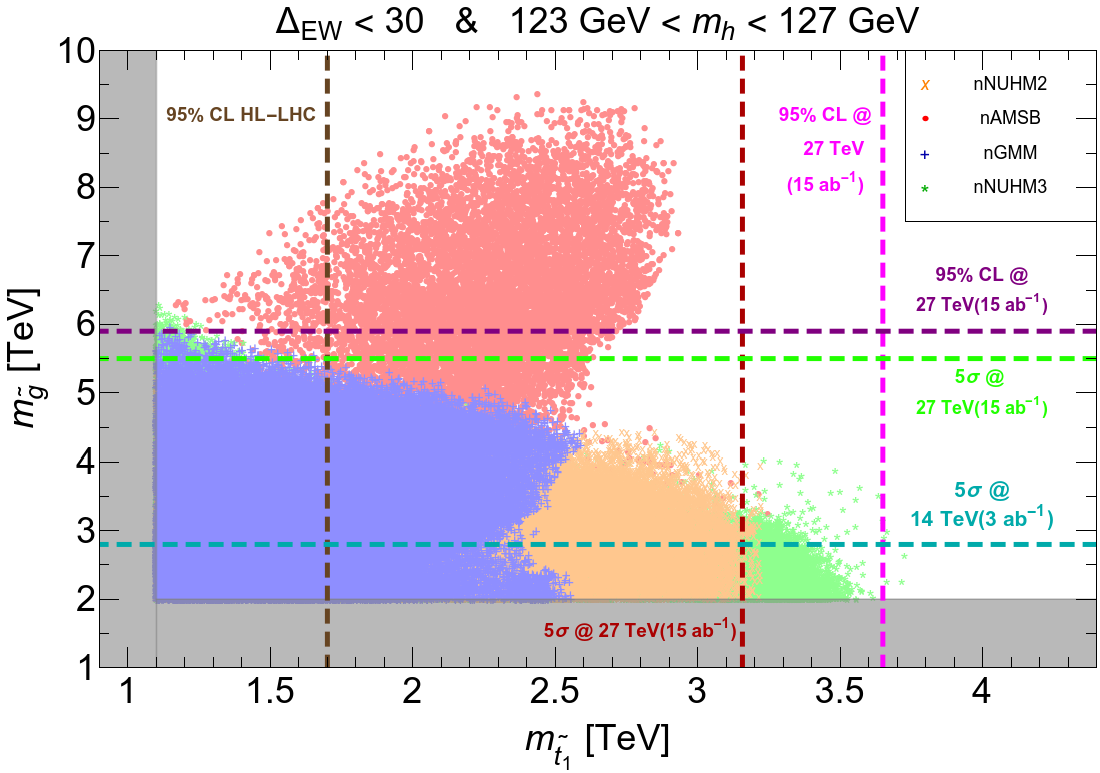}
\caption{Plot of points in the $m_{\tst_1}$ vs. $m_{\tg}$ plane from a
  scan over nNUHM2, nNUHM3, nGMM and nAMSB model parameter space.  We
  compare to recent search limits from the ATLAS/CMS experiments and show
  the projected reach of HL- and HE-LHC. The gray-shaded regions are already
excluded by LHC gluino and top-squark searches.
\label{fig:mt1mgl}}
\end{center}
\end{figure}

In Fig.~\ref{fig:mt1mgl} we exhibit the gluino and top-squark reach
values in the $m_{\tst_1}$ vs. $m_{\tg}$ plane.  We compare the reach of
HL- and HE-LHC to values of gluino and stop masses (shown by the dots)
in a variety of natural SUSY models defined to have $\Delta_{\rm EW} <
30$~\cite{ltr,rns},
including the two- and three-extra parameter
non-universal Higgs models~\cite{nuhm2} (nNUHM2 and nNUHM3),
natural generalized mirage mediation~\cite{nGMM} (nGMM) and
natural anomaly-mediation~\cite{nAMSB} (nAMSB).
These models all allow
for input of the SUSY $\mu$ parameter at values $\mu\sim 100-350$ GeV which
is a necessary (though not sufficient) condition for naturalness in the
MSSM. 

The highlight of this figure is that at least one of the gluino or the
stop should be discoverable at the HE-LHC.  We also see that in natural
SUSY models (with the exception of nAMSB), the highest values of
$m_{\tg}$ coincide with the lowest values of $m_{\tst_1}$ while the
highest top squark masses occur at the lowest gluino masses.  Thus, a
marginal signal in one channel (due to the sparticle mass being near
their upper limit) should be accompanied by a robust signal in the other
channel.  Over most of the parameter range of weak scale natural SUSY
there should be a $5\sigma$ signal in {\it both} the top-squark and
gluino pair production channels at HE-LHC.

\subsubsection{LHC wino pair searches}

\begin{figure}[tbp]
\includegraphics[height=0.2\textheight]{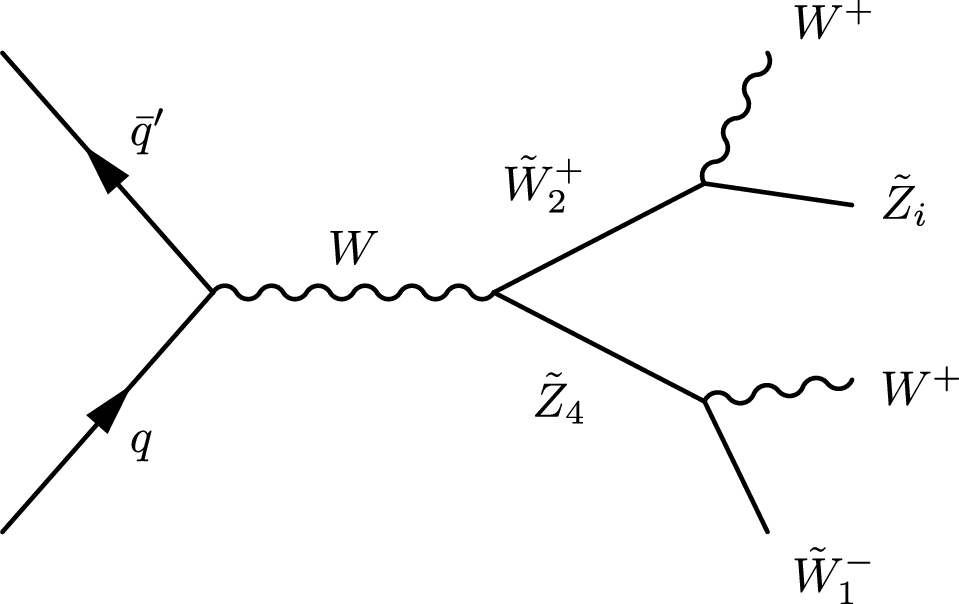}
\caption{Feynman diagram for $pp\to \tw_2^+\tz_4$ production
followed by $\tw_2^+\to W^+\tz_i$ and $\tz_4\to W^+\tw_1^-$
leading to the clean same-sign diboson signature.
\label{fig:w2z4}}
\end{figure}

The wino pair production reaction
$pp\to\tilde{\chi}_2^\pm\tilde{\chi}_4^0$ can occur at observable rates
for SUSY models with light higgsinos.  The decays
$\tilde{\chi}_2^\pm\rightarrow W^\pm\tilde{\chi}_{1,2}^0$ and
$\tilde{\chi}_4^0\rightarrow W^\pm\tilde{\chi}_1^\mp$ lead to final
state dibosons which half the time give a relatively jet-free same-sign
diboson signature (SSdB) which has only tiny SM
backgrounds~\cite{Baer:2013yha,Baer:2013xua,Baer:2017gzf}:
see Fig. \ref{fig:w2z4}.

\begin{figure}[tbp]
\begin{center}
\includegraphics[width=0.5\textwidth]{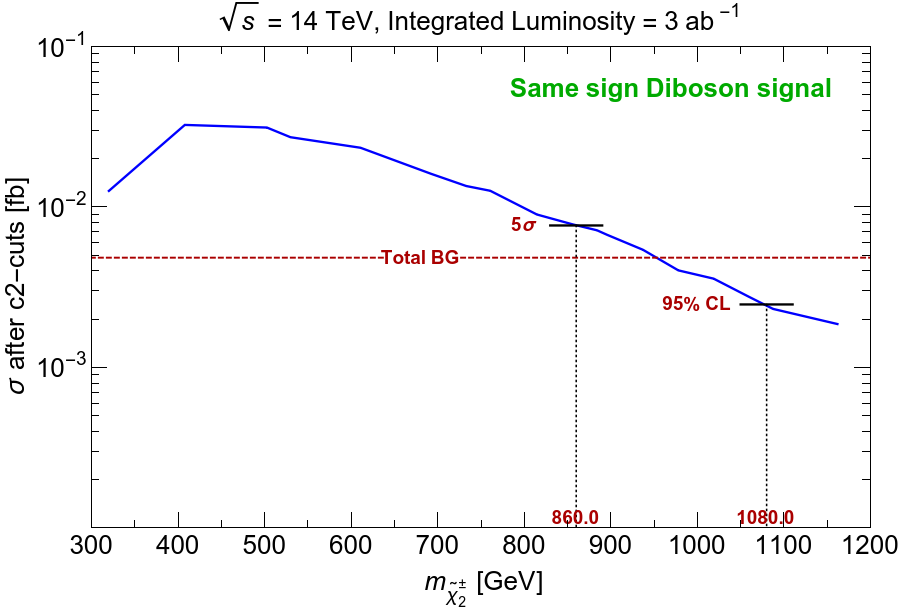}
\caption{Cross section for SSdB production 
(after C2 cuts as delineated in Ref.~\cite{Baer:2017gzf}) 
versus wino mass at the LHC with $\sqrt{s}=14$ TeV. 
We show the $5\sigma$ and 95\% CL reach 
assuming a HL-LHC integrated luminosity of 3 ab$^{-1}$.
\label{fig:ssdb}}
\end{center}
\end{figure}

We have computed the reach of HL-LHC for the SSdB signature 
in Fig. \ref{fig:ssdb} including $t\bar{t}$, $WZ$, $t\bar{t}W$,
$t\bar{t}Z$, $t\bar{t}t\bar{t}$, $WWW$ and $WWjj$ backgrounds. 
We see that for LHC14 with 3 ab$^{-1}$ of integrated luminosity,
the $5\sigma$ reach extends to $m(wino)\sim 860$ GeV while
the $95\%$ CL exclusion extends to $m(wino)\sim 1080$ GeV.
In models with unified gaugino masses, these would correspond to 
a reach in terms of $m_{\tg}$ of 2.4 (3) TeV respectively.
These values are comparable to what LHC14 can achieve via gluino pair
searches with 3 ab$^{-1}$. The SSdB signature is distinctive for the case of
SUSY models with light higgsinos. 

While Fig. \ref{fig:ssdb} presents the HL-LHC reach for SUSY in the
SSdB channel, the corresponding reach of HE-LHC has not yet 
been computed.
The SSdB signal arises via  EW production, and  
the signal rates are expected to rise by a factor of a few 
by moving from $\sqrt{s}=14$ TeV to $\sqrt{s}=27$ TeV. In contrast,
some of the QCD backgrounds like $t\bar{t}$
production will rise by much larger factors.
Thus, it is not yet clear whether the reach for SUSY in the SSdB 
channel will be increased by moving from HL-LHC to HE-LHC.
We note though that other signals channels from wino decays to higgsinos
plus a $W$, $Z$ and Higgs boson may offer further SUSY detection
possibilities.

\subsubsection{LHC higgsino pair searches}

\begin{figure}[tbp]
\includegraphics[height=0.3\textheight]{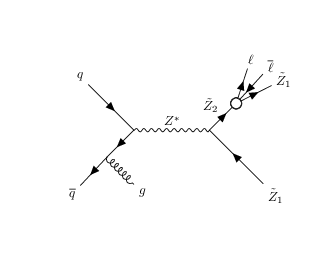}
\caption{Feynman diagram for $pp\to \tz_1\tz_2$ production
followed by $\tz_2\to\ell^+\ell^-\tz_1$ plus radiation of
a gluon jet from the initial state.
\label{fig:z1z2}}
\end{figure}

The four higgsino-like charginos $\tw_1^\pm$ and neutralinos $\tz_{1,2}$
are the only SUSY particles required by naturalness to lie near to the
weak scale at $m_{weak}\sim 100$ GeV.  In spite of their lightness, they
are very challenging to detect at LHC.  The lightest neutralino
evidently comprises just a portion of dark matter~\cite{Baer:2018rhs},
and if produced at LHC via $pp\to\tz_1\tz_1,\ \tilde{\chi}_1^\pm
\tilde{\chi}_1^\mp\ {\rm and}\ \tilde{\chi}_1^\pm\tilde{\chi}_{1,2}^0$
could escape detection. This is because the decay products of $\tilde{\chi}_2^0$
and $\tilde{\chi}_{1}^\pm$ are expected to be very soft, causing the
signal to be well below SM processes like $WW$ and $t\bar{t}$
production.  The monojet signal arising from initial state QCD radiation
$pp\to\tz_1\tz_1 j$, $\tilde{\chi}_1^\pm\tilde{\chi}_1^\mp j$ and
$\tilde{\chi}_1^\pm\tilde{\chi}_{1,2}^0 j$ has been evaluated in
\cite{Baer:2014cua} and was found to have similar shape
distributions to the dominant $pp\to Zj$ background but with background
levels about 100 times larger than signal. 
However, at HE-LHC harder monojet cuts may be possible~\cite{Han:2018wus}.

\begin{figure}[tbp]
\begin{center}
\includegraphics[width=0.5\textwidth]{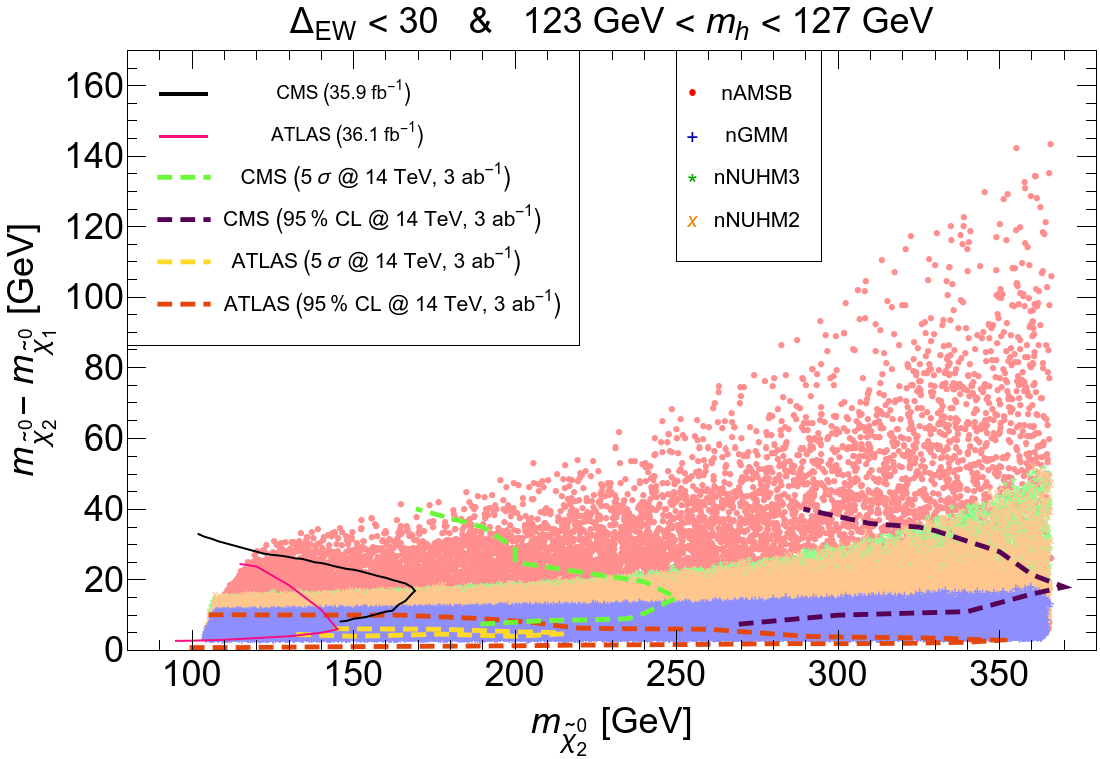}
\caption{Plot of points in the $m_{\tz_2}$ vs. $m_{\tz_2}-m_{\tz_1}$ plane 
from a scan over nNUHM2, nNUHM3, nGMM and nAMSB model parameter space.
We compare to recent search limits from the ATLAS/CMS experiments
and to future reach contours for HL-LHC (from Ref. \cite{Baer:2019xww}).
\label{fig:mz2mz1}}
\end{center}
\end{figure}

A way forward has been proposed via the $pp\to\tz_1\tz_2 j$ channel
where $\tz_2\to\ell^+\ell^-\tz_1$\cite{Baer:2011ec}: 
a soft OS dilepton pair recoils against a hard initial state jet 
radiation which serves as a trigger~\cite{Han:2014kaa,Baer:2014kya}:
see Fig. \ref{fig:z1z2}.
Recent searches in this $\ell^+\ell^-j+MET$ 
channel have been performed by CMS~\cite{CMS:2017fij} 
and by ATLAS~\cite{Aaboud:2017leg}.\footnote{
The ATLAS collaboration has recently completed an updated study of this
reaction using 139 fb$^{-1}$ of data\cite{Aad:2019qnd}.}
Their resultant reach contours are shown as solid black and red contours 
respectively in the $m_{\tz_2}$ vs. $m_{\tz_2}-m_{\tz_1}$ plane in Fig. \ref{fig:mz2mz1}. 
These searches have indeed probed a portion of promising parameter space
since the lighter $m_{\tz_2}$ masses are preferred by naturalness. 
The ATLAS\cite{ATLAS:2018jjf} and CMS experiments\cite{CidVidal:2018eel}
have computed some $5\sigma$ and $95\%$ CL projected reach 
contours for HL-LHC with 3 ab$^{-1}$ as the yellow, green,
purple and red dashed contours. 
We see these contours can probe considerably more parameter space 
although some of natural SUSY parameter space (shown by dots for the
same set of models as in Fig.~\ref{fig:mt1mgl}) might lie beyond these
projected reaches. So far, reach contours for HE-LHC in this search 
channel have not been computed but it is again anticipated that HE-LHC will 
not be greatly beneficial here since $pp\to\tz_1\tz_2$ is an electroweak
production process so the signal cross section will increase only marginally
while QCD background processes like $t\bar{t}$ production will increase
substantially. 

It is imperative that future search channels try to squeeze
their reach to the lowest $m_{\tz_2}-m_{\tz_1}$ mass gaps which are 
favored to lie in the 3-5 GeV region for string landscape 
projections~\cite{Baer:2017uvn} of SUSY mass spectra. 
The ATLAS red-dashed contour appears to go a long way in this regard, 
though the corresponding $5\sigma$ reach is considerably smaller.

\subsubsection{Conclusions for natural SUSY at HL- and HE-LHC:} 

We have delineated the reach of the HE-LHC and compared it to the
corresponding reach of the HL-LHC for SUSY models with light higgsinos,
expected in a variety of natural SUSY models. While the HL-LHC increases
the SUSY search range (and may probe the bulk of natural SUSY parameter space
at 95\% CL in the soft $\ell^+\ell^- j+MET$ channel), 
it appears that the HE-LHC will definitively
probe natural SUSY models with $\Delta_{\rm EW} < 30$ via a $5\sigma$
discovery of at least one of the top squark or the gluino (likely even
both), possibly also with signals in other channels.
Thus, we strongly recommend the construction of an upgraded or new  hadron 
collider with $\sqrt{s}\sim 27-100$ TeV in order to fully
test natural weak scale SUSY.
 
\subsection{ILC searches}

The International Linear $e^+e^-$ Collider, or ILC, is a proposed linear
$e^+e^-$ collider to be built in Japan at an initial energy of 
$\sqrt{s}=250$ as a Higgs factory. It is expected to be upgradable 
at later stages to $\sqrt{s}=500$ and perhaps even 1000 GeV.

\subsubsection{Precision measurements at a Higgs factory}

The goal of the initial stage of ILC operating at $\sqrt{s}=250$ GeV is
to make detailed precision measurements of the properties of the 
newly discovered Higgs boson with $m_h\simeq 125$ GeV, mainly via
$e^+e^-\to Zh$ production.
While greater precision on the Higgs boson mass and spin quantum numbers is
always welcome, a more tantalizing avenue towards new physics will be
precision measurement of the various Higgs boson decay modes and branching
fractions. The presence of new particles, or else virtual effects and modified
couplings from physics beyond the Standard Model, are expected to modify
the quantities $\kappa_{\tau,b}$, $\kappa_t$, $\kappa_{W,Z}$ $\kappa_g$ 
and $\kappa_\gamma$ which parametrize the ratio of the 
measured Higgs coupling to SM particles as compared to 
the coupling as expected from the SM: {\it e.g.} 
$\kappa_b\equiv g_{hb\bar{b}}/g_{hb\bar{b}}(SM)$.

In Ref. \cite{Bae:2015nva}, 
a detailed study of expected values of the $\kappa_i$
was made for natural SUSY models with $\Delta_{EW}<30$ and where the models
also obeyed LHC8 sparticle and heavy Higgs mass constraints, $m_h=125\pm 2$ GeV 
and bounds from $B\to X_s\gamma$ and $B_s\to \mu^+\mu^-$ decay rates. 
The presence of two Higgs doublets in the MSSM leads to modified Higgs 
couplings while the presence of superpartners can modify couplings
such as $hgg$ and $h\gamma\gamma$ which occur via loop effects.
In that work, it was typically found that the bulk of allowed natural SUSY
parameter space leads to very SM-like Higgs couplings since the required 
rather heavy SUSY particles (except higgsinos) largely decouple and 
Higgs mixing effects are small. 
If these results are updated to include LHC Run 2 search results then 
the expected Higgs couplings will become even more SM-like. 
Furthermore, in the string landscape picture where soft terms and hence 
sparticle masses (other than higgsinos) are drawn to large values, then the
$\kappa_i$ values become even more SM-like. While exceptions can occur, 
for instance if $m_{A,H,H^\pm}$ are in the few hundred GeV range and $\tan\beta$
is small, the general expectation for landscape SUSY 
is that the ILC Higgs factory precision measurements will see a 
very SM-like Higgs boson.

\subsubsection{Higgsino pair production}

While the string landscape is expected to pull soft SUSY breaking
terms to large values (subject to not-too-large of contributions to
generating a weak scale with $m_{weak}\sim 100$ GeV), 
the same is not true of the SUSY preserving $\mu$ parameter 
which sets the mass of the lightest higgsino-like electroweakinos. 
Thus, these latter particles 
$\tz_{1,2}$ and $\tw_1^\pm$ offer lucrative targets for an $e^+e^-$ 
collider operating with $\sqrt{s}>2m(higgsino)\sim 500-600$ GeV\cite{Baer:2011ec}. 
The energy upgrade of the International Linear Collider (ILC) is such a 
machine\cite{Fujii:2017ekh}. 

\begin{figure}[tbp]
\begin{center}
\includegraphics[width=12cm,clip]{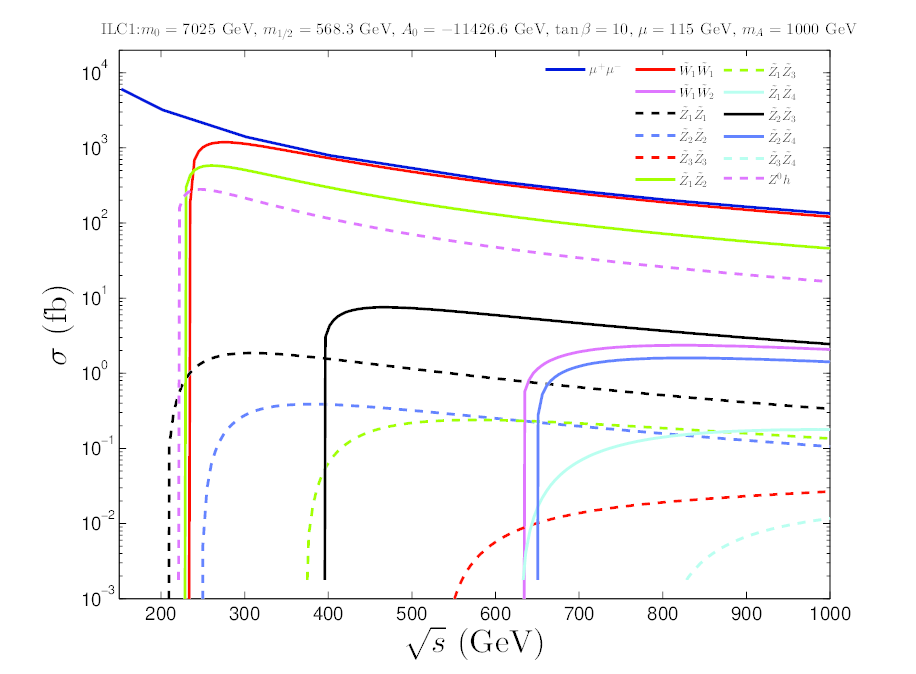}
\caption{Sparticle production cross sections vs. $\sqrt{s}$ for
  unpolarized beams at an $e^+e^-$ collider for the benchmark point
labelled as ILC1 in Ref. \cite{Baer:2014yta}.}
\label{fig:xsec}
\end{center}
\end{figure}

In Fig.~\ref{fig:xsec}, we show the total production cross sections for
 a variety of SUSY signal reactions along with dominant SM backgrounds
for a typical SUSY mass spectrum from radiative natural SUSY with 
$\mu\sim 115$ GeV vs. $\sqrt{s}$ of an $e^+e^-$ collider.
We see that once the energy threshold $\sqrt{s}=2m(higgsino)$ is passed,
then there is a rapid rise in the production cross sections for
$e^+e^-\to \tz_1\tz_2$ and $\tw_1^+\tw_1^-$. Since the mass gaps
$m_{\tz_2}-m_{\tz_1}$ and $m_{\tw_1}-m_{\tz_1}$ are typically $5-15$ GeV, then
most of the beam energy goes into making the dark matter mass $2m_{\tz_1}$
and the visible decay products of $\tz_2\to f\bar{f}\tz_1$ and
$\tw_1^\pm\to f\bar{f}^{\prime}\tz_1$ (the $f$ and $f^{\prime}$ are SM fermions) 
are quite soft. 
Nonetheless, the clean operating environment of an $e^+e^-$ collider will 
have no trouble picking out such new physics signal events from more 
energetic SM backgrounds. Detailed analyses are presented in 
Refs. \cite{Baer:2014yta} and \cite{Baer:2019gvu}.
For these reactions, precision measurements of the difermion invariant mass
and energy distributions will allow ILC measurement of $m_{\tw_1}$, 
$m_{\tz_1}$ and $m_{\tz_2}$ to percent level precision. This will also allow
the SUSY $\mu$ parameter to be measured. The higgsino mass splittings 
are sensitive to the gaugino masses $M_1$ and $M_2$ so these can be 
extracted as well. Extrapolation of the measured gaugino masses to high
energy using the RGEs will allow for tests of gaugino mass unification at the
GUT scale or at some intermediate mass scale as expected in mirage 
mediation\cite{Baer:2019gvu}.
Also, extraction of the gaugino vs. higgsino content of the
light electroweakinos will allow for insights into the 
dark matter content of the universe.

\subsection{Search for SUSY via lepton flavor violation (LFV)}

A complementary way to search for SUSY is via SUSY virtual effects
on rare, lepton-flavor violating processes. Such processes include
{\it i}). search for $\mu\to e\gamma$ decay, {\it ii}). search
for $\tau\to \mu\gamma$, {\it iii}). search for $\mu\to eee$ decay, 
and search for $\mu\to e$ conversion via $\mu N\to e N$ where $N$
denotes a nuclear target. These various processes have been evaluated
for the case of natural SUSY with $\Delta_{EW}<30$ 
in Ref. \cite{Baer:2019xug} (for related work, see Ref. \cite{Han:2020exx}). 
The results depend strongly on the assumed form of the
neutrino Yukawa matrix ${\bf f}_{\nu}$. For large mixing similar to 
the PMNS mixing matrix, then these processes are typically observable while for
small mixing similar to the CKM matrix, then the expected rates are typically
below projected sensitivity of upcoming experiment like MEG-II, Belle-II
and Mu3e. The results also depend sensitively on whether the SUSY scalars
obey a normal mass hierarchy with $m_0(1,2)\ll m_0(3)$ or an 
inverted scalar mass hierarchy $m_0(3)\ll m_0(1,2)$ as expected 
from the string landscape and mini-landscape. In the former case, where
smuons are lighter, then rates are more promising whilst in the latter case
where smuons and muon sneutrinos inhabit the tens-of-TeV regime, 
then again rates are suppressed.

\subsubsection{$(g-2)_\mu$}

The above results recall the light/heavy smuon controversy which arises from
$(g-2)_\mu$. Current data matched to Standard Model theory predictions 
find a more than
$3\sigma$ discrepancy between these values\cite{Davier:2019can}. 
This discrepancy could be explained by the presence of light smuons with mass
$m_{\tmu}\sim 0.1-1$ TeV (although so far, LHC has seen no sign of these). 
However, a recent ab initio lattice evaluation of the leading order hadronic
vacuum polarization produce theory predictions in close alignment with
the measured $(g-2)_\mu$ value\cite{Borsanyi:2020mff}. 
These latter results would be in accord with
our expectations for SUSY from the string theory landscape, where one
expects smuons in the tens-of-TeV regime, and hence close alignment between
SM theory and experiment.

\section{Mixed axion plus higgsino-like WIMP dark matter}
\label{sec:DM}

We have seen that solving the weak scale naturalness problem requires the 
introduction of weak scale SUSY while solving the QCD naturalness problem
requires the PQWW invisible axion\cite{pqww,ksvz,dfsz}. 
The SUSY DFSZ axion naturally solves the SUSY $\mu$ problem while yielding
a Little Hiararchy $\mu\ll m_{soft}$. A gravity-safe axionic solution to
the strong CP problem can emerge from  a strong enough anomaly-free 
discrete $R$-symmetry ${\bf Z}_{24}^R$. In that case, both $U(1)_{PQ}$ and
$R$-parity emerge as accidental, approximate symmetries from the more
fundamental discrete $R$ symmetry which in turn may emerge from
compactification of 10-d string theory to 4-d. In this very attractive scenario,
then dark matter is expected to consist of two particles:
a higgsino-like WIMP which is LSP and a SUSY DFSZ axion.
Typically, the higgsino-like WIMPs are thermally underproduced
with $\Omega_{\tz_1}^{TP}h^2\sim (0.1-0.2)\times 0.12$ so that the bulk of 
dark matter is made of SUSY DFSZ axions. However, now one must include as 
well the axion superpartners axino $\ta$ and saxion $s$ into the relic 
density calculation (along with gravitinos).

\subsection{Relic density of mixed axion-higgsino-like WIMP dark matter}

\begin{figure}[tbp]
\begin{center}
\includegraphics[height=0.3\textheight]{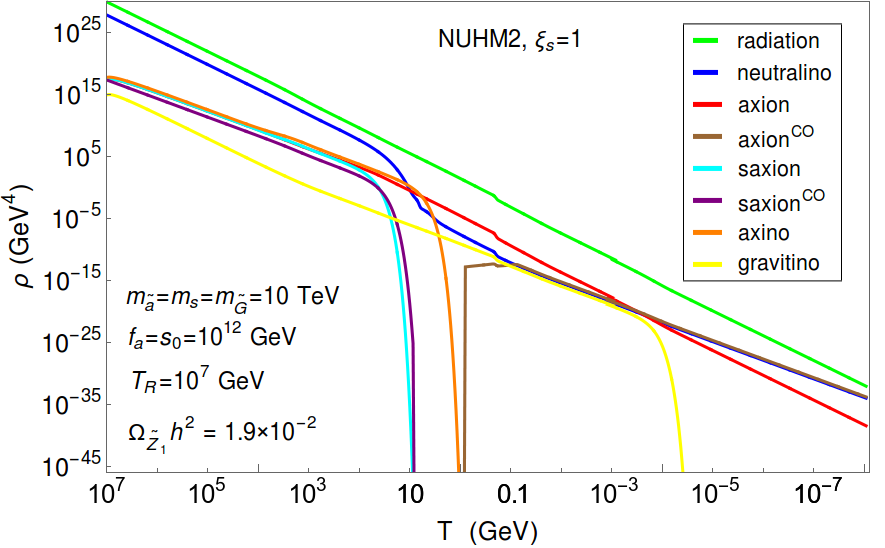}\\
\caption{A plot of various energy densities $\rho$ vs. 
temperature $T$ starting from $T_R=10^7$ GeV until the era of entropy conservation from our eight-coupled Boltzmann equation solution to the mixed
axion-neutralino relic density in the SUSY DFSZ model for 
a natural SUSY benchmark point. We take $\xi_s=1$.
\label{fig:rhoT}}
\end{center}
\end{figure}

To calculate the relic density of mixed axion-WIMP dark 
matter\cite{Baer:2011hx}, 
now one must solve eight coupled Boltzmann equations starting at the 
temperature of re-heat (at the end of inflation) $T_R$ until the era of 
entropy conservation\cite{Baer:2011uz}. 
The coupled Boltzmann equations track the energy densities
of radiation (SM particles), WIMPs, axinos, saxions, gravitinos and axions.
Tracking of coherent oscillation (CO) produced axions and saxions and thermal
and decay produced axions and saxions are treated separately.
The results of such a calculation for the SUSY DFSZ model\cite{Bae:2013hma,Bae:2014rfa} 
are shown in Fig. \ref{fig:rhoT} from Ref. \cite{Bae:2017hlp}.

\begin{figure}
\begin{center}
{\includegraphics[width=9.0cm]{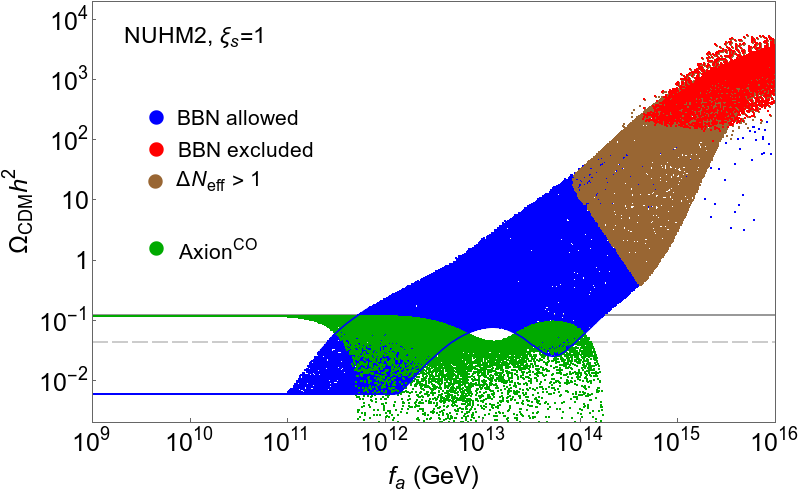}}
\caption{We plot the relic density of DFSZ axions and 
higgsino-like WIMPs from a natural SUSY 
benchmark model using a scan over PQMSSM parameters in the 
SUSY DFSZ axion model. The dashed line corresponds to 50$\%$
of observed Dark Matter relic density.
\label{fig:relic}}
\end{center}
\end{figure}

As the PQ-breaking scale $f_a$ increases, then presumably CO-produced axion
abundance increases although this can be compensated for by a small
axion mis-alignment angle. However, as $f_a$ increases, then axinos
and saxions produced in the early universe decay after WIMP freeze-out and 
give non-thermal contributions to both the WIMP and axion abundance.
At too large of $f_a$ values, then mixed WIMP-axion dark matter is overproduced.
The result of such a calculation from a scan over PQ parameters 
is shown in Fig. \ref{fig:relic}. The green dots correspond to
the axion relic density while the blue dots correspond to the WIMP
relic density. The brown and red dots are excluded by dark radiation
constraints ($\Delta N_{eff}>1$) and BBN constraints, respectively.
Values of $f_a\agt 10^{14}$ GeV are completely excluded by overproduction
of WIMP dark matter. 
%\textbf{Explain the dip at $f_a$ = $10^{13}$ GeV}
%

\subsection{Direct higgsino-like WIMP searches}

\begin{figure}[tbp]
\includegraphics[height=0.3\textheight]{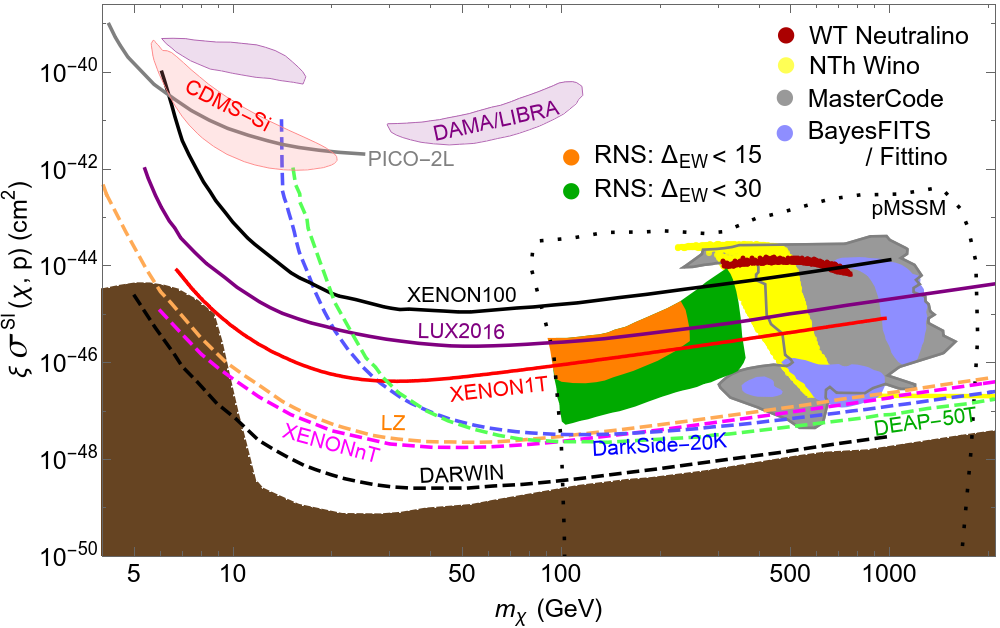}
\caption{Plot of rescaled spin-independent WIMP detection rate $\xi \sigma^{SI}(\chi, p)$ 
versus $m_{\chi}$ from several published results versus current and 
future reach (dashed) of direct WIMP detection experiments. 
$\xi =1$ ({\it i.e.} it is assumed WIMPs comprise the totality of DM) 
for the experimental projections and for all models {\it except} RNS and pMSSM. 
The brown region shows the so-called neutrino floor.
\label{fig:SI}}
\end{figure}

Even if higgsino-like WIMPs may make up only a fraction of the dark matter, 
they still may be detected by spin-independent (SI) 
WIMP direct detection (DD) experiments.
In fact, their coupling to Higgs $h$ turns out to be a product of 
gaugino times higgsino components and is never small since while the WIMPs
are mainly higgsino, the naturalness requirement keeps the gaugino component
from never going to zero. However, detection rates must be multiplied
by the factor $\xi\equiv \Omega_{\tz_1}h^2/0.12$ since now there are fewer
WIMPs floating around as they make up only a {\it portion} of the dark matter. 
The rates for SI DD are shown in Fig. \ref{fig:SI} for 
radiatively-driven natural SUSY (RNS) along with projections from several 
other SUSY models (updated from Ref.~\cite{Baer:2016ucr}). 
The predicted theory rates are compared against
current WIMP detection limits (solid lines) and future projected limits
(dashed lines). While current limits only exclude a portion of natural SUSY 
parameter space (orange and green regions), 
the entire natural SUSY parameter space will be explored ultimately
by multi-ton noble liquid SI DD experiments. Thus, if no signal is seen
by multi-ton SI DD experiments, this basic natural SUSY scenario will be 
ruled out.

The spin-dependent (SD) DD experiments can also probe natural 
SUSY parameter space, but must also be multiplied by the fractional relic
density parameter $\xi$. Current limits from IceCube barely touch the 
natural SUSY parameter space. 
Future experiments such as  Xenon-nton\cite{Aprile:2015uzo}, LZ\cite{Akerib:2015cja} and
PICO-500\cite{Amole:2015cca} will probe only a small portion of natural SUSY parameter 
space. For plots, see Ref.~\cite{Baer:2016ucr}.

Finally, we remark here that the DAMA/LIBRA annual modulation signal
and also the gamma-ray excess from the galactic center hint at rather light
WIMPs in the 10 GeV regime\cite{Gelmini:2016emn}.
Such light WIMPs are difficult to reconcile with natural SUSY where the
higgsino-like WIMP is required in the 100-350 GeV regime. 
If such light WIMPs exist, they should soon be revealed by a bevy of direct 
(as shown in Fig. \ref{fig:SI}), indirect and collider search experiments.

\subsection{Indirect higgsino-like WIMP searches}

It is also possible to search for WIMP-WIMP annihilation into $\gamma$s
and anti-matter at various indirect WIMP detection (IDD) experiments
such as Fermi-LAT, HESS, CTA and AMS-II. The theory projections for 
these searches must all be rescaled by a factor of $\xi^2$ since now 
one is looking for WIMP-WIMP annihilation. The $\xi^2$ factor typically
moves the theory projections to regions well below projected sensitivities
of the various ID experiments (see Fig. 3 of Ref. \cite{Baer:2016ucr}).

\subsection{SUSY DFSZ axion searches}

A further possibility for dark matter detection in SUSY models with a 
DFSZ solution to the strong CP and SUSY $\mu$ problems is the 
detection of relic axions. Microwave cavity experiments are currently making
inroads in the $m_a$ vs. $g_{a\gamma\gamma}$ (axion-photon effective coupling)
parameter space. 
The idea here is that relic axions can interact with microwave photons in a super-cooled microwave cavity chamber, and then convert to photons with energy
equal to the axion mass. One then searches for bumps in the photon spectra
within the cavity.

Usually experiments such as ADMX plot their reach results in the $m_a$ vs.
$g_{a\gamma\gamma}$ plane vs. the KSVZ and (non-SUSY) DFSZ axion models. 
However, in the case of SUSY DFSZ assumed here, the higgsinos also circulate
in the $a\gamma\gamma$ anomaly loop. Since the higgsinos necessarily have 
opposite-sign PQ charge from matter fermions, they will cancel against
SM triangle diagrams in the $a\gamma\gamma$ coupling\cite{Bae:2017hlp}. 
Along with the anomaly contribution to the $a\gamma\gamma$ coupling, there is
a chiral contribution depending on the up- and down-quark masses.
In the SUSY DFSZ model, there is a nearly complete cancellation between these two contributions so that the $g_{a\gamma\gamma}$ coupling is highly suppressed.
Also, one must multiply by the fractional axion abundance $\xi_a\equiv \Omega_ah^2/0.12$.

\begin{figure}[tbp]
\begin{center}
\includegraphics[height=0.3\textheight]{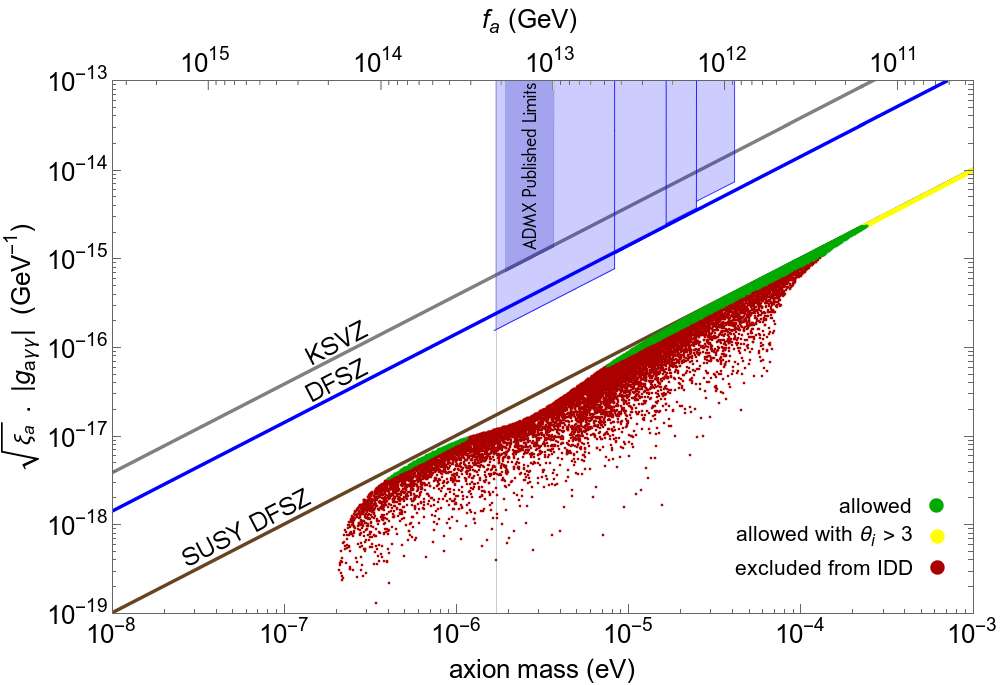}
\caption{Axion detection rates at microwave cavity experiments in terms of the
axion coupling $|g_{a\gamma\gamma}|$ vs. $m_a$. The vertical axis includes a factor
$\sqrt{\xi_a}$ where $\xi_a\equiv \Omega_ah^2/0.12$ to account for the depleted abundance of axions.
The green points are allowed from natural SUSY while red points are excluded by Fermi-LAT
constraints on higgsino-like WIMP annihilation into gamma rays. We also plot lines of 
SUSY and non-SUSY coupling strengths and current and projected ADMX search regions.
The yellow dots are regarded as unnatural since they would require an
axion misalignment angle $\theta_i>3$.
\label{fig:axion}}
\end{center}
\end{figure}

The situation is shown in Fig. \ref{fig:axion}\cite{Bae:2017hlp}.
There, we see that the SUSY DFSZ axion model line is well below current
ADMX limits, thus rendering at least for now the SUSY DFSZ axion as 
{\it back to invisible}. The green dots show the allowed theory prediction
from a scan over NUHM2 model space. 
Some range of $m_a$ (and correspondingly $f_a$) is already excluded by
WIMP IDD)! 
This occurs for large enough $f_a$ values that non-thermal 
production of WIMPs occurs due to late time axino and saxion decays. 
Then the models have large $\xi (WIMP)$ values and actually are
excluded by Fermi-LAT searches.

\section{Scenarios for baryogenesis in natural SUSY}
\label{sec:baryo}

One of the main mysteries of particle physics concerns how the
matter-antimatter asymmetry arose in the early universe. Starting
with Big Bang cosmology, the goal is to explain one number: 
the baryon-to-photon ratio
\be
\eta_B\equiv\frac{n_B}{n_\gamma}\simeq(6.2\pm 0.5)\times 10^{-10}\ \ \ \ (95\%\ CL).
\ee
Production of the baryon asymmetry of the universe or BAU requires
mechanisms which satisfy Sakharov's three criteria: 1. baryon number violation, 
2. $C$ and $CP$ violation and 3. a departure from thermal equilibrium.
In the SM, it is possible to generate the baryon asymmetry via a
first order electroweak phase transition, 
but only if the Higgs mass $m_H\alt 50$ GeV, which is obviously excluded. 
Thus, to produce the measured BAU, new physics is required.

SUSY theories offers a number of different mechanisms for generating the BAU.
These include:
\bi
\item Electroweak baryogenesis: for a strong enough first order EW phase
transition, then evidently $m_h\alt 113$ GeV with $m_{\tst_R}\alt 115$ GeV
is required unless very heavy values of $m_A\agt 10$ TeV are allowed.
Such heavy $m_A$ values violate our naturalness conditions where
$m_A\simeq m_{H_d}$ and from Eq. \ref{eq:mzs} then $m_{H_d}\alt \sqrt{m_Z^2/2}\tan\beta\alt 4-8$ TeV\cite{cw}. 
Thus, we expect EW baryogenesis in SUSY to be highly implausible.
\item Thermal leptogenesis (THL)\cite{thl,thl_susy,thl_reviews}: 
this mechanism occurs if right-hand-neutrinos
can be thermally produced at re-heat temperatures $T_R\agt 1.5\times 10^9$ GeV,
just below upper limits of $T_R\alt 10^{10}$ GeV to avoid overproduction 
of gravitinos, and consequent overproduction of dark matter or disruptions in 
Big Bang nucleosynthesis (BBN). The thermally produced right-hand neutrinos (RHNs) 
would decay asymmetrically into leptons versus antileptons and then the lepton
asymmetry would wash into the baryon asymmetry via sphaleron effects.
\item Non-thermal leptogenesis (NTHL)\cite{nthl}: Here it is assumed the production of
RHNs takes place via inflaton decay in the early universe. In this case,
re-heat temperatures of just $T_R\agt 10^6$ GeV are required.
\item Leptogenesis from oscillating sneutrino decay (OSL)\cite{osl}: 
In this case, the heavy {\it sneutrinos} are produced via coherent oscillations 
and their decays generate the lepton asymmetry which is again washed 
into the baryon asymmetry via sphalerons.
\item Affleck-Dine leptogenesis (ADL)\cite{adl}: 
Usual Affleck-Dine baryogenesis\cite{ad} is 
afflicted by various problems such as $Q$-ball production. 
However, Affleck-Dine {\it leptogenesis}\cite{adl} is perfectly viable. 
ADL uses the $LH_u$ flat direction in the SUSY scalar potential 
to generate a condensate carrying non-zero lepton number. 
The condensate oscillates and then decays asymmetrically
to generate the lepton asymmetry which is again converted to the 
baryon asymmetry via the sphaleron.
\ei

\begin{figure}[tbp]
\begin{center}
\includegraphics[height=0.29\textheight]{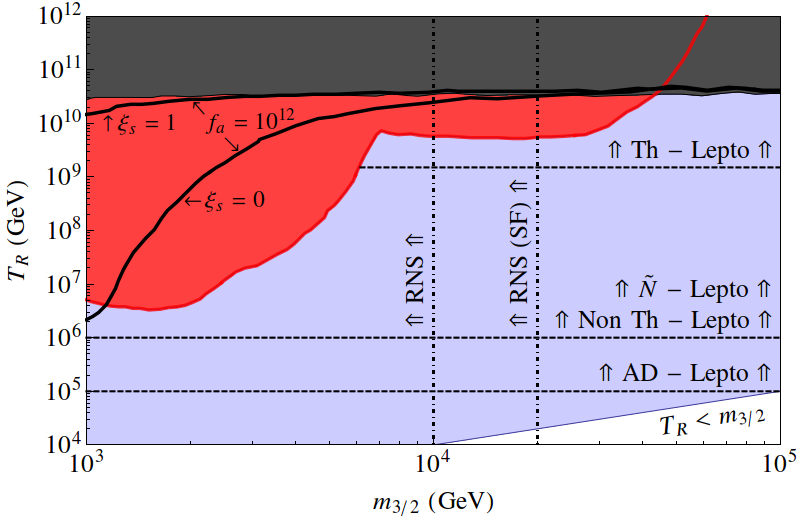}\\
\caption{
Plot of allowed regions in $T_R$ vs. $m_{3/2}$ plane in the SUSY DFSZ axion 
model for $f_a=10^{11}$ and $10^{12}$ GeV
for $\xi_s =0$.
For $f_a=10^{11}$, $T_{R}>10^{11}$ is forbidden to avoid PQ symmetry restoration.
We take $m_s=m_{\ta}\equiv m_{3/2}$ (from Ref. \cite{yfz}).
\label{fig:baryo}}
\end{center}
\end{figure}

For natural SUSY models with a $\mu$ parameter generated via the SUSY DFSZ
axion sector, then the baryon asymmetry relies on the SUSY 
soft breaking scale $m_{3/2}$, the re-heat temperature $T_R$ and the PQ sector
parameters such as $f_a$, the axino mass $m_{\ta}$, the saxion mass $m_s$
and $\xi_s$ which governs whether the saxion decays to axinos and axions
($\xi_s=1$) or not ($\xi_s =0$). The viable regions for the different 
mechanisms are shown in Fig. \ref{fig:baryo}\cite{yfz} for $f_a=10^{11}$ GeV or
$10^{12}$ GeV and for $\xi_s=0$ or 1. The upper black-shaded region is excluded
by overproduction of WIMPs from gravitino decay. The red-shaded region is 
excluded by disruption of BBN. The various allowed regions are labeled
as are the regions that accommodate radiatively-driven natural SUSY (RNS)
with universal or split families (SF). From the plot, it can be seen 
that only a small region of THL is allowed, but in contrast large regions of
parameter space are allowed for successful baryogenesis via NTHL, ADL or OSL.
Finally, for the lower-right region with $T_R<m_{3/2}$, then none
of the examined mechanisms would apply and perhaps some sort of
alternative baryogenesis mechanism would be required
(see {\it e.g.} Ref.~\cite{Cui:2015eba} for a WIMP baryogenesis alternative).

\section{Conclusions}
\label{sec:conclude}

In this midi-review, we have sought to outline the status of weak scale
supersymmetry\cite{WSS} as it stands after LHC13 Run 2 with 139 fb$^{-1}$ of data
and after first results from ton-scale noble liquid direct WIMP searches.
At first sight, the lack of WIMP signals along with the seemingly severe 
sparticle mass limits from LHC, as compared to early naturalness-derived 
upper bounds on sparticle masses, has led much of the HEP community 
to a rather pessimistic attitude towards the vitality  
of the weak scale SUSY paradigm.

However, as noted in the Introduction, the latest experimental limits are 
usually compared against an early cartoonish picture as to how weak scale
SUSY would manifest itself. 
Several important developments in the 21st century
have required a change in the WSS paradigm. These include:
\bi
\item a clarification of the notion of weak scale naturalness in SUSY 
(a summary Table \ref{tab:Delta} is provided which presents each naturalness
measure, its definition, motivation and some principle consequences),
\item the influence of including a (axionic) solution to the strong CP
problem into the SUSY paradigm
\item the emergence of discrete $R$-symmetries and their role in the SUSY $\mu$ problem, suppression of proton-decay, and as a source for the emergence of
the accidental, approximate $R$-parity and gravity-safe global PQ symmetry,
\item the emergence of the string theory landscape and its role in solving the
cosmological constant problem and setting the scale for SUSY breaking
and electroweak symmetry breaking, its role in solving the SUSY flavor 
and CP problems, and the implications of stringy naturalness.
\ei
\begin{table}[!htb]
\renewcommand{\arraystretch}{0.8}
\begin{center}
\begin{tabular}{|c|c|c|c|}
\hline
measure & definition & motivation & consequences \\
\hline
 $\Delta_{BG}$ & $max_i |\frac{p_i}{m_Z^2}\frac{\partial m_Z^2}{\partial p_i}|$ & 
measure fine-tuning of & favors soft terms \\
 & & effective theory parameters $p_i$ & at or around $m_{weak}$\\
 & & to obtain measured $m_Z$ & \\
\hline
$\Delta_{HS}$ & $\delta m_{H_u}^2/m_h^2$ & require small change in & $m_{\tst_{1,2}},m_{\tb_1}\alt 500$ GeV \\
 & & running contribution & small $A_t$ \\
 &  & to $m_h$ & \\
\hline
 $\Delta_{EW}$ & $|largest\ cont.\ to\ m_Z^2/2|$/ $m_Z^2/2$ & 
parameter indep. measure & require $\mu\sim 100-300$ GeV and \\
 & & based on practical & highly mixed stops $m_{\tst_1}\alt 3$ TeV; \\
 & & naturalness  & allow radiatively-driven naturalness \\
\hline
$stringy$ & largest contribution & string landscape & pull sparticles \\
 & to $m_{weak}^{PU}$ $<(2-5)m_{weak}^{OU}$ & & beyond LHC limits\\
 &  with power-law draw & &  with $m_h\to 125$ GeV; \\
 & to large soft terms & & radiatively-driven naturalness \\
\hline
\end{tabular}
\caption{Summary of naturalness measures along with definition, 
motivation and some principle consequences. The superscripts $PU$ stands
for {\it pocket universe} while $OU$ stands for {\it our universe}.
Tabular formatting precludes us from adding a fifth column on 
{\it limitations} for each measure, so we include this information here.
$\Delta_{BG}$: parameter, scale and model dependent; 
$\Delta_{HS}$: oversimplified, breaks measure into {\it dependent} terms;
$\Delta_{EW}$: model-independent within MSSM, but may require additional 
terms for extended models;
$stringy$: depends on string multiverse/landscape paradigm. 
}
\label{tab:Delta}
\end{center}
\end{table} 

We presented here a midi-review of recent work that seeks to update the WSS
paradigm by addressing these concerns. 
The emergent picture of weak scale BSM physics includes the following.
\bi
\item Retention of WSS to stabilize the Higgs mass and retain the
successful agreement between virtual effects within the MSSM and 
data including 
1. measured strengths of weak scale gauge couplings and 
gauge coupling unification within the MSSM,
2. the measured value of $m_t$ and its role in seeding a radiative breakdown of EW symmetry,
3. the measured value of $m_h\simeq 125$ GeV and its consistency with MSSM
predictions including highly-mixed, TeV-scale top squarks and 
4. precision EW measurements of $m_W$ vs. $m_t$ which favor soft terms
in the multi-TeV range.
\item Requirement of {\it practical naturalness} wherein weak scale SUSY
contributions to the magnitude of the weak scale are comparable 
to the weak scale. This requires the SUSY-conserving $\mu$ parameter
not too far from $m_{weak}\sim m_{W,Z,h}\sim 100$ GeV while soft SUSY breaking 
terms, which enter the weak scale via loop-supressed terms, can range into 
the TeV or even tens of TeV regime. The higgsinos are then the lightest
superpartners and one expects a mainly higgsino-like LSP. This has major
consequences for both collider and dark matter signatures.
\item Inclusion of a (gravity-safe) PQ sector to solve the strong CP problem.
This may involve a Kim-Nilles solution to the SUSY $\mu$ problem with a 
Little Hierarchy $\mu\ll m_{soft}$ which is still natural. The gravity-safe
$U(1)_{PQ}$ and $R$-parity could both emerge from a more fundamental
anomaly-free discrete $R$-symmetry such as ${\bf Z}_{24}^R$ which in turn
is interpreted as the discrete remnant of compactification of 10-d 
Lorentz symmetry down to 4-dimensions. The discrete $R$ symmetry also 
plays a role in suppressing dangerous dimension 5 proton decay operators.
\item The inclusion of the string landscape allows for Weinberg's
anthropic solution to the cosmological constant problem. Under rather
general stringy considerations, the landscape should also 
statistically favor soft SUSY breaking terms as large as possible subject
to the condition that contributions to the weak scale are comparable
to the weak scale (within a factor 2-5\cite{Agrawal:1997gf}). This leads to a 
statistical pull on $m_h\to 125$ GeV whilst pulling most sparticle masses
to beyond LHC limits\cite{Baer:2017uvn}. 
In fact, under stringy naturalness, a 3 TeV gluino is more natural
than a 300 GeV gluino\cite{Baer:2019cae}!
The exceptions to TeV-level sparticle masses 
are the light higgsinos whose mass
term is SUSY conserving and arises from whatever mechanism solves the
SUSY $\mu$ problem (such as the gravity-safe hybrid CCK models based on
${\bf Z}_{24}^R$ symmetry).
\ei

While the emergent WSS paradigm includes solutions to a host of problems
which were typically previously neglected, it also leads to new 
collider signatures. While an LHC upgrade to at least $\sqrt{s}\sim 27$ 
TeV may be needed to access gluinos and top squarks, a corroborative 
signature emerges in SUSY with light higgsinos; the ultimate appearance of 
same-sign $W$-boson pairs arising from wino pair production followed by
decay to higgsinos. However, the most lucrative signature for
natural landscape SUSY appears to be the soft OS/SF dilepton plus jet
signature arising from direct higgsino pair production\cite{Baer:2011ec}. 
HL-LHC may be able to explore a sizable chunk of natural SUSY parameter
space via this novel signature, which should slowly emerge as more and
more integrated luminosity accrues. The OS/SF dilepton invariant mass is
bounded by the inter-higgsino mass gap
$m(\ell^+\ell^-)< m_{\tz_2}-m_{\tz_1}\sim 5-10$ GeV which makes for challenging
searches for very soft dileptons at ATLAS and CMS.

In the updated WSS paradigm, we can also understand why WIMPs have not yet 
been detected. We now expect mixed axion with higgsino-like-WIMP dark matter
where the WIMPs typically make up only 10-20\% of the dark matter whilst
axions make up the remainder. Multi-ton noble liquid dark matter detectors
will be needed to probe the entire predicted parameter space.
Indirect WIMP detection seems rather unlikely in the near future since
detection rates are suppressed by the square of the fractional WIMP abundance.
Axion detection via microwave cavity experiments also seem unlikely in 
the near-term since the presence of higgsinos in the $g_{a\gamma\gamma}$ 
coupling leads to cancellations and consequently suppressed axion couplings to
photons\cite{Bae:2017hlp}.

Overall, the updated weak scale SUSY paradigm-- as manifested in 
natural landscape SUSY--  predicts that LHC at this time should see
a Higgs boson with $m_h\sim 125$ GeV but as yet no signals from sparticles.
Indeed, updated experimental facilities-- a higher energy LHC 
with $\sqrt{s}\sim 27-100$ TeV and/or a
$\sqrt{s}>2m(higgsino)$ linear collider may be needed for SUSY discovery.
As well, we may have to await a full exploration of relic WIMP
parameter space by multi-ton noble liquid detectors for 
verification or falsification of the presence of WIMPs from weak scale SUSY.

{\it Acknowledgments:} 

We thank our many collaborators for their dedicated contributions to 
this midi-review. 
This material is based upon work supported by the U.S. Department of Energy, 
Office of Science, Office of High Energy Physics
under Award Number DE-SC-0009956.

%%%%%%%%%%%%%%%%%%%%%%%%%%%%%%%%%%%%%%%%%%%%%%%%%%%%%%

%
\end{document}